\documentclass[useAMS,usenatbib]{mn2e}
\usepackage{psfig}
\usepackage{color}
\definecolor{turquoise}{rgb}{0.1,0.5,0.5}

\newcommand{\Mpc}{\rm\,Mpc}
\newcommand{\Msol}{\rm\,M_{\odot}}

\newcommand{\gtrsim}{\rm\,\ga}
\newcommand{\ncenexp}{\rm\,\langle N_{cen}(M)\rangle}
\newcommand{\nsatexp}{\rm\,\langle N_{sat}(M)\rangle}

\title[The red galaxy fraction as a function of environment]{More than just halo mass: Modelling how the red galaxy fraction depends on multiscale density in a HOD framework}
\author[S. Phleps, D.~J. Wilman, S. Zibetti, and T. Budav{\'a}ri]{S. Phleps$^{1}$\thanks{E-mail: sphleps@mpe.mpg.de}, D.~J. Wilman$^{1,2}$, S. Zibetti$^{3,4}$, and T. Budav{\'a}ri$^{5}$\\
$^{1}$Max-Planck-Institut f{\"u}r extraterrestrische Physik,  Giessenbachstra{\ss}e,
85748 Garching, Germany\\
$^{2}$Universit{\"a}tssternwarte M{\"u}nchen, Ludwig-Maximilians-Universit{\"a}t, Scheinerstra{\ss}e 1, 81679 M{\"u}nchen, Germany\\
$^{3}$Dark Cosmology Centre, University of Copenhagen, Juliane Maries Vej 30, 2100 Copenhagen, Denmark \\
$^{4}$INAF -- Osservatorio Astrofisico di Arcetri, Largo Enrico Fermi 5, 50125 Firenze, Italy\\ 
$^{5}$Department of Physics and Astronomy, The Johns Hopkins University, 3701 San Martin Drive, Baltimore, MD 21218, USA\\
}

\begin{document}

\date{Accepted 2013 December 3. Received 2013 December 03; in original form 2013 February 8}

\pagerange{\pageref{firstpage}--\pageref{lastpage}} \pubyear{2013}

\maketitle

\label{firstpage}

\begin{abstract}
The fraction of galaxies with red colours depends sensitively on environment, and on the way in which environment is measured. To distinguish competing theories for the quenching of star formation, a robust and complete description of environment is required, to be applied to a large sample of galaxies. The environment of galaxies can be described using the density field of neighbours on multiple scales -- the {\it multiscale density field}. We are using the Millennium simulation and a simple HOD prescription which describes the multiscale density field of Sloan Digital Sky Survey DR7 galaxies to investigate the dependence of the fraction of red galaxies on the environment. Using a volume limited sample where we have sufficient galaxies in narrow density bins, we have more dynamic range in halo mass and density for satellite galaxies than for central galaxies. Therefore we model the red fraction of central galaxies as a constant while we use a functional form to describe the red fraction of satellites as a function of halo mass which allows us to distinguish a sharp from a gradual transition. While it is clear that the data can only be explained by a gradual transition, an analysis of the multiscale density field on different scales suggests that colour segregation within the haloes is needed to explain the results. We also rule out a sharp transition for central galaxies, within the halo mass range sampled.
\end{abstract}

\section{Introduction}
It has long been known that the properties of galaxies (e.g. colour, luminosity, star-formation rate, gas content, metallicity, morphology, ...) depend on their environment (\citealp{Abell65,Melnick77,Dressler80,Goto03,Kauffmann04,Balogh04,Wilman05,Blanton05,Croton05,Baldry06,Cooper06,Park07,Elbaz07,Ball08,Wilman08,Cowan08,OMill08,Wilman09,Tasca09,Ellison09,Peng10,Peng12,Hoyle12,Wilman12}). It is also established that there is a clear bimodality in the colour of galaxies (see e.g. \citealp{Strateva01,Hogg02,Blanton03,Bell04}). While red, early type galaxies (the spectra of which are dominated by old stars) tend to live in denser environments and inhabit more massive dark matter halos, late-type, star-forming and therefore blue galaxies live primarily in the lower density ``field'' regions. The reason for this environment/halo mass dependent colour segregation is however not yet clear. The quenching of star formation in galaxies requires that gas accretion is suppressed. This can happen after mergers and during feedback from an AGN; when a galaxy experiences stripping of its cold and/or hot gas components; or, it might happen that the accretion of gas and its ability to cool onto the galaxy and subsequent form stars is suppressed for some other reason. Whichever mechanism(s) are acting, they result in the observed correlations between galaxy properties and environment.

A related, practical problem is that the term ``environment'' is not well-defined. Some authors construct group catalogues to investigate properties as a function of host halo mass, while others measure their dependence on local galaxy overdensity. Overdensity can be measured on different scales or quantified within the distance to the $n^{th}$ nearest neighbour (for a review and comparison of the different methods see e.g.  \citet{Haas12} and \citet{Muldrew12}, and references therein). These are all local measurements computed for each galaxy. 

In contrast, the spatial correlation function measures the expectation value of the number of galaxy pairs with a given separation (scale) relative to a random distribution for a complete sample of galaxies. To examine the dependence on any galaxy property, the sample is first binned on that property.

None of these methods can simultaneously distinguish between the influence on galaxy properties of environment on different scales, while probing the full distribution of environment which is computed for each individual galaxy. \citet{Wilman10} (hereafter WZB10) have proposed a new, practical method to disentangle the effects of galaxy density on different scales on the fraction of red galaxies.

Annuli of variable inner and outer radii are centred on each galaxy in a volume limited sample, and the surface density of neighbouring galaxies within these annuli and within $\delta v=1000$\,km~s$^{-1}$ is measured. They then explore the dependence of the red galaxy fraction on these multiscale densities. Measuring the densities on different scales and combining them opens up another dimension and can give insight into the underlying covariances.

In this paper we use the halo model as a tool to investigate the dependence of the fraction of red galaxies on the distributions of multiscale density as explored by WZB10. In its most general form, the halo model assumes that halos can be populated with galaxies in a predictable fashion which only depends on simple halo parameters (e.g. mass). This  can explain the observed dependencies of clustering statistics on galaxy properties. Models of the halo occupation distribution (HOD in the following) start by dividing the galaxy population into two: those sitting in the centre of a halo, dubbed ``central galaxies'', and those which have been accreted onto the halo, and are now orbiting within it, called ``satellite galaxies''. Differences between the properties of the two populations can be expected, since central galaxies have always lived at the bottom of their potential well, whereas satellites are subject to stripping processes. Different processes are likely to be actively quenching star formation in these two galaxy types. For a review of HOD models, see \citet{Cooray02}.

The halo model requires the spatial distribution of dark matter halos; the halo mass function and bias; the HOD; and the spatial and velocity distribution of galaxies within halos. The HOD describes the expected number of galaxies (central and satellite) per halo using a parameteric form as a function of host halo mass.

In this paper, we utilize a halo catalogue extracted from the $N$-body Millennium simulation \citep{Springel05}. This circumvents problems establishing halo catalogues on small (non-linear) scales which is a difficult analytic problem. An analytic description of the halo bias, mass function and clustering would facilitate a faster computation of the model, however, it is currently not possible to model the highly nonlinear scales we are interested in accurately enough even with renormalised perturbation theory (\citealp{Crocce06a,Crocce06b,Jeong09}). The Millennium simulation also provides a catalogue of associated subhalos which we use to distribute the galaxies within halos. We then populate halos with (red and blue) galaxies according to the HOD model of \citet{Zehavi11}, and while keeping the fraction of red central galaxies per halo constant, describe the fraction of red satellite galaxies using a simple function with two free parameters we can fit, a transition mass and a slope.

Using the halo model to investigate the dependence of the red galaxy fraction on multiscale density, we are able to distinguish a scenario where the red fraction of satellites makes a sharp transition as a function of halo mass from one which makes a more gradual transition. This provides important constraints for physical models of galaxy formation and evolution. By adopting a model in which the red galaxy fraction is merely a function of halo mass we are testing the simplest hypothesis. However, we will show that information on different scales can provide both information on the dependence on halo mass and constrain the detailed way in which environment tracks the residual red fraction after accounting for this halo mass dependence.

This paper is structured as follows: In Section \ref{Method} we explain in more detail how density on different scales is measured, and the two-dimensional parameter spaces we use to characterize the density field. In Section \ref{Data} we describe the sample of SDSS galaxies we use for the study. In Section \ref{Model} we describe how we populate halos with galaxies. To do this we use the functional form of the HOD by \citet{Zehavi11}, for which we derive the appropriate parameters for the galaxies in our sample from their analysis of the clustering of SDSS galaxies; this is described in detail in Appendix \ref{Appendix}.Once the haloes have been populated with galaxies, in Section \ref{redgalfracsection} we parametrise the fraction of red galaxies in the model as a function of halo mass using a tanh function with two free parameters. This recipe is used to allocate colours to galaxies. We can then compute the red fraction as a function of multiscale density for models with different parameters. We go on to discuss the feasibility of alternative results, highlighting the regions of our parameter space which rule them out. In Section \ref{Discussion} we discuss our results in the context of the literature, and room for future improvements. Finally, we conclude in Section \ref{Conclusion}. Throughout the paper we assume the cosmological parameters: $(\Omega_m,\Omega_\Lambda,h)=(0.238,1-\Omega_m,0.75)$.

\section[]{Method}\label{Method}
It is a matter of debate whether halo mass or galaxy (over-) density on sub- or even super-halo scales drives galaxy evolution and the quenching of star formation. This is likely to be different for central and satellite galaxies. The scale-dependence of quenching can be hard to disentangle by comparing ``one-dimensional'' quantities like estimated halo mass or measured overdensity. Using multi-scale densities, WZB10 find no evidence for any {\it positive} correlation of red fraction with density on scales $>1$\,Mpc, once density on smaller scales are taken into account, consistent with results from \citet{Kauffmann04,Blanton06,Blanton07}.

\citet{Reid09} used a technique based on counts-in-cylinders to constrain the contribution of satellites to the HOD of luminous red galaxies (LRGs) in SDSS \citep{Eisenstein01}. We take this approach a step further, using {\it complementary scale pairs} of counts-in-cylinders (or rather, cylindrical annuli) to investigate the dependence of the fraction of red (central and satellite) galaxies on halo mass. More detailed information on environment and its influence on galaxy properties is available using {\it densities on different scales}.

We use the multiscale approach first outlined by WZB10. They select complementary pairs of annular scales (a circular inner aperture and an adjacent but non-overlapping outer annulus) which are used to trace the density field on these scales and the covariance between them. Within this parameter space, they explore how the fraction of red galaxies depends on multiscale density, using a volume-limited sample with $-21.5\leq M_r\leq -20.0$ selected from SDSS-DR5. We note that galaxies living in different topological regions (e.g. clusters, voids, filaments...) live in different parts of the multiscale density parameter space. This allows one to e.g. separate galaxies living in the outskirts of clusters from those living in thin but dense filaments. In the following we will describe in a bit more detail how the density-density histograms are measured.

\subsection{Multiscale Density}\label{Multiscales}
In order to measure density on multiple scales, cylindrical annuli are centered on each galaxy in the sample, and the local projected number densities of neighbouring galaxies inside these are measured. For an annulus with inner radius $r_i$ and outer radius $r_o$, the surface density $\Sigma_{r_i,r_o}$ is given by
\begin{eqnarray}\label{Sigma}
\Sigma_{r_i,r_o}=\frac{N_{r_i,r_o}^n}{\pi(r_o^2-r_i^2)}~,
\end{eqnarray}
where $N_{r_i,r_o}^n$ is the (where necessary completeness corrected) number of neighbours brighter than our limit, within the restframe velocity range $\pm \delta v$ and with a projected distance $r$ from the primary galaxy, where $r_i \leq r\leq r_o$. As in WZB10 we choose neighbours brighter than the volume limited sample depth of $M_r=-20.0$ and within $\pm \delta v=1000$\,km~s$^{-1}$ (which corresponds roughly to a redshift space cylinder of length $\pm 14$\,Mpc at the redshift under consideration). We calculate projected surface densities for the following annuli: $(r_i,r_o)= (0.,0.5), (0.5,1.), (0.,1.), (1.,2.), (0.,2.), (2.,3.), (0.,3.), (3.,5.)$ Mpc. In order to understand how galaxy properties depend on the overdensities measured on different scales, we combine pairs of non-overlapping annuli, in which case the densities  are only correlated due to correlations in the underlying dark matter density field, not because the same galaxies are counted multiple times. For each pair of annular scales, $(r_i,r_o)^1$ and $(r_i,r_o)^2$, we divide the galaxies into two-dimensional logarithmically-spaced bins of density $\Sigma_{{r_i,r_o}}^1$ and $\Sigma_{{r_i,r_o}}^2$. In this paper we choose the following pairs: $\left[(r_i,r_o)^1,(r_i,r_o)^2\right]=\left[(0.,0.5), (0.5,1.)\right], \left[(0.,1.), (1.,2.)\right], \left[(0.,2.), (2.,3.)\right]$, and $\left[(0.,3.), (3.,5.)\right]$.  For each 2d density bin for each of these pairs, we can compute the (completeness-weighted) fraction of red real and model galaxies, which can then be used to test the ability of our model to reproduce the observed population.

\begin{figure*}
\centerline{\psfig{figure=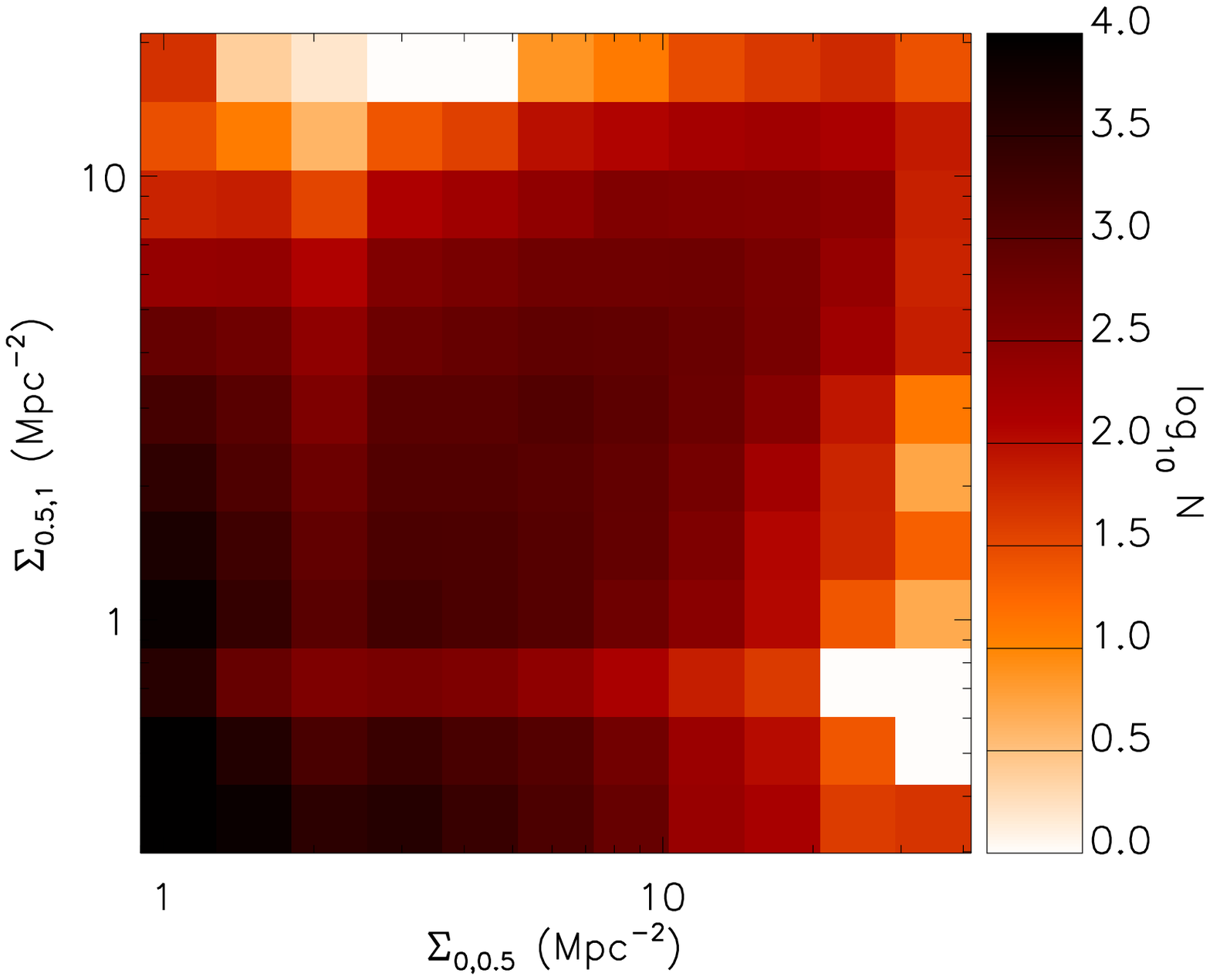,clip=t,width=8.cm}\psfig{figure=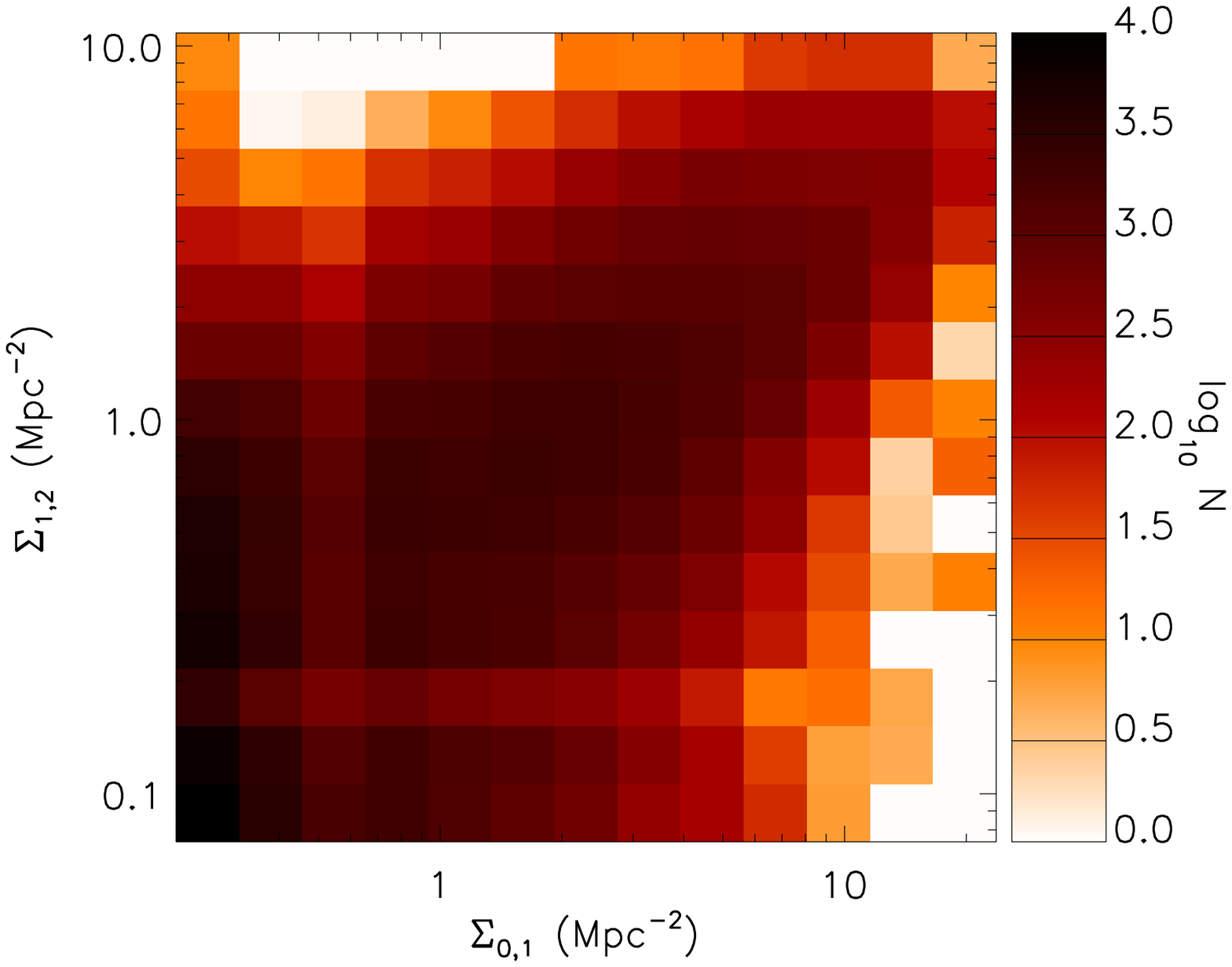,clip=t,width=8.cm}}
\centerline{\psfig{figure=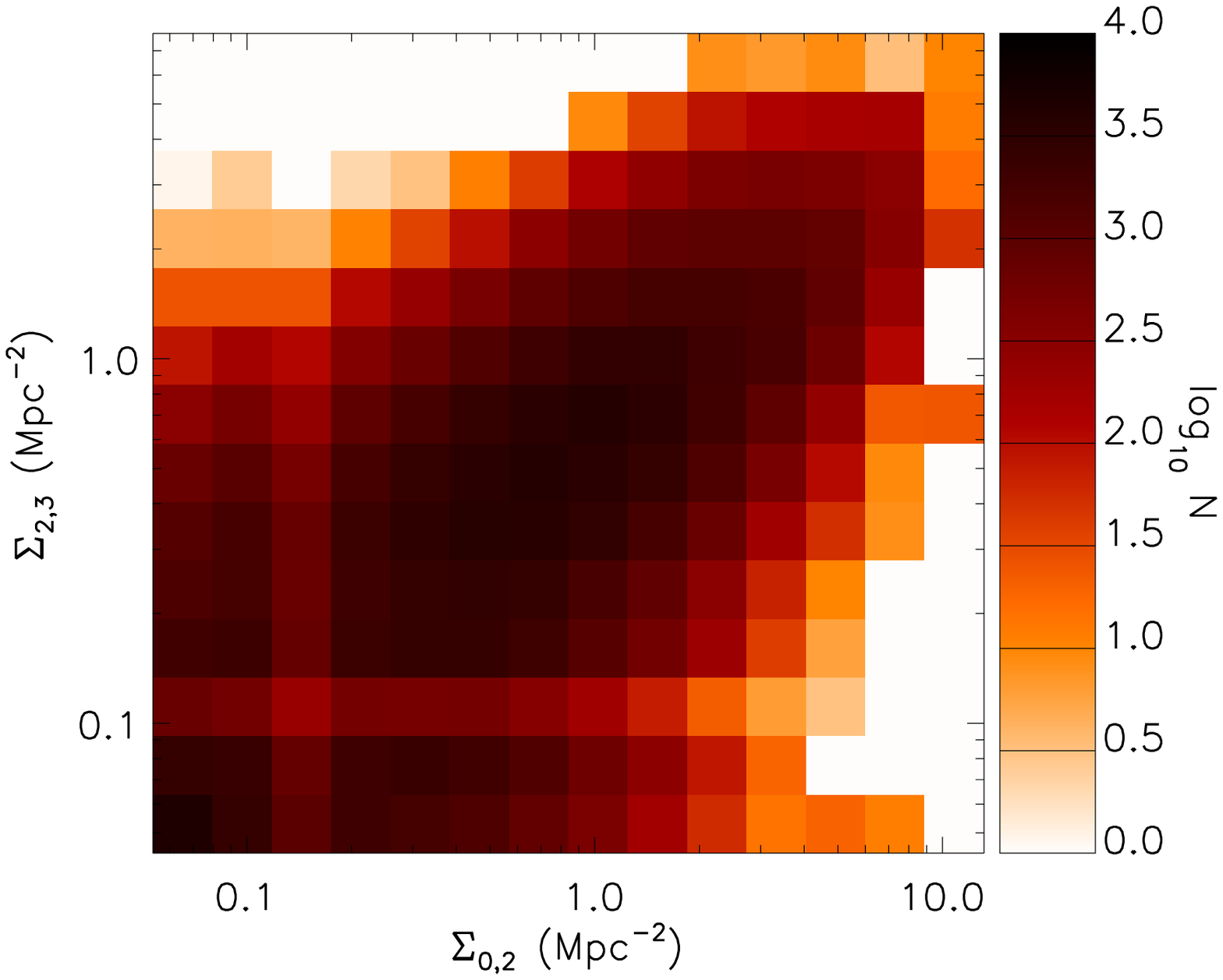,clip=t,width=8.cm}\psfig{figure=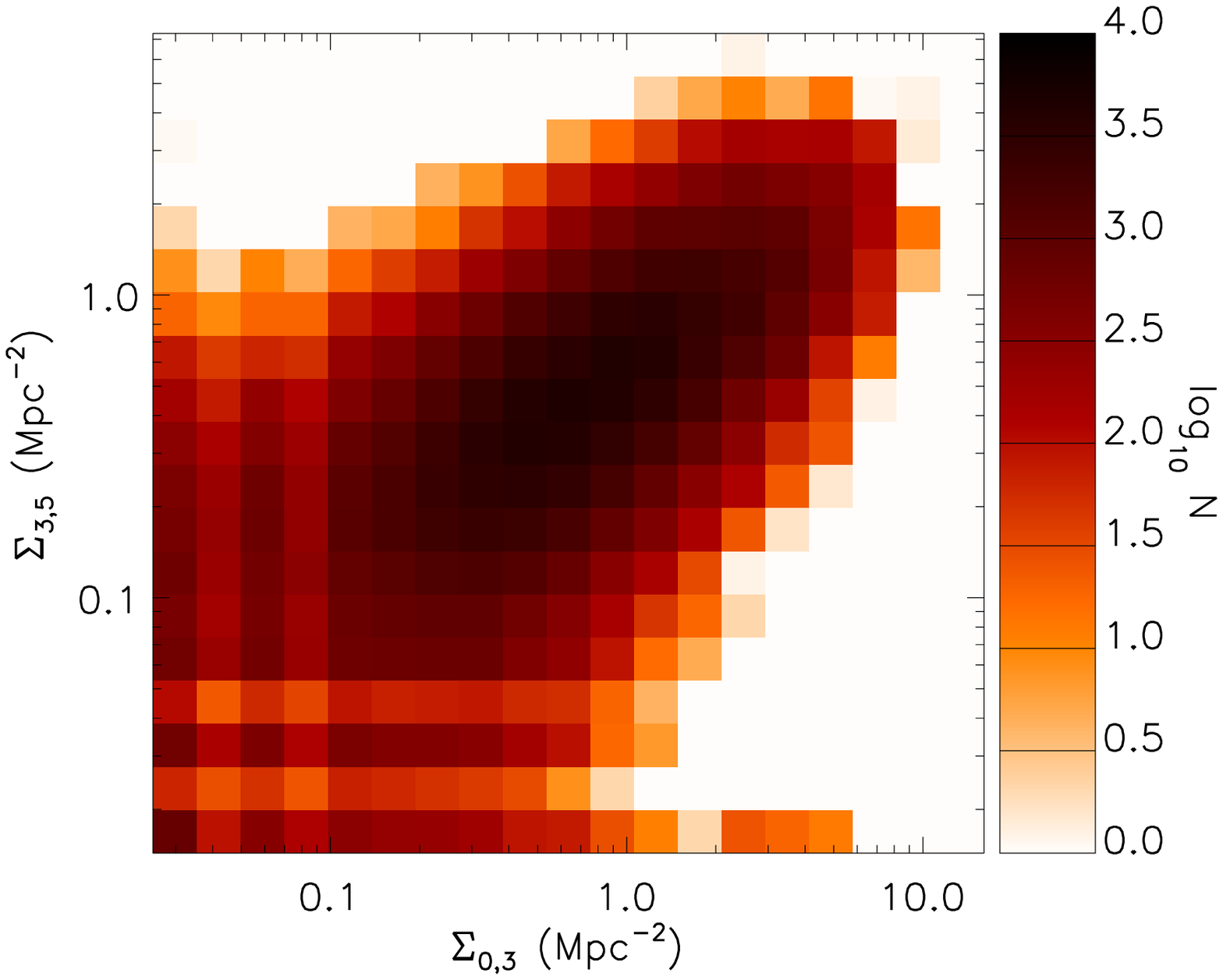,clip=t,width=8.cm}}
\caption[ ]{Two-dimensional histograms of the logarithmic galaxy density on different scales (for SDSS DR8 galaxies with $0.015\leq z \leq 0.08$.}\label{Datamultiscales}
\end{figure*}

Fig.~\ref{Datamultiscales} shows the 2d multiscale density histograms (binned number of galaxies) for the SDSS galaxy sample described in Section \ref{Data}. The binning in density is logarithmic with two exceptions: The first bin on both scales contains galaxies with no neighbour, while the last bin on both scales contains galaxies with densities higher than a given threshold density. These are $\log_{10} \Sigma_{0,0.5}^{max}=2.17$, $\log_{10}\Sigma_{0.5,1}^{max}=1.83$, $\log_{10}\Sigma_{0,1}^{max}=1.90$, $\log_{10} \Sigma_{1,2}^{max}=1.50$, $\log_{10} \Sigma_{0,2}^{max}=1.60$, $\log_{10}\Sigma_{2,3}^{max}=1.37$, $\log_{10} \Sigma_{0,3}^{max}=1.38$, and $\log_{10} \Sigma_{3,5}^{max}=1.07$.  In each of the four histograms there are columns and rows in which the number of galaxies is much lower than in the adjacent columns and rows. This happens because the number of neighbours (uncorrected for incompleteness) is always an integer, and these do not fall into all logarithmically spaced bins. However, since the observed numbers have been completeness corrected (see Section \ref{completenesscorrection}), some (fewer) galaxies cross into these columns/rows.

\begin{figure*}
\centerline{\psfig{figure=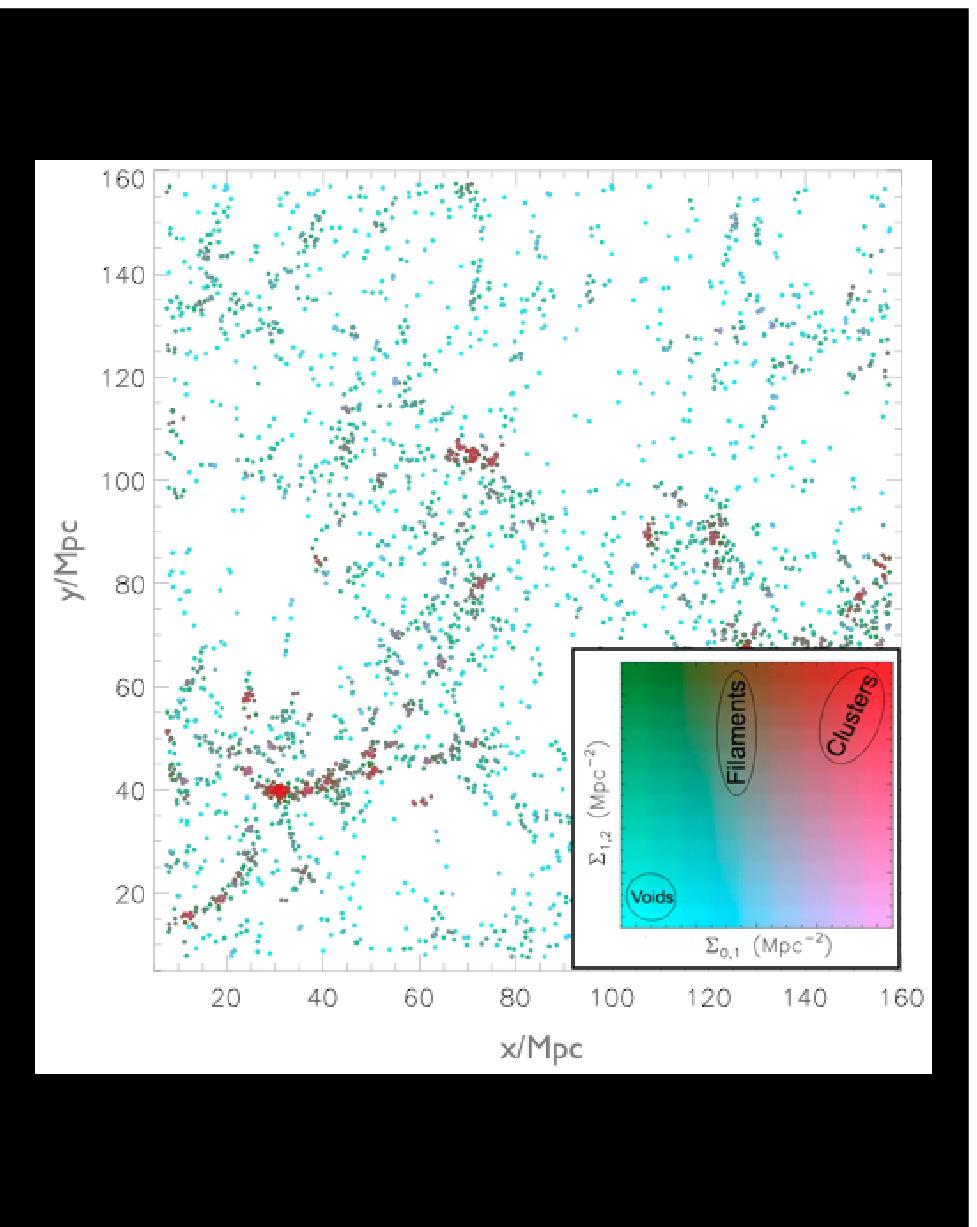,clip=t,width=10.cm}}
\caption[ ]{The spatial distribution of mock galaxies in a thin slice ($\Delta z=30$\,Mpc) of a section of the Millennium, colour coded by their position in the 2d histogram of densities on scales of $\left[(r_i,r_o)^1,(r_i,r_o)^2\right]=\left[(0.,1.), (1.,2.)\right]$\,Mpc, as indicated in the plot in the lower right corner.}\label{spatialdistribution}
\end{figure*}

One of the advantages of combining the measurement of the density field on different scales is that there is a clear relation between the location of a galaxy in the 2d density histograms and the kind of environment it lives in. Not only is it possible to distinguish cluster from field (or ``void'') environments, but also to discriminate between filaments and infall regions. Fig.~\ref{spatialdistribution} shows the spatial distribution of the simulated galaxies in a cut-out of the Millennium simulation ($\Delta x=160$\,Mpc, $\Delta y=160$\,Mpc, and $\Delta z=30$\,Mpc) in redshift space. Every point is a galaxy, colour coded by its position in the 2d histogram of densities. We show the example of the histogram where we combine density measurements on scales $\left[(r_i,r_o)^1,(r_i,r_o)^2\right]=\left[(0.,1.), (1.,2.)\right]$\,Mpc. 

While isolated galaxies in sparsely populated regions (mostly light cyan) live in the lower left corner of the 2d histogram (low density on both scales), galaxies in very dense, cluster environments (red) live in the upper right corner, filamentary structures show up in dark green (upper part, left of center, where the density on the smaller scale is intermediate and the density on the larger scale high), and infall regions show up in a brownish colour (upper part, between red and green).  

This demonstrates nicely how our understanding of ``environment'' and its influence on galaxies can be enhanced by investigating galaxy properties as a function of the density on different scales.

\section{Observational data}\label{Data}
For our investigation we use spectra from the seventh and photometry from the eight data release (DR7 \citep{Abazajian09} and DR8\footnote{DR8 contains all spectroscopy from DR7 plus additional photometry}  (\citealp{Aihara11a,Aihara11b})) of the Sloan Digital Sky Survey (part of SDSS-II \citep{York00}, and SDSS-III \citep{Weinberg07}, respectively). The selection criteria for our sample match that used by WZB10. A volume limited sample is created by first selecting galaxies with valid spectroscopic redshift determination, which are part of the so-called `main galaxy sample' (all extended sources with Petrosian dereddened magnitude $m_r\leq 17.77$ and mean surface brightness within half-light radius $\mu_r \leq 23.0$\,mag arcsec$^{-2}$, see \citet{Strauss02}), and then further restricting the selection to the redshift range $0.015 \leq z \leq 0.08$, and Petrosian k-corrected\footnote{K-corrections are applied using the IDL code {\it Kcorrect} version 4.1.2 \citep{Blanton07kcor}.} luminosity $M_r\leq-20$.

It is impossible to find all neighbours for galaxies close to the edges of the SDSS photometric footprint. Therefore we measure the fraction of the area contained within the footprint for $3\Mpc$ and $5\Mpc$ circles centred on each galaxy using the method and code developed by \citet{Budavari10}. Our further analysis is only conducted for the 132\,291 (129\,897) galaxies in our volume-limited sample for which the photometric completeness is more than 99.35\% on $r\leq 3 \Mpc$ ($r\leq 5 \Mpc$) scales, where the larger sample is then used for all scales $r_o\leq 3 \Mpc$.

\subsection{Spectroscopic Completeness}\label{completenesscorrection}
Since SDSS spectra are acquired using a multiple fibre spectrograph, the diameter of the fibres puts a lower limit on the angular separation between two spectra: if two target galaxies are closer to each other than $\theta_{lim}=55"$, it is only possible to take a spectrum for one of them in a single pass. This leads to a -- certainly scale and density dependent -- spectroscopic incompleteness of the number of galaxies in our annuli. Our correction for this incompleteness is two-fold.

When measuring the density in each annulus, the number of neighbours is weighted by the sprectroscopic completeness correction factor:
\begin{eqnarray} 
C_{r_{i},r_{o}}^{s}=\frac{N_{r_{i},r_{o}}^p}{N_{r_{i},r_{o}}^s}~,
\end{eqnarray}
where $N_{r_{i},r_{o}}^p$ is the total number of photometrically identified spectroscopic targets within the annulus, and $N_{r_{i},r_{o}}^s$ is the number with measured redshift. This upweights the numbers of neighbours in dense regions such that in Equation \ref{Sigma} $N_{r_i,r_o}^n= C_{r_{i},r_{o}}^{s} \times N_{r_{i},r_{o}}^{un}$, where $N_{r_{i},r_{o}}^{un}$ is the number of uncorrected neighbours within our radial, velocity and luminosity constraints.

Secondly, our HOD description requires the full distribution of a volume-limited sample of galaxies in multiscale density parameter space. This requires correcting the number of galaxies in each density bin of complementary scales for spectroscopic incompleteness. We weight them according to their own completeness on the smallest scale, $0.5\Mpc$.

\section{Model}\label{Model}
\subsection{The Halo Catalogue}
We take our halo catalogue from a numerical $N$-body simulation. This provides accurate halo clustering and growth on non-linear scales. In order to have sufficient resolution to be able to populate small enough halos over a large volume, we chose to use the Millennium halo catalogue \citep{Springel05}. Since the Millennium run was simulated with a larger value of $\sigma_8=0.9$ than is commonly accepted today ($\sigma_8\approx 0.8$, see e.g.  \citet{Komatsu11}), we follow  suggestion of \citet{LiWhite09} and \citet{Angulo10} and use the  output at $z=0.32$, treated as if it represents the local Universe.
\subsection{Populating the halos with galaxies}
The Halo Occupation Distribution (HOD) model describes the expected number of galaxies per halo, as a function of halo mass. We follow the description of \citet{Zehavi11}, and split the HOD into separate descriptions for central and satellite galaxies, such that:
\begin{eqnarray}\label{HODcen}
\ncenexp=0.5\left[1.+\mathrm{erf}\left(\frac{\log(M)-\log(M_{min})}{\sigma}\right)\right]~
\end{eqnarray}
where $\ncenexp$ is the expectation value for the number of central galaxies hosted by a halo of mass $M$; $M_{min}$ is defined as the halo mass at which a halo has a 50\% probability of containing a central galaxy; and $\sigma$ describes how steeply the function is rising towards unity at low mass. In practice, non-integral values of $\ncenexp$ are applied by randomly allocating our central galaxies to halos of mass $M$ with a probability $\ncenexp$.

The HOD for the satellite galaxies scales as a power law with index $\alpha$, cut-off mass scale $M_0$, and normalisation $M^\prime_1$:
\begin{eqnarray}\label{HODsat}
  \nsatexp=\nonumber
  0.5\left[1.+\mathrm{erf}\left(\frac{\log(M)-\log(M_{min})}{\sigma}\right)\right]\left(\frac{M-M_0}{M^\prime_1}\right)^\alpha
\end{eqnarray}

Again, for small mass halos the scatter is incorporated by means of a simple Monte-Carlo method. For the larger masses, when $\nsatexp > 1$, we include the scatter by randomly drawing a number from the Poisson distribution function which is determined by the expectation value predicted by the HOD, $P_{\nsatexp}(n_{sat})=\frac{\nsatexp^{n_{sat}}}{n_{sat}!}{\mathrm{e}}^{-\nsatexp}$, hence we populate a given halo with $n_{sat}$ satellites.

The general shape of the HOD is justified by results of hydrodynamical cosmological simulations as well as high-resolution collisionless simulations (e.g. \citealp{Kauffmann97,Kauffmann99,Benson00,Berlind03,Kravtsov04,Zheng05,Conroy06,Wetzel10}) for volume limited samples.

In Section \ref{redgalfracsection}, requirements for a minimum sample size per $0.5$mag luminosity bin mean we need to limit our sample to $-21.5\leq M_r\leq -20.0$. A single HOD describes the full population down to a luminosity limit -- e.g. $M_r\leq -20.0$ for our full sample. To describe the $-21.5\leq M_r\leq -20.0$ population we also need a HOD description of the $M_r\leq -21.5$ {\it bright} population. This can then be subtracted from the full sample, which contains all our (completeness-corrected) galaxies in the volume and luminosity-limited ($M_r\leq-20$), photometrically complete sample. We define the difference between the full and bright samples (representing galaxies in the luminosity range $-21.5\leq M_r\leq -20.0$) as the {\it select sample}. \citet{Zehavi11} have infered HOD parameters as a function of luminosity from the projected correlation function of SDSS DR7 galaxies; in Appendix \ref{Appendix} we describe in detail how we determine the appropriate HOD parameters for our samples from their measurements.

For each FOF (friends-of-friends) halo of a given top-hat mass $M_{th}$ in the Millennium simulation we now have the number of central (zero or one) and satellite galaxies in the halo. The central galaxy lives by definition in the centre of the halo, so it is allocated the coordinates of the parent halo. A catalogue of subhalos for each halo has been generated using the subhalo finder {\tt SUBFIND} \citep{Springel01}. Where any exist, their coordinates are allocated to the satellites. If the HOD predicts more satellite galaxies than there are subhalos, we distribute the remaining galaxies according to the NFW profile \citep{NFW}, with the appropriate concentration parameter $c=0.971-0.094\log(M/10^{12})$ \citep{Maccio08} and virial radius given by the halo mass. These galaxies can be thought of as satellites which have lost their own dark matter halos via tidal stripping and follow the dark matter density distribution of the parent FOF halo.

\subsection{Moving to redshift space: peculiar velocities}
Galaxies have peculiar velocities, they move inside the deep potential of massive galaxy clusters, and galaxies in smaller halos are accreted onto larger structures. The Doppler shift due to their peculiar velocity component along the line of sight adds to the cosmological redshift ($z_{obs}=z_{cosmo}+v_{pec}(1+z_{cosmo})/c$, where $z_{obs}$ is the observed redshift, $z_{cosmo}$ the cosmological redshift, $v_{pec}$ the peculiar velocity of the object, and $c$ is the speed of light). It is impossible to disentangle the individual contributions to the measured redshift. Therefore we need a model of peculiar velocities to render the measurement of the densities performed in the simulation more realistic. Since we work at $z\sim0$ very low redshift, we ignore the $1+z_{cosmo}$ term in the equation above.

The above described case of coherent infall onto large clusters leads to a compression of spherical structures on large scales \citep{Kaiser87}. In order to take this effect into account, we use the redshift space coordinates of the halos in the Millennium simulation, which are calculated from the real space coordinates and the peculiar velocity component in one direction, which we define as the ``line of sight'', so the comoving coordinate of a halo becomes $\vec{r}=(x_1,x_2,x_3^s)$ with $x_3^s=x_3+v_3/H_0$. 

Satellite galaxies orbit within the potential of their host dark matter halo. The contribution of these peculiar velocities also leads to a distortion of the shape of a galaxy cluster in redshift space: it appears to be elongated along the line of sight. This is why we measure the projected densities in cylinders (which have a length that corresponds approximately to the width of the line of sight velocity distributions of massive clusters, $\sigma_v\approx 1000$\,km s$^{-1}$). For the satellite galaxies for which we use the coordinates of subhalos, we also directly use their redshift space coordinates as they are given by the Millennium data base. For the others, distributed inside the halos following an NFW profile, we shift the real space coordinates of each satellite galaxy by an amount which corresponds to the redshift space distortion due to the typical velocity of a galaxy in a halo with the given mass.  We calculate these values using the relation between halo mass and velocity dispersion by \citet{Evrard08},
\begin{eqnarray}
\sigma_{DM}(M,z)=\sigma_{DM,15}\left[\frac{h(z)M_{200}}{10^{15}M_\odot}\right]^\alpha~,
\end{eqnarray}
where $\sigma_{DM,15}$ is the normalization at mass $10^{15} h^{-1}M_\odot$ and $\alpha$ is the logarithmic slope. \citet{Evrard08} have determined the parameters from a fit to a an ensemble $\Lambda$CDM simulations to be $\sigma_{DM,15}=1082.9\pm 4$\,km~s$^{-1}$ and $\alpha=0.3361\pm 0.0026$. We assume spherical symmetry of the velocity dispersion and divide by $\sqrt{3}$ in order to receive the width of the velocity distribution function along the line of sight, convert this value to a physical distance, then randomly draw a value $\delta x_3^s$ from the corresponding distribution function (which we assume to be Gaussian), and add it to the redshift space coordinate of the halo in which the galaxy under consideration lives. Thus the final redshift space coordinate (along the line of sight) of a simulated satellite galaxy is $x_3^s=x_3+v_3/H(z)+\delta x_3^s$.

\section{The fraction of red galaxies}\label{redgalfracsection}
\subsection{The red fraction of SDSS galaxies}\label{redgalfracdatasection}
It is always a matter of debate which observable is a good proxy for the ``passiveness'' of a galaxy. Estimated starformation rate or the strength of the 4000\AA~break D4000 determined from the galaxies' spectra would of course be an obvious choice, however, since the SDSS-I/II fibers have a diameter of only $3''$, for a large fraction of objects the aperture is smaller than the extent of the galaxy, so the spectrum will not reflect their overall SEDs. Aperture corrections introduce an additional uncertainty. Therefore we decided to use the broad-band observations to distinguish passive from starforming galaxies, as they are the most direct and most unbiased observables. Also we consider in this paper only one population integrated over the observed luminosity range, because we focus on the density dependence. Subsamples split by mass and luminosity can be investigated in future work.

\begin{figure*}
\centerline{\psfig{figure=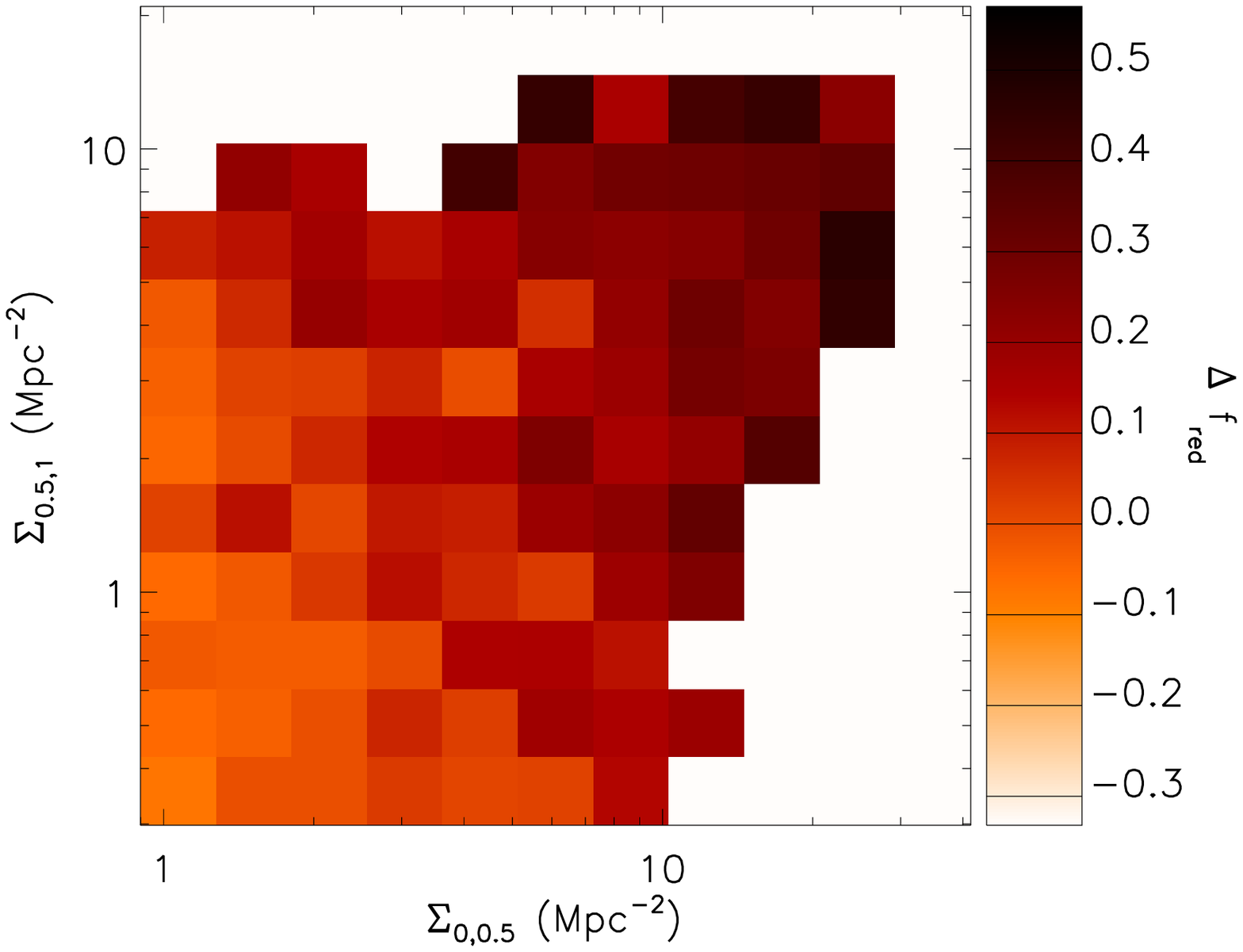,clip=t,width=8.cm}\psfig{figure=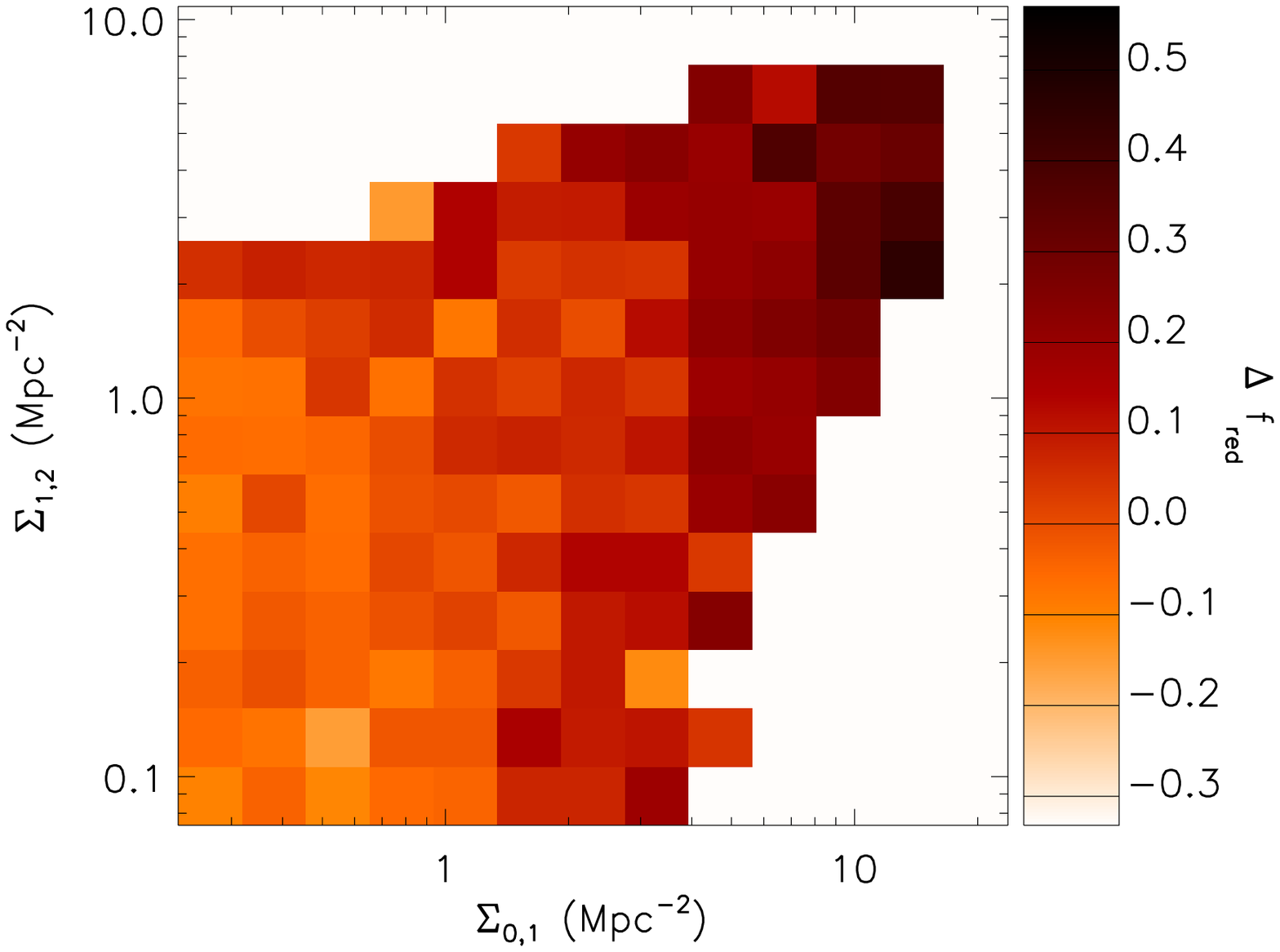,clip=t,width=8.cm}}
\centerline{\psfig{figure=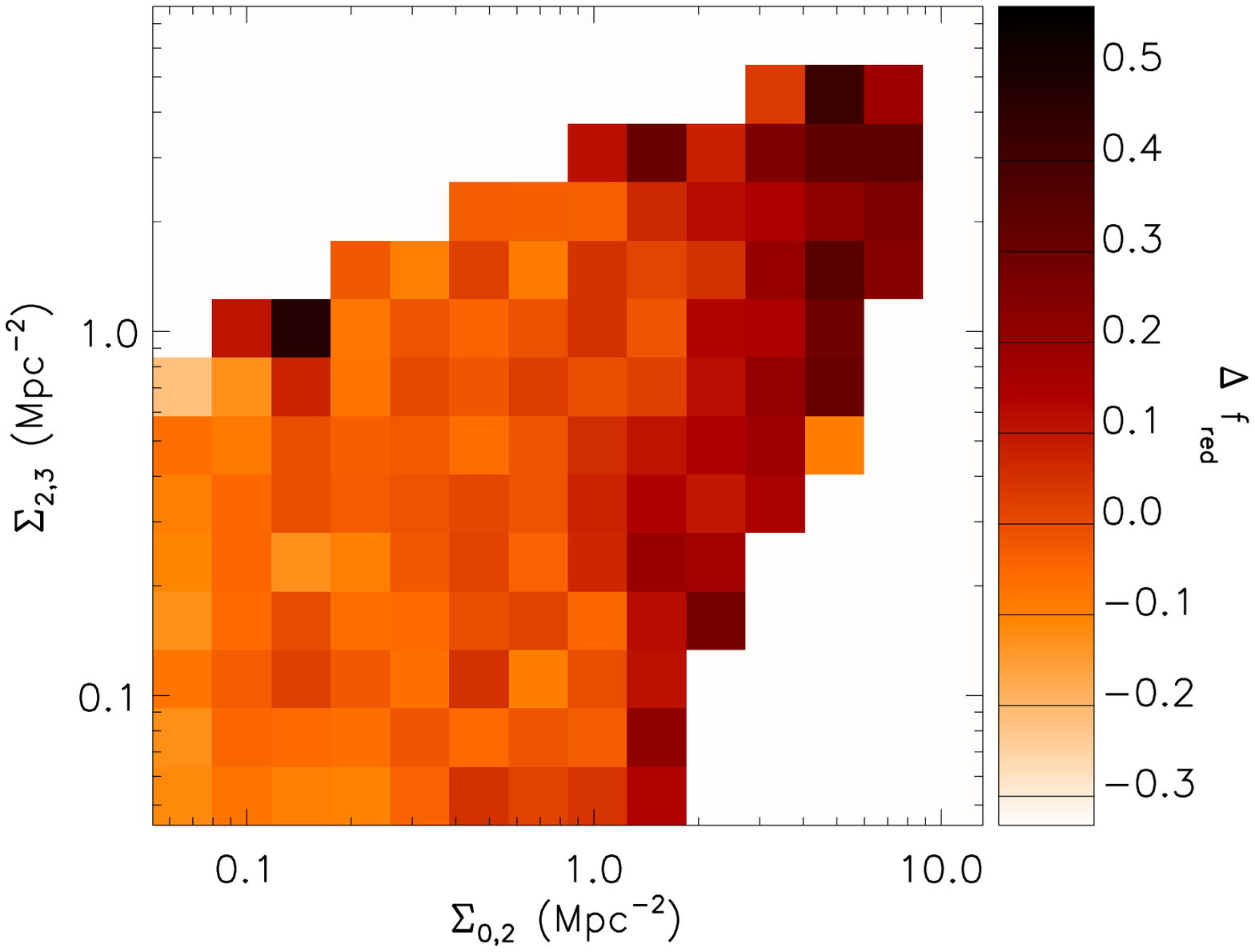,clip=t,width=8.cm}\psfig{figure=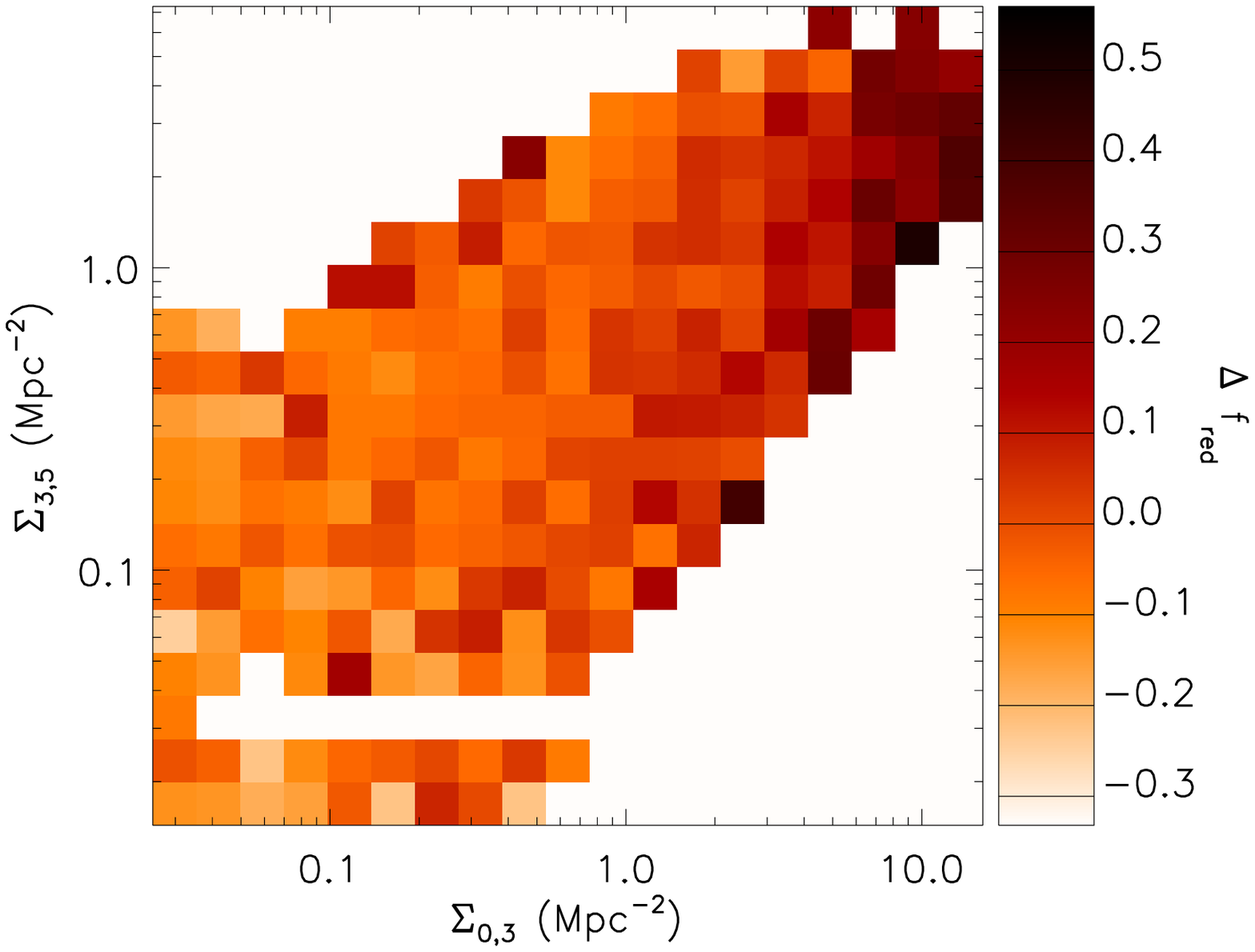,clip=t,width=8.cm}}
\caption[ ]{The excess of the fraction of red galaxies over the mean red fraction as a function of multiscale density for the SDSS sample with $-20.0 \le M_r\le -21.5$. Scales and binning are the same as in Fig.~\ref{Datamultiscales}.}\label{densdensredfracdata}
\end{figure*}

We determine the fraction of red SDSS galaxies in each 2d density bin for each pair of scales using the exact same procedure described by WZB10. For each subset of galaxies, the bimodal $u-r$ model colour distribution is fit using a double Gaussian function as first proposed by \citet{Baldry04}. The parametrisation of the colour distribution with a double gaussian (and subsequent integration of the fit to derive the fraction of red galaxies) also takes the intrinsic scatter between the two populations into account, and describes the overlap of the wings of the two gaussians. For more details of the exact fitting procedure we refer to WZB10. There are five free parameters to the fits: the relative fraction of galaxies in each (red and blue) peak, and the mean colour and width of each peak. As found by WZB10, the peak means and widths depend upon both density and luminosity. Therefore we bin galaxies not only by density but also by luminosity in bins of $\Delta M_r=0.5$ width. We require at least 25 galaxies for a robust fit: the most luminous galaxies (Petrosian magnitudes $M_r\le -21.5$) cannot be subsampled with sufficient resolution in density and luminosity. Thus we only consider galaxies in the luminosity range $-20.0 \le M_r\le -21.5$ as in WZB10. From each fit we extract the fraction of galaxies in the red peak. We then compute the weighted mean red fraction of all luminosity bins within a given density bin, weighted by the total number of galaxies per luminosity bin. The final red fraction per 2d density bin is then subtracted from the total red fraction of galaxies (which is $f^{red}_{tot}=0.46$)  to produce the residual dependence of red fraction on multiscale density, $\Delta f_{red}$. The resultant plots of $\Delta f_{red}$ vs density for our four  pairs of scale are shown in Fig.~\ref{densdensredfracdata}.

\subsection{The red fraction of model galaxies}\label{modelredfrac}
Having populated the halos using the HOD parameters derived from the measurement of \citet{Zehavi11}, we can assign a colour (red or blue) to the model galaxies, according to a description which reflects the different scenarios we want to probe. As different physical processes act on central and satellite galaxies, and the fraction of passive, red galaxies has a different dependence on halo mass \citep[e.g.][]{vandenBosch08}, we treat them differently.

\begin{figure*}
\centerline{\psfig{figure=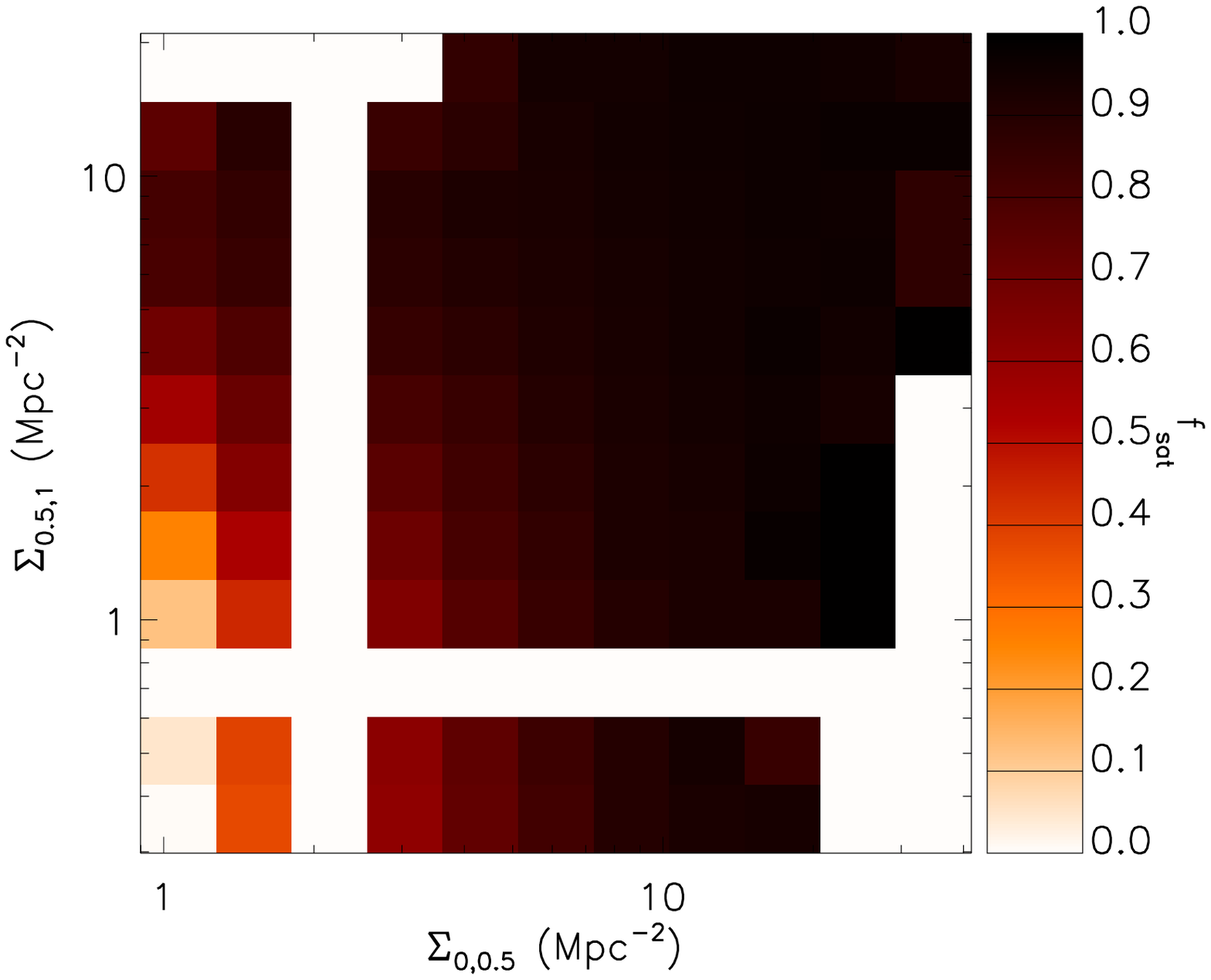,clip=t,width=8.cm}\psfig{figure=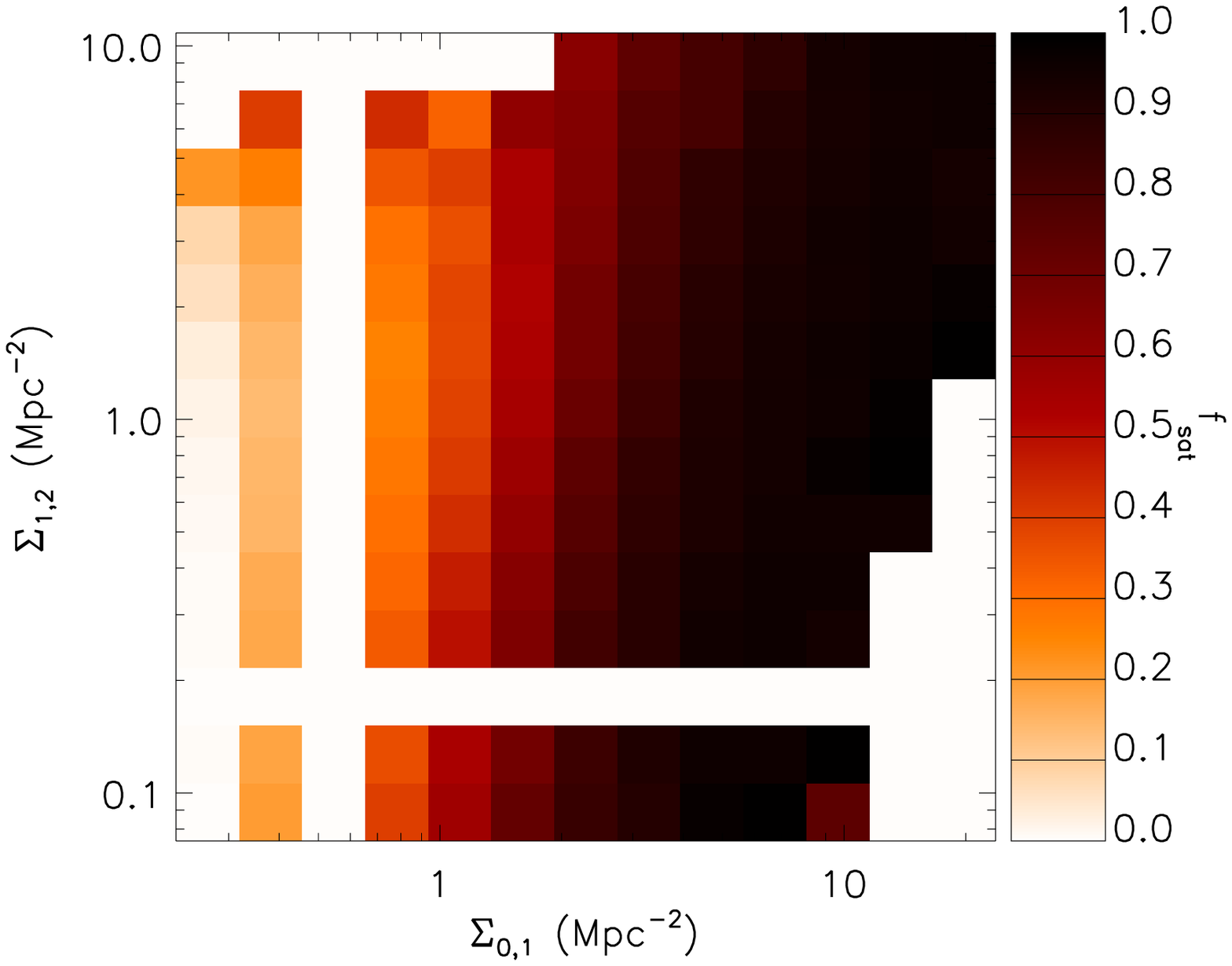,clip=t,width=8.cm}}
\centerline{\psfig{figure=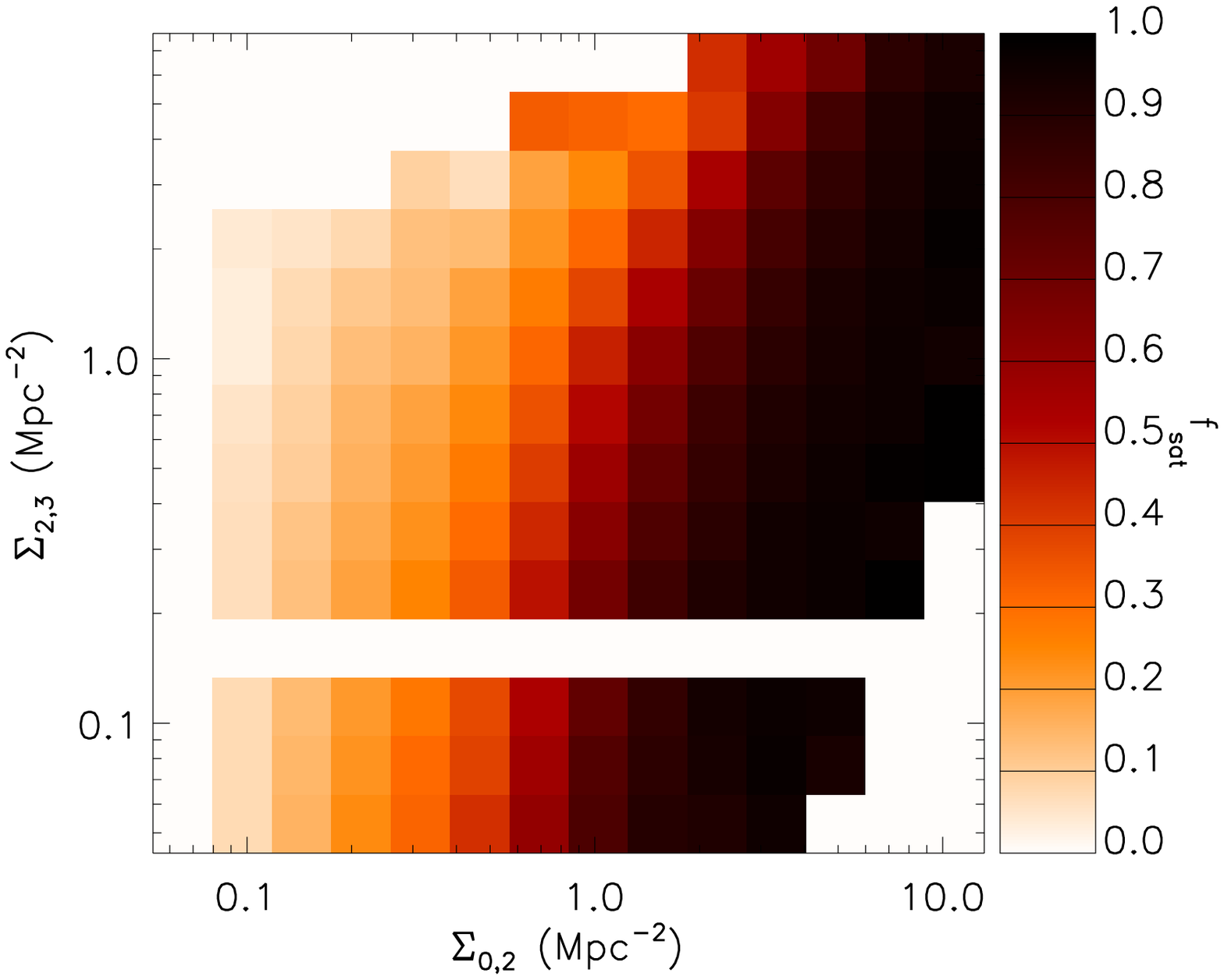,clip=t,width=8.cm}\psfig{figure=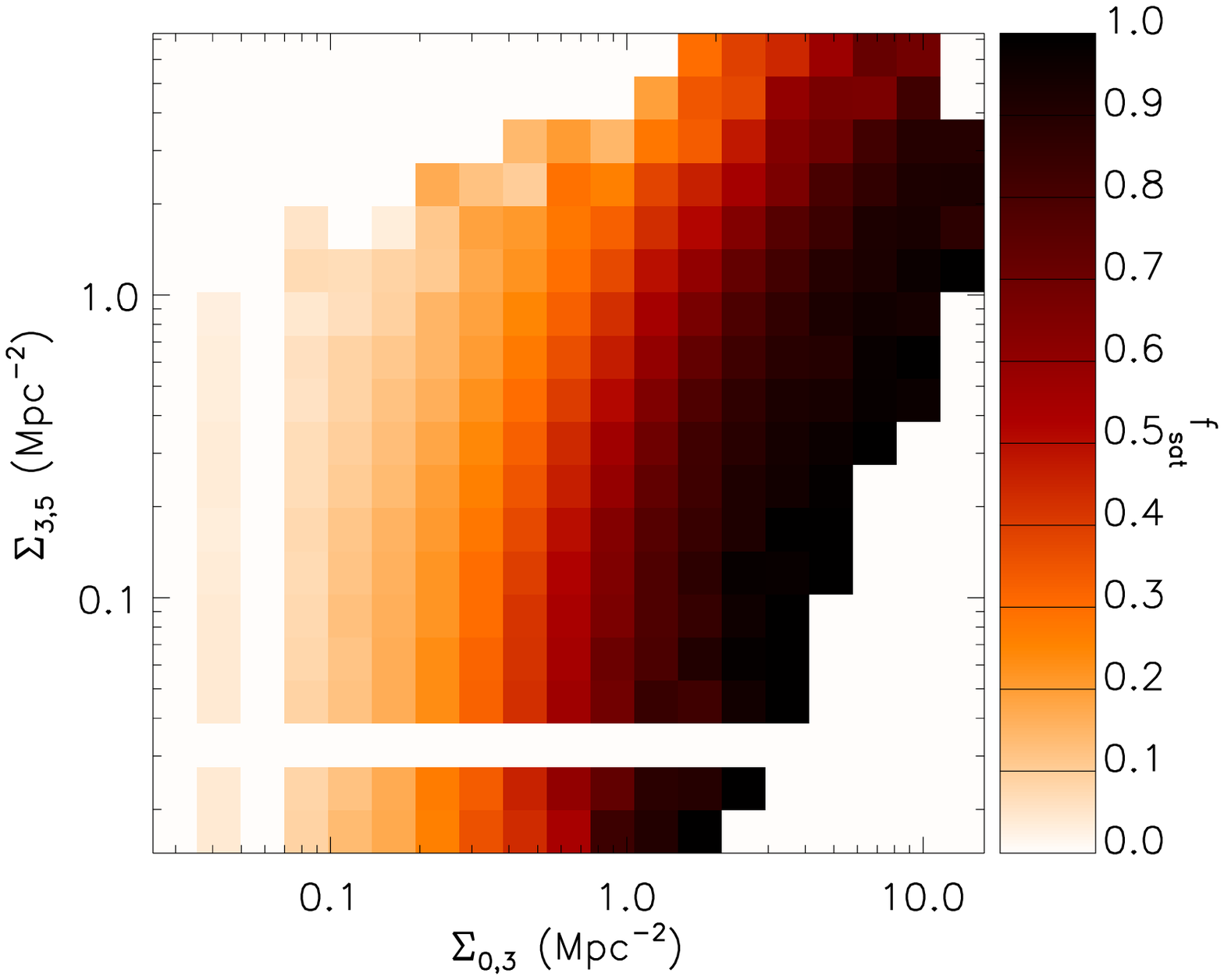,clip=t,width=8.cm}}
\caption[ ]{The fraction of model satellite galaxies for the HOD parameters describing the select sample.}\label{satfrac}
\end{figure*}

Fig.~\ref{satfrac} shows the fraction of galaxies in the select sample of model galaxies (best fits) which are satellites as a function of density for our four pairs of scales. Although most galaxies are centrals, these are heavily focused to low density, especially on small scales. This means that much of our multiscale density parameter space is dominated by satellite galaxies. Combine this fact with the greater halo mass range of satellites (Fig.~\ref{HODplot}), and it is clear that our data and method offers more dynamic range to explore the imprint of environment on satellite galaxy properties than on central galaxies. However, our parameter space is also useful to find central galaxies: galaxies with no neighbours, especially on larger scales, are clearly the isolated, central galaxies of lowish mass halos.

\subsubsection{Central galaxies}\label{redbluecen}
The total number of central galaxies in the select sample is greater than the total number of satellites. However, central galaxies in this sample are found in a relatively narrow range of halo mass -- $2.5\times10^{11}\Msol\leq M\leq2.5\times10^{12}\Msol$. As these halos contain $\le 1$ satellite (figure~\ref{HODplot}) the dynamic range in the number of neighbours is limited. Therefore centrals are located at low density only, and most 2d bins of our multiscale density parameter space are dominated by satellite galaxies (Fig.~\ref{satfrac}). For example, it is difficult to distinguish $3\times10^{11}\Msol$ halo central galaxies from $2\times10^{12}\Msol$ halo central galaxies, and therefore to say anything about the correlation between halo mass and central galaxy properties.

This being the case, we adopt the simplest strategy: a constant red fraction at all halo masses. The value of $f^{red}_{cen}$ can be determined directly from Fig.~\ref{densdensredfracdata}: the galaxies having the lowest density on the largest scales (no neighbours even in large annuli) are all central galaxies. This sets $f^{red}_{cen}=0.38$.
\subsubsection{Satellite galaxies}\label{redbluesat}
We parameterize the dependence of the fraction of red satellite galaxies on halo mass using a hyperbolic tangent form which assymptotically approaches minimum and maximum values at low and high halo mass respectively, with a symmetry around the {\it transition mass}, $M_t$:

\begin{eqnarray}\label{redcenfraceq}
f^{red}_{sat}\left(M_{halo}\right)=\nonumber\\
\frac{1}{2}\left(f^{red}_{max}-f^{red}_{min}\right)\cdot\left[1.+\tanh\left(\frac{\log M-\log M_t}{\Gamma}\right)\right]+f^{red}_{min}~,
\end{eqnarray}

\begin{figure*}
\centerline{\psfig{figure=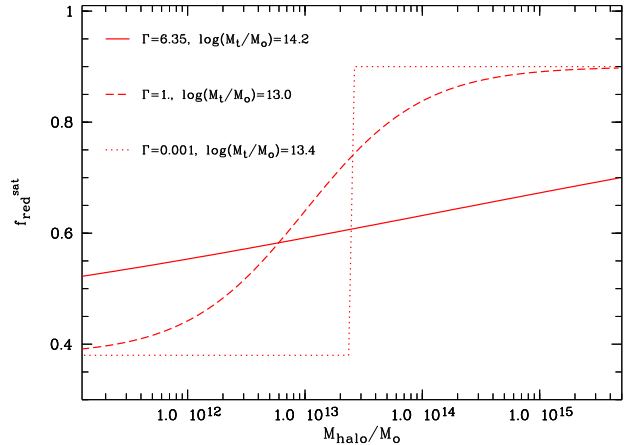,angle=270,clip=t,width=9.cm}}
\caption[ ]{The fraction of red satellite galaxies as a function of halo mass, for three example cases, parameterised by Equation~\ref{redcenfraceq}. A larger $\Gamma$ indicates a more gradual rise in red satellite fraction with halo mass, while a very small value of $\Gamma$ ($\Gamma=0.001$, dotted line) indicates a sharper transition in red fraction at the transition mass $M_t$. The total fraction of red satellites is the same in all cases.}\label{redsatfracplot}
\end{figure*}

The red fraction of satellites increases with halo mass from the minimum value $f^{red}_{min}$ up to the maximum value $f^{red}_{max}$. The parameter $\Gamma$ controls how steeply the function rises (i.e. how rapid is the transition from $f^{red}_{min}$ to $f^{red}_{max}$). Fig.~\ref{redsatfracplot} shows three example cases with $\Gamma=0.001$,$1$ and $6.35$. With a small value of $\Gamma$ the transition is very sharp at $M=M_t$ while with large $\Gamma$ the transition is much more gradual. In Fig.~\ref{redsatfracplot} the parameters are chosen to conserve the same total fraction of red satellite galaxies -- a large change in $\Gamma$ corresponds to a small change in $M_t$ (see Section~\ref{constraintstotalredfrac}).

We assume the maximum red satellite fraction is equal to the red fraction of galaxies in the highest density bin on the smallest scales which is dominated by the satellites of massive halos: $f^{red}_{max}=0.9$. To estimate the minimum red satellite fraction, we assume it cannot be lower than the red central fraction. All galaxies are formed as centrals, and those which are now satellites have been since accreted onto a more massive halo. Also, whatever the effect of becoming a satellite is, it can not turn galaxies which have already been red before accretion into blue ones. 

We need to define the fraction of galaxies which were quenched as satellites, accounting for the fact that some will have been quenched in processes which affect central galaxies. \citet{Wetzel13} nicely demonstrate in their Fig.~7 that the fraction of galaxies which have their star formation quenched as satellites, and its dependence on stellar mass, is sensitive to the definition of {\it quenched as satellites}. There are two possible definitions. The first is all satellite galaxies quenched since infall (\citet{Wetzel13} green line in their Fig.~7) -- to estimate this fraction one needs the fraction of galaxies which were quenched as central galaxies {\it before infall}, which is substantially different from the $z=0$ quenched central fraction. The second definition (our choice) is those satellite galaxies quenched since infall {\it which would not have been quenched anyway as central galaxies} (\citet{Wetzel13} dotted line in their Fig.~7). This is {\it defined} to be the difference between the quenched fraction of satellites and centrals at $z=0$. Indeed, if the processes affecting central galaxies are still active once a galaxy is a satellite, it represents that fraction of satellites which have been quenched by satellite-specific processes (for an equivalent definition, see \citet{vandenBosch08}). 

We therefore choose to examine the red satellite fraction halo mass dependence as a {\it net} difference with respect to the central red fraction. Therefore we set $f^{red}_{min}=f^{red}_{cen}=0.38$.

With these two parameters fixed, we have two free parameters:  $M_t$ and $\Gamma$. We will test our sensitivity to our assumed values of $f^{red}_{max}$ in Section~\ref{fredminmax}.

\subsection{Fitting $M_t$ and $\Gamma$}\label{fitredfrac}
To find the best fitting values of the free parameters $M_t$ and $\Gamma$, we compute $\chi^2$ using the full covariance matrix to test the fit of the model to the observed red fraction for a grid of models with $12.0\leq \log M_t\leq 16.8$ in 24 steps of $\Delta \log M_t=0.2$ and $0.001\leq \Gamma \leq 19.801$ in 99 steps of $\Delta \Gamma=0.2$. 

To calculate the covariance matrix for each parameter combination we geometrically divide the Millennium simulation into 64 ($=4^3$) boxes and calculate the covariance of the {\it numbers} of red and blue galaxies ($N_{red}$ and $N_{blue}$) in each 2d density bin on each pair of scales where there are at least four model galaxies and the 25 SDSS galaxies required to compute the red fraction. We fit the actual numbers of red and blue galaxies, not only because we use the model and not the data to calculate the covariance matrices (and there the numbers are the observables), but also because fitting the red fraction would require complicated and unnecessary error propagation from the numbers to the fractions. By using the actual numbers instead of the red fraction, we can also include the cross-correlation between red and blue galaxies in the covariance matrix. We then calculate the Moore-Penrose pseudo inverse $C^{+}$, and finally $\chi^2=\mathbf{v}^t C^{+} \mathbf{v}$, where $\mathbf{v}$ is a vector which contains the difference of both the numbers of red and blue data and model galaxies in the 2d density bins, $\mathbf{v}=\left(\mathbf{N}_{red}^{data}-\mathbf{N}_{red}^{model},\mathbf{N}_{blue}^{data}-\mathbf{N}_{blue}^{model}\right)$. Since in the case of the data we do not count red and blue galaxies, but derive the red fraction per bin from a fit to the colour distribution (see Section \ref{redgalfracdatasection}), we calculate the numbers assuming the HOD describes the total number of the galaxies in the haloes perfectly, such that in each bin
\begin{eqnarray}
N_{red}^{data}=f_{red}^{data}\left[N_{red}^{model}+N_{blue}^{model}\right]\nonumber\\
N_{blue}^{data}=(1-f_{red})^{data}\left[N_{red}^{model}+N_{blue}^{model}\right]~.
\end{eqnarray}

\begin{figure*}
\centerline{\psfig{figure=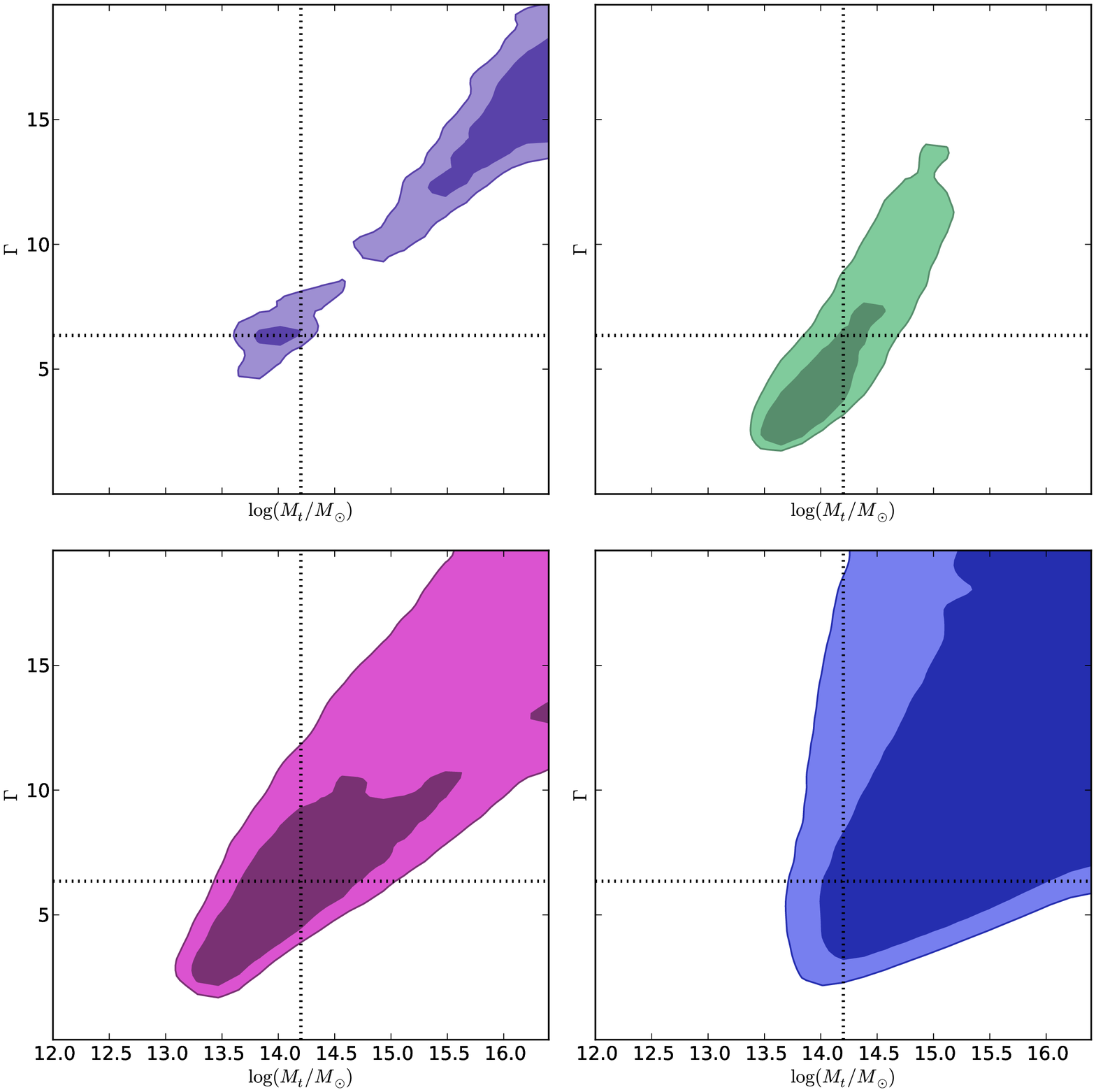,clip=t,width=9.cm}}
\caption[ ]{$\chi^2$ as a function of $\log M_t$ and $\Gamma$ for the four combinations of scales we are examining in this paper, upper left: for the combination of two radii $(r_i,r_o)_{1,2}=(0.,0.5), (0.5,1.)$, upper right: for $(r_i,r_o)_{1,2}=(0.,1.), (1.,2.)$, lower left: $(r_i,r_o)_{1,2}=(0.,2.), (2.,3.)$, and lower right:$(r_i,r_o)_{1,2}=(0.,3.), (3.,5.)$\,Mpc, respectively. The contours are the 68\% and 90\% confidence limits. The dotted lines indicate the point where all four 68\% confidence regions overlap.}\label{chisquared_logMtGamma}
\end{figure*}

Fig.~\ref{chisquared_logMtGamma} shows $\chi^2$ as a function of $\log M_t$ and $\Gamma$ for the four combinations of scales we are examining in this paper. The contours indicate the the 68\% and 90\% confidence limits. To guide the eye, we indicate the point where all four 68\% confidence regions overlap (at $\log M_t=14.2$ and $\Gamma=6.35$).

The shape of the $\chi^2$ distribution is different for the four different combinations of scales we examine in this paper, and the confidence limits of all four overlap in a very small region. In particular the values we derive from the smallest scales are different from the others, for which the confidence limits are much more similar. One possible explanation is that our model does not take any colour segregation within the haloes into account, so the fact that red galaxies tend to live in the inner regions of massive clusters while bluer galaxies live in the outskirts (\citealp{Melnick77,Balogh00,Girardi03,deLucia12}) might not be reflected properly when the densities of the simulated galaxies are measured on scales comparable to or smaller than halo radii: if the red galaxy fraction is a function of galactocentric radius and red galaxies thus concentrated in the centre of massive haloes, that would create very high red fractions at the highest densities on the smallest scales, while we would not see these high fractions on scales comparable to or larger than the halos.

\begin{figure*}
\centerline{\psfig{figure=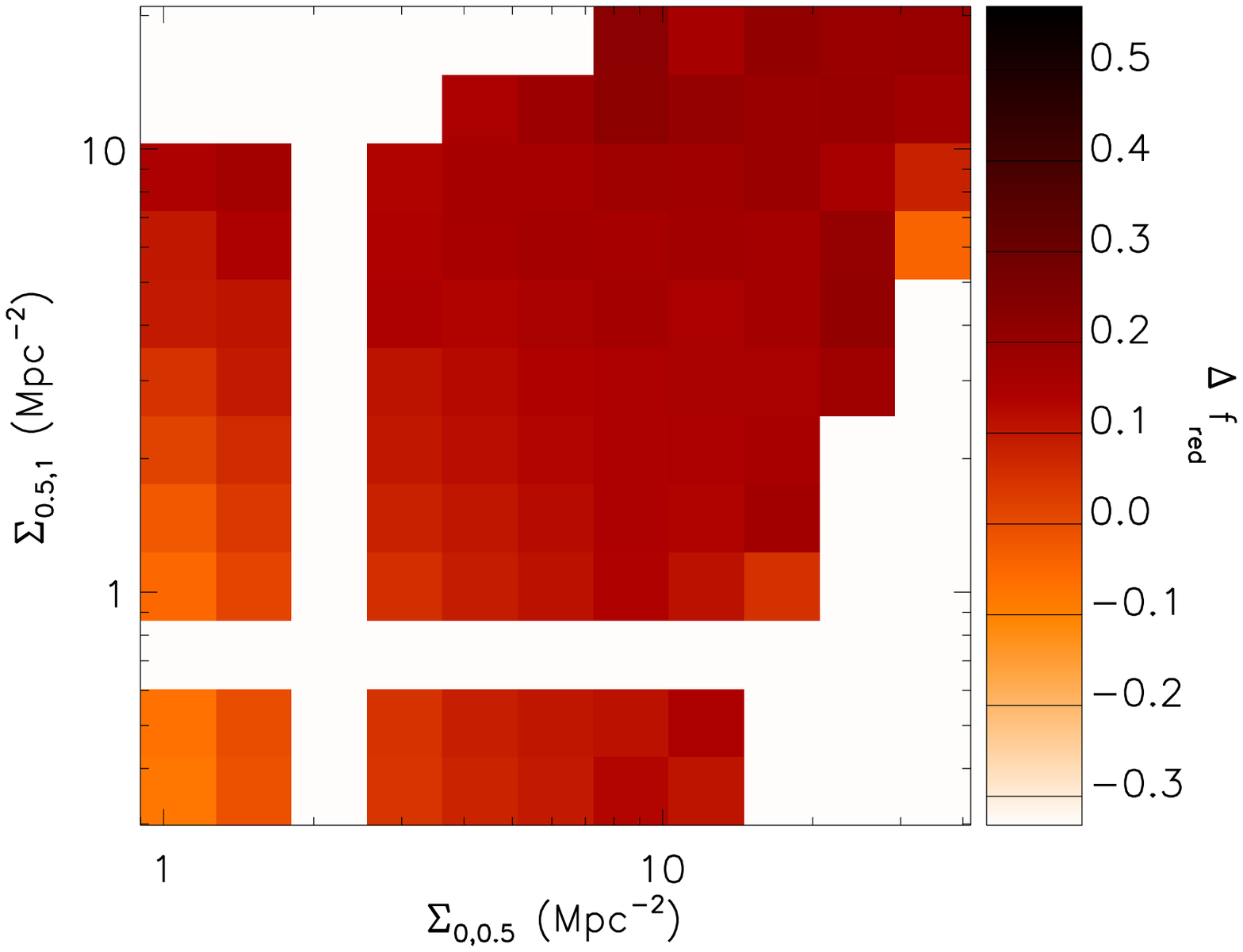,clip=t,width=8.cm}\psfig{figure=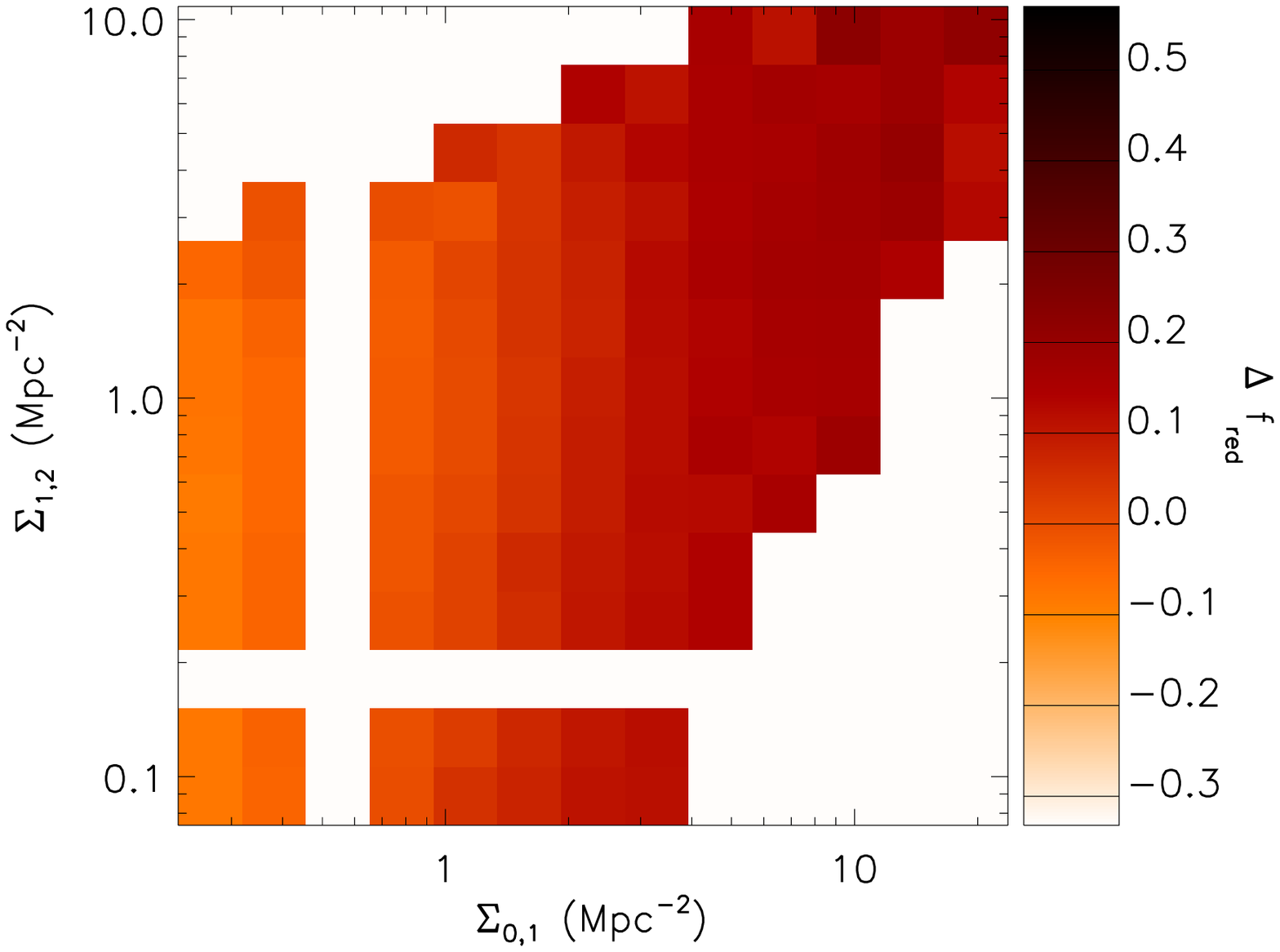,clip=t,width=8.cm}}
\centerline{\psfig{figure=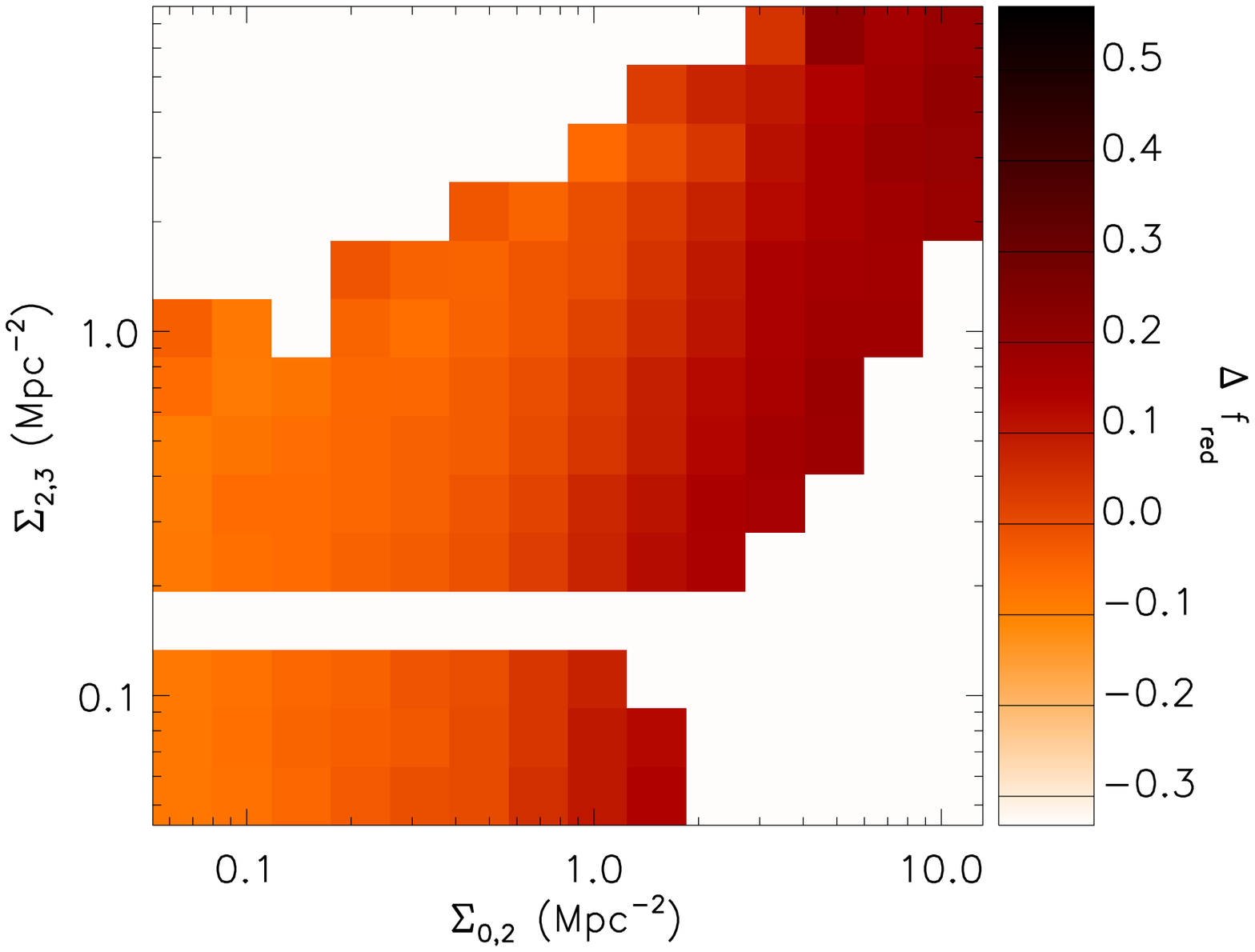,clip=t,width=8.cm}\psfig{figure=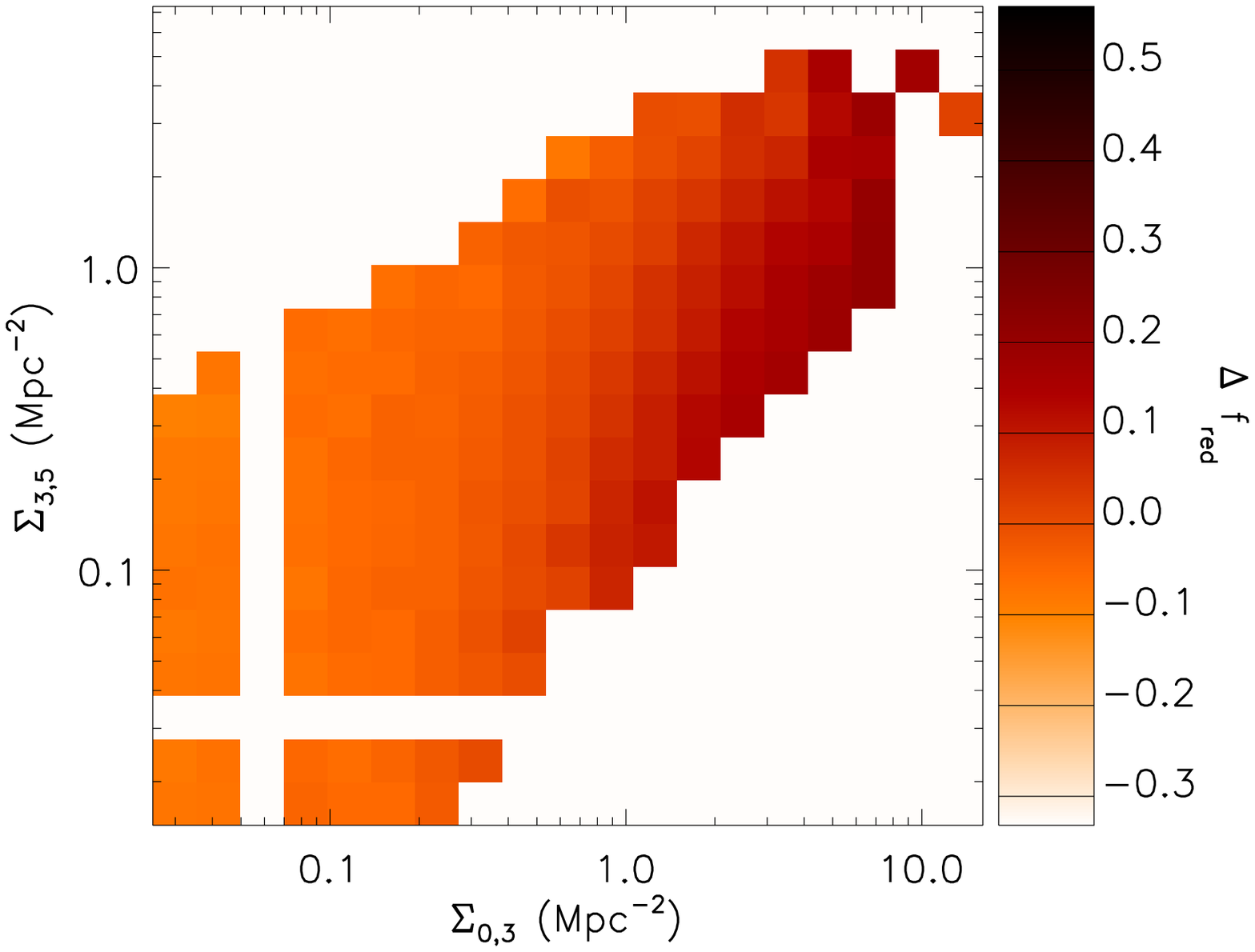,clip=t,width=8.cm}}
\caption[ ]{Model of the (excess) red galaxy fraction, for $\log M_t=14.2$ and $\Gamma=6.35$. Areas which are populated by galaxies in the model but not in the observations are omitted.}\label{fredmodel}
\end{figure*}

\begin{figure*}
\centerline{\psfig{figure=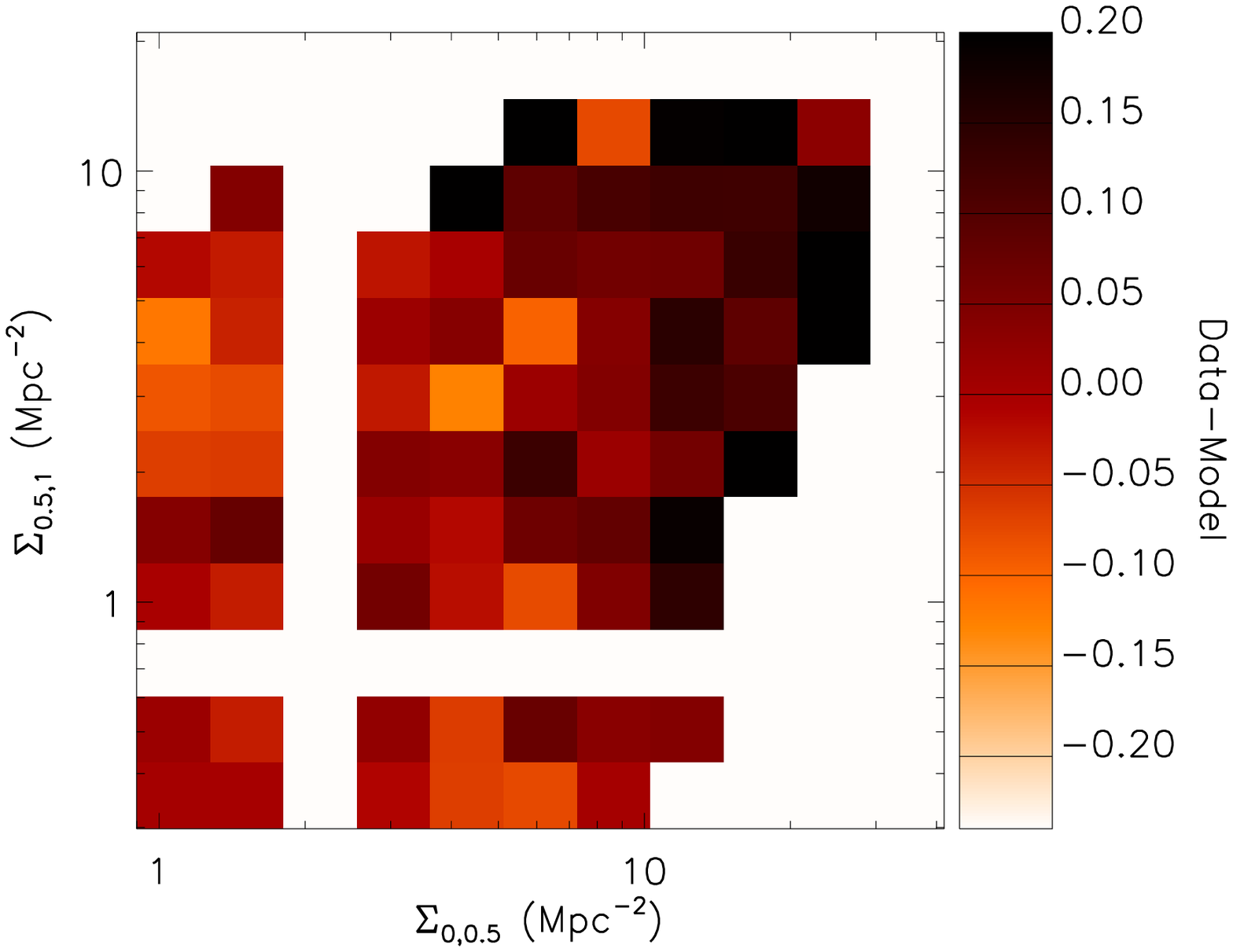,clip=t,width=8.cm}\psfig{figure=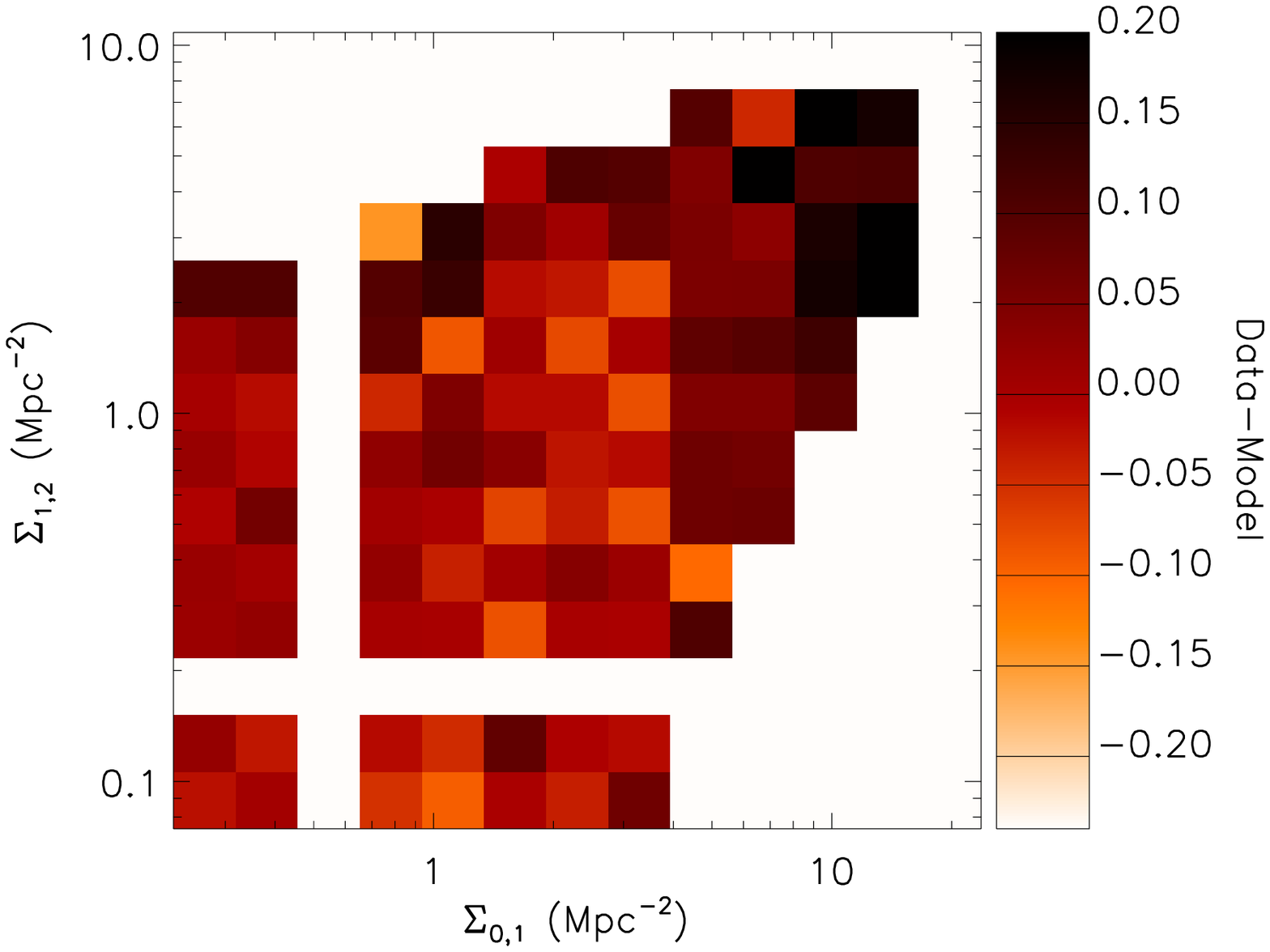,clip=t,width=8.cm}}
\centerline{\psfig{figure=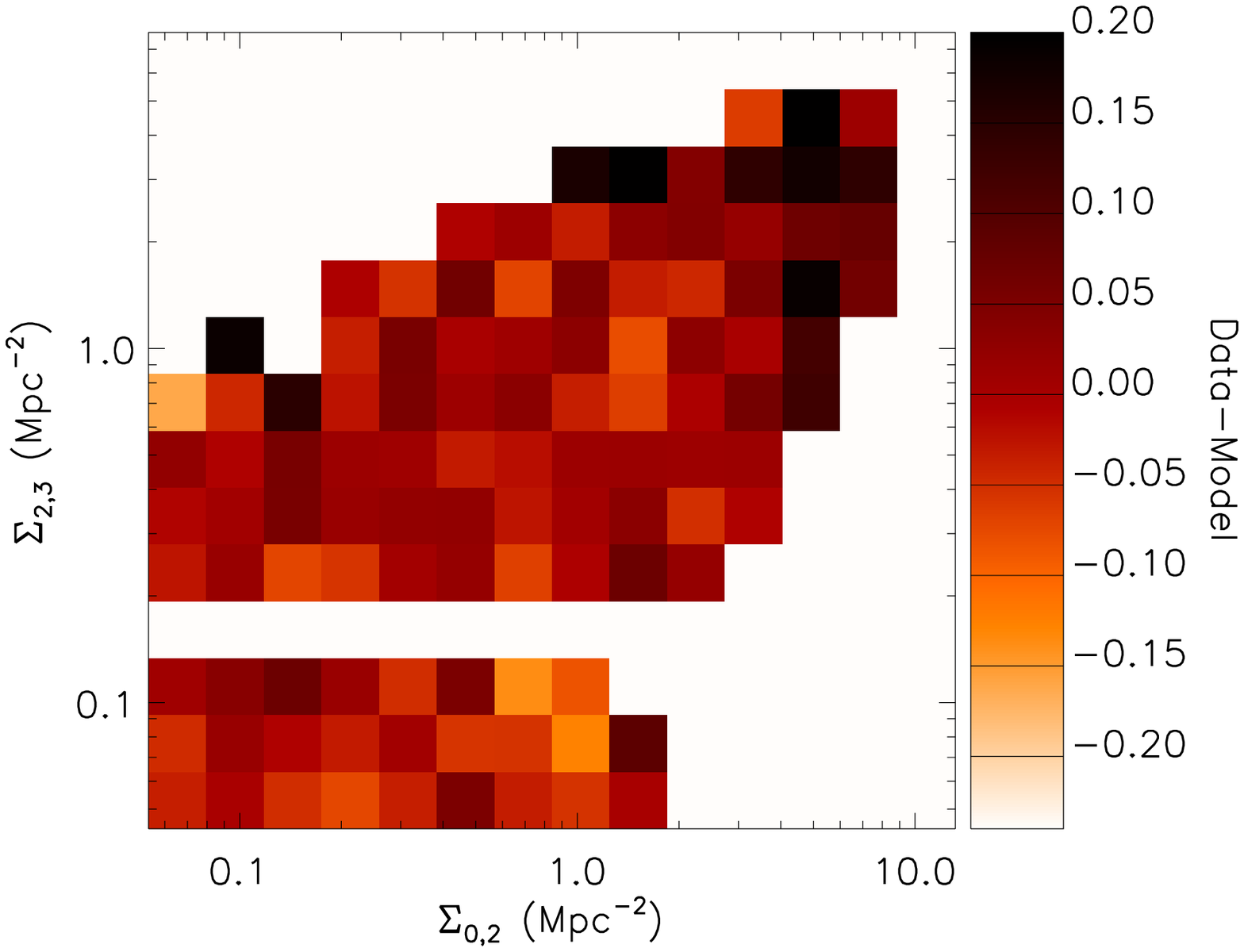,clip=t,width=8.cm}\psfig{figure=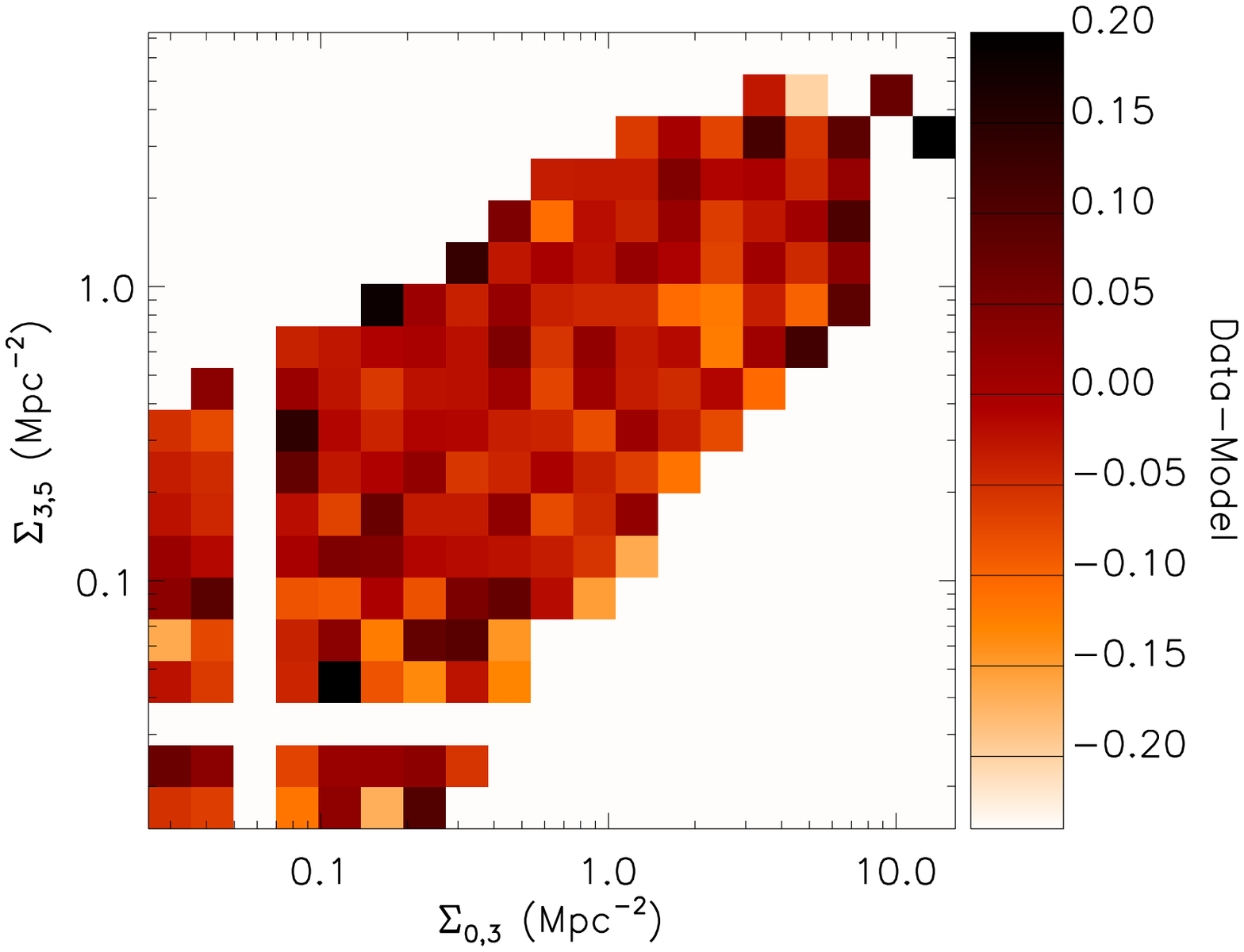,clip=t,width=8.cm}}
\caption[ ]{Distribution of the difference between observed and modelled red fraction, for $\log M_t=14.2$ and $\Gamma=6.35$.}\label{fredresiduals}
\end{figure*}

Figs.~\ref{fredmodel} and~\ref{fredresiduals} respectively show the excess red galaxy fraction ($\Delta f_{red}$ as in Fig.~\ref{densdensredfracdata}) of a model for which we assume $\log M_t=14.2$ and $\Gamma=6.35$, and the difference between this  model and the observed red fraction in the multiscale density parameter space. The model demonstrates the characteristic positive correlation of $f_{red}$ with density on scales up to $\sim 1$\,Mpc, and anti-correlations with density at larger scales with fixed smaller scale (interior) density, as pointed out by WZB10. Across most of the multiscale density parameter space, the residuals remain small, however on the smallest scales the model can not well reproduce the observed gradients and produces too few red galaxies in high density regions. As outlined above, this might be a hint that colour segregation within the haloes should be taken into account in a future analysis, and demonstrates the power of the method (considering multiple scales independently) to constrain dependencies which exceed a simple dependence of the red satellite fraction on halo mass.

\subsubsection{Dependence of $\log M_t$ and $\Gamma$ on $f^{red}_{max}$}\label{fredminmax}
\begin{figure*}
\centerline{\psfig{figure=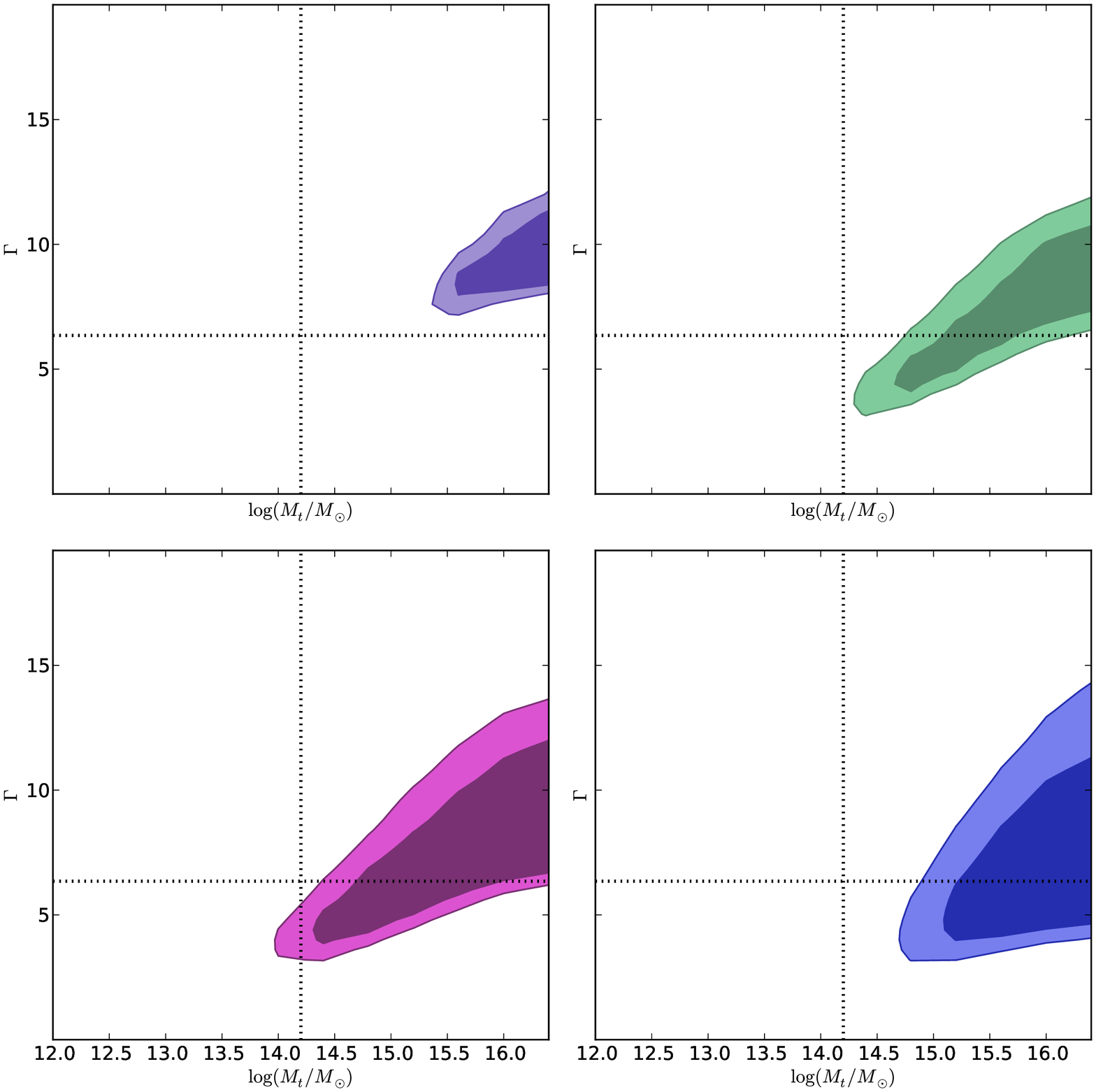,clip=t,width=9.cm}}
\caption[ ]{$\chi^2$ as a function of $\log M_t$ and $\Gamma$, as in Fig.~\ref{chisquared_logMtGamma}, but with $f^{red}_{max}=1.0$. The dotted lines indicate where the 68\% confidence areas overlapped in the $f^{red}_{max}=0.9$ case.}\label{chisquared_logMtGamma_frmax1}
\end{figure*}

The transition mass $M_t$ and the slope $\Gamma$ depend on the values of $f^{red}_{max}$, $f^{red}_{min}$ and $f^{red}_{cen}$. Our default values are simply the minimum and maximum values of $f_{red}$ in the lowest and highest bins of density respectively, but $f^{red}_{max}$ in particular is subject to statistical and systematic errors. Background and foreground objects which are scattered into our cylinders (due to their peculiar velocities) can dilute the measurement. For example, blue field galaxies can be found at very high density if they are projected within a cylinder containing the dense core of a galaxy cluster: the true maximum red satellite fraction may be larger than we measure. To examine this possibility, we investigate the case where all satellites in very massive halos are red, $f^{red}_{max}=1.0$ (while keeping all other parameters (including $f^{red}_{min}=f^{red}_{cen}=0.38$) fixed). Fig.~\ref{chisquared_logMtGamma_frmax1} shows the distribution of $\chi^2$ as a function of $\log M_t$ and $\Gamma$ for $f^{red}_{max}=1.0$. 

\begin{figure*}
\centerline{\psfig{figure=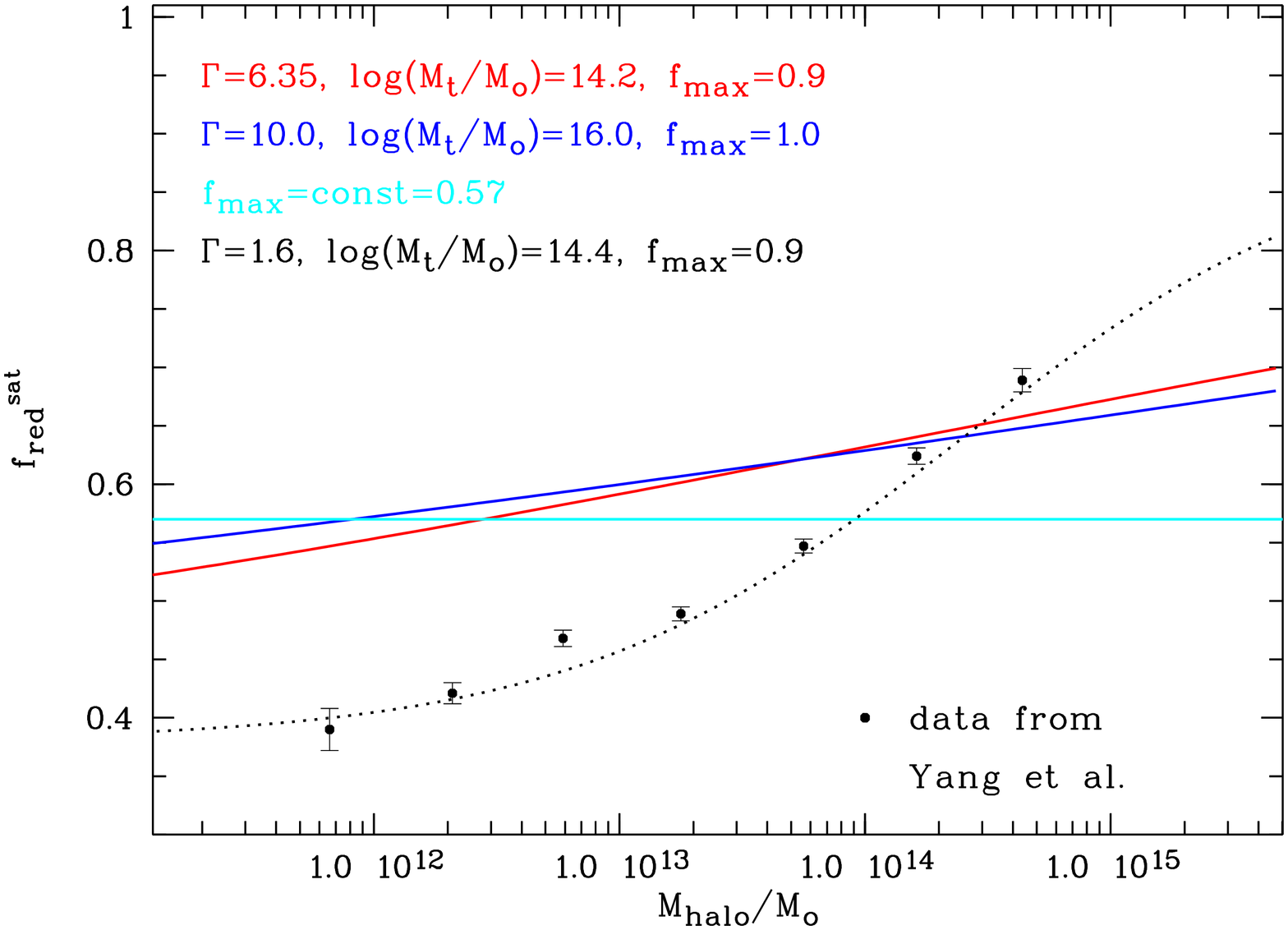,angle=270,clip=t,width=9.cm}}
\caption[ ]{The fraction of red satellite galaxies as a function of halo mass, for $f^{red}_{max}=0.9$, $\log M_t=14.2$, and $\Gamma=6.35$ (red line), and a case in which all satellite galaxies become red at high halo masses (blue line). The cyan line shows the case of a constant red satellite fraction which would result in the same number of red galaxies ($f^{red}_{sat}=0.575$), as discussed in Section \ref{flatsatellites}.}\label{frmaxdemo}
\end{figure*}

Fig.~\ref{frmaxdemo} demonstrates the similarity of the two solutions in the observed range of halo masses: although the ``best-fitting'' values are different for the two cases (the values of $\log M_t$ become larger and the values of $\Gamma$ become slightly smaller as compared to the case in which $f^{red}_{max}=0.9$), the choice of parameters $f^{red}_{max}=0.9$, $\log M_t=14.2$, and $\Gamma=6.35$ (red line) produces a very similar course of the function as one with  $f^{red}_{max}=1.0$, $\log M_t=16.0$, and $\Gamma=10.00$ (where all four $\chi^2$ distributions overlap), blue line. Neither function reaches the maximum value $f^{red}_{max}$ within the range of observed halo masses, in fact, assuming the maximum existing halo mass is $M_{max}\approx 3\cdot 10^{15}$, we find $f^{red}_{max}\la 0.7$ in both cases. Since the fit is dominated by the densities on larger scales (and is able to describe them well), and the value of $f^{red}_{max}$ read from the highest densities on the smallest scales, the measured high fraction of red galaxies found on these small scales can only be explained if colour segregation in massive haloes leads to an agglomeration of red galaxies towards the cluster centres. Hence, these high fractions do exist, but can only be measured on small scales. 

\subsection{Constraints from the total red fraction}\label{constraintstotalredfrac}

\begin{figure*}
\centerline{\psfig{figure=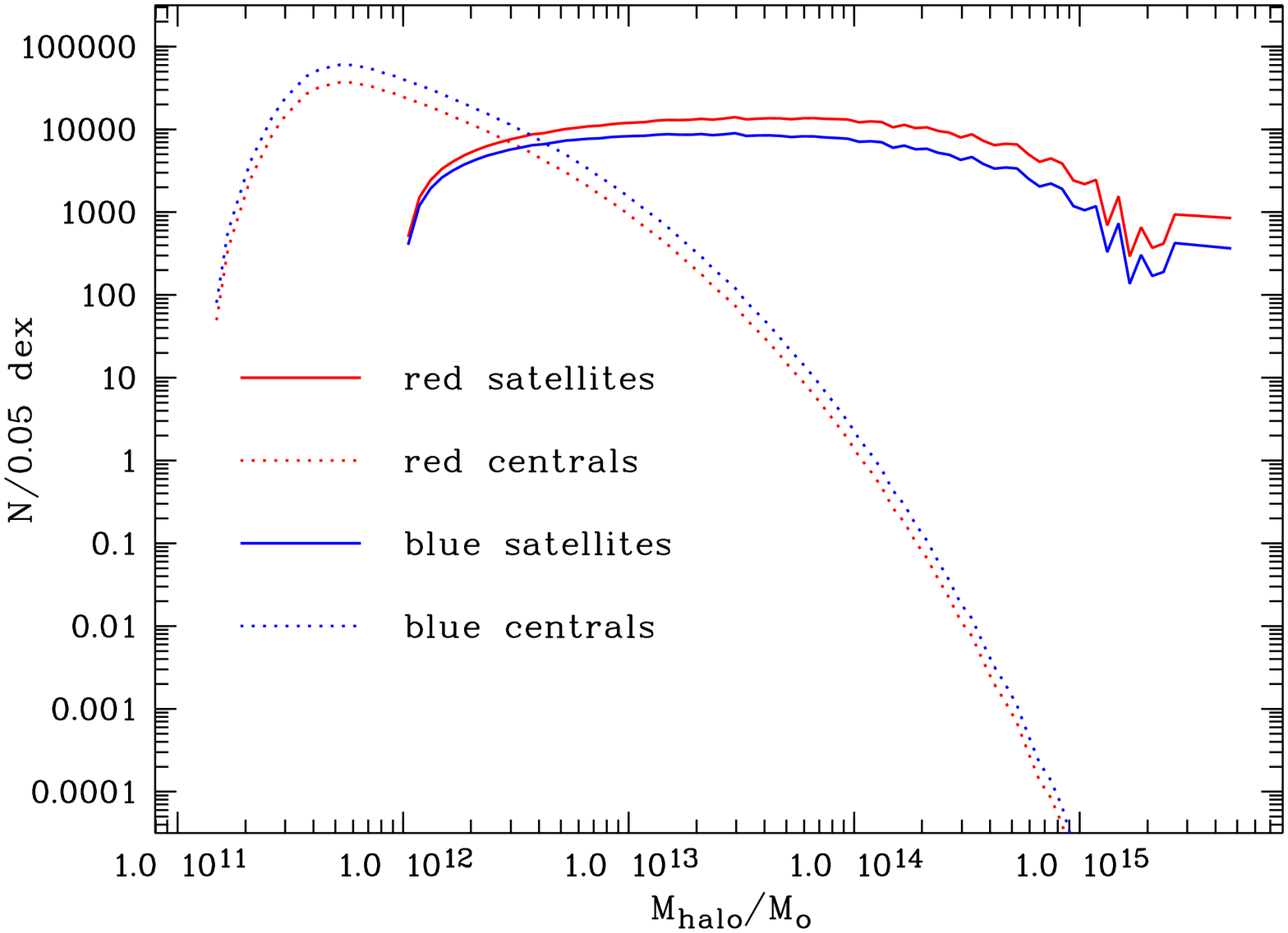,angle=270,clip=t,width=9.cm}}
\caption[ ]{The total number of red and blue galaxies per 0.05~dex in halo mass, as a function of halo mass. The halo mass function is measured from the Millennium simulation, and multiplied by our select sample HOD and the parameterized fraction of red and blue galaxies, given the HOD parameters derived from \citet{Zehavi11}, $\log M_t=14.2$ and $\Gamma=6.35$, and our default values of $f^{red}_{max}$, $f^{red}_{min}$, and $f^{red}_{cen}$ as determined from the data.}\label{Nred}
\end{figure*}

Fig.~\ref{Nred} shows the total number of red and blue galaxies as a function of halo mass for the select sample. The total number of halos per d($\log~M$)$=0.05~dex$ in mass is measured directly from the Millennium simulation. HOD parameters are taken from \citet{Zehavi11}, see Appendix \ref{Appendix}. The total number of red central and satellite galaxies per d($\log~M$) are then:

\begin{eqnarray}
  \frac{dN^{red}(M)}{d(\log~M)}&=&\frac{dn_{halo}(M)}{d(\log~M)}.N^{select}(M).f^{red}(M)
\end{eqnarray} 
with the HOD for the select sample, $N^{select}(M) = N^{all}(M) - N^{bright}(M)$. This equation is applied separately to the central and satellite populations to produce figure~\ref{Nred}. 

The interplay between the dark matter halo mass function, satellite HOD, and red satellite fraction results in an almost constant total number of red satellite galaxies per log halo mass bin over halo masses of mass $M_{halo}\ga 2\cdot 10^{12} M_\odot$. At the highest masses ($M_{halo}>10^{15} M_\odot$) the number of halos is so small that the measurement becomes noisy.

\begin{figure*}
\centerline{\psfig{figure=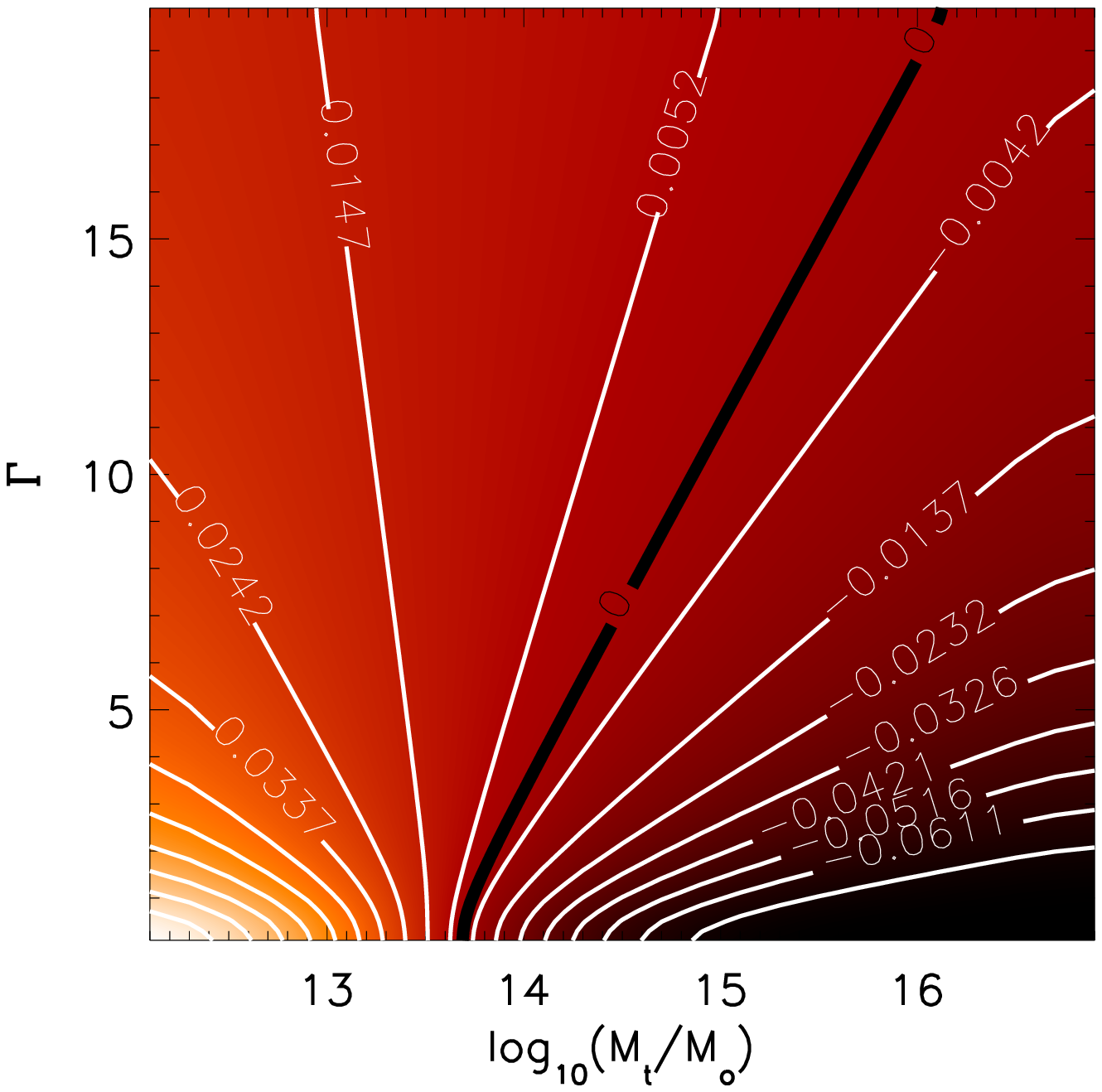,clip=t,width=8.cm}}
\caption[ ]{Difference between the total red fraction produced by the model as a function of $\log M_t$ and $\Gamma$ and the total red fraction measured from the data, $f^{red}_{tot}=0.46$.}\label{constraintsfromfredtot} 
\end{figure*}

Fig.~\ref{constraintsfromfredtot} shows the difference between the {\it total} red fraction produced by the HOD model as a function of $\log M_t$ and $\Gamma$ and the total red fraction measured from the data ($f^{red}_{tot}=0.46$). The combinations of $\log M_t$ and $\Gamma$ which produce the correct total red fraction are those which sit on the line for which the difference is zero (the solid black line). The course of this line is consistent with the shape of the $\chi^2$ distribution (Fig.~\ref{chisquared_logMtGamma}. $\log M_t$ and $\Gamma$ are not well constrained from the total red fraction alone, but the dependence of red fraction on multiscale density is a more sensitive diagnostic as we shall illustrate in section~\ref{sharpsatellites}. 

\subsection{Alternative scenarios}
\subsubsection{A sharp transition in the fraction of red satellites?}\label{sharpsatellites}
In section~\ref{fitredfrac} we have infered a very shallow rise in the fraction of red satellites with halo mass. To constrain this behaviour we fit the red fraction as a function of multiscale densities, which are sensitive to the total red fraction and to its dependence on different types of environment. In section~\ref{constraintstotalredfrac} we found that the total red fraction alone can help constrain the transition mass, $M_t$, but not the steepness of the transition, $\Gamma$. Figure~\ref{constraintsfromfredtot} shows that the total red fraction could equally well be matched with a very steep transition. This would occur if the processes suppressing star formation in satellite galaxies occur {\it only} in halos above a certain mass. 

\begin{figure*}
\centerline{\psfig{figure=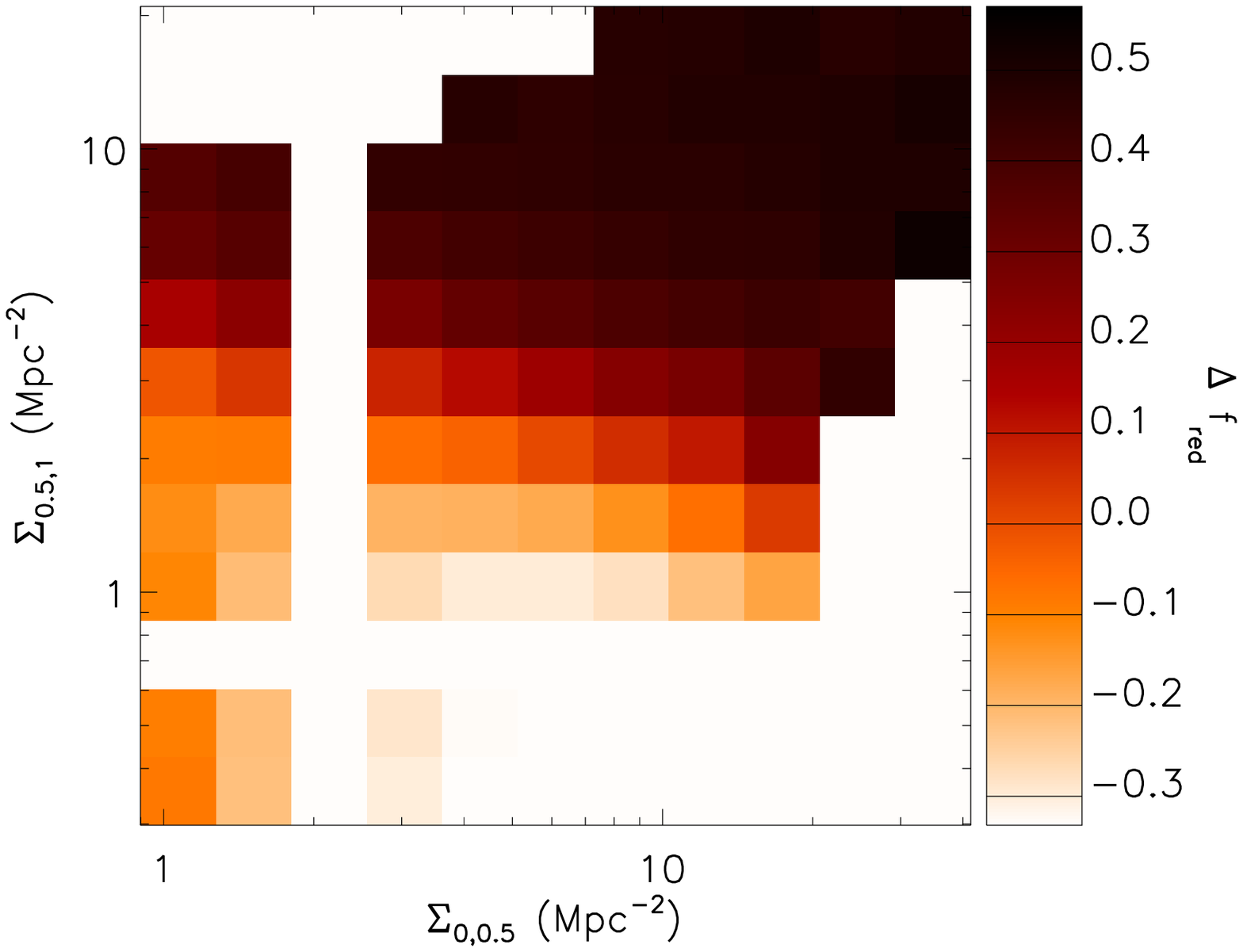,clip=t,width=8.cm}\psfig{figure=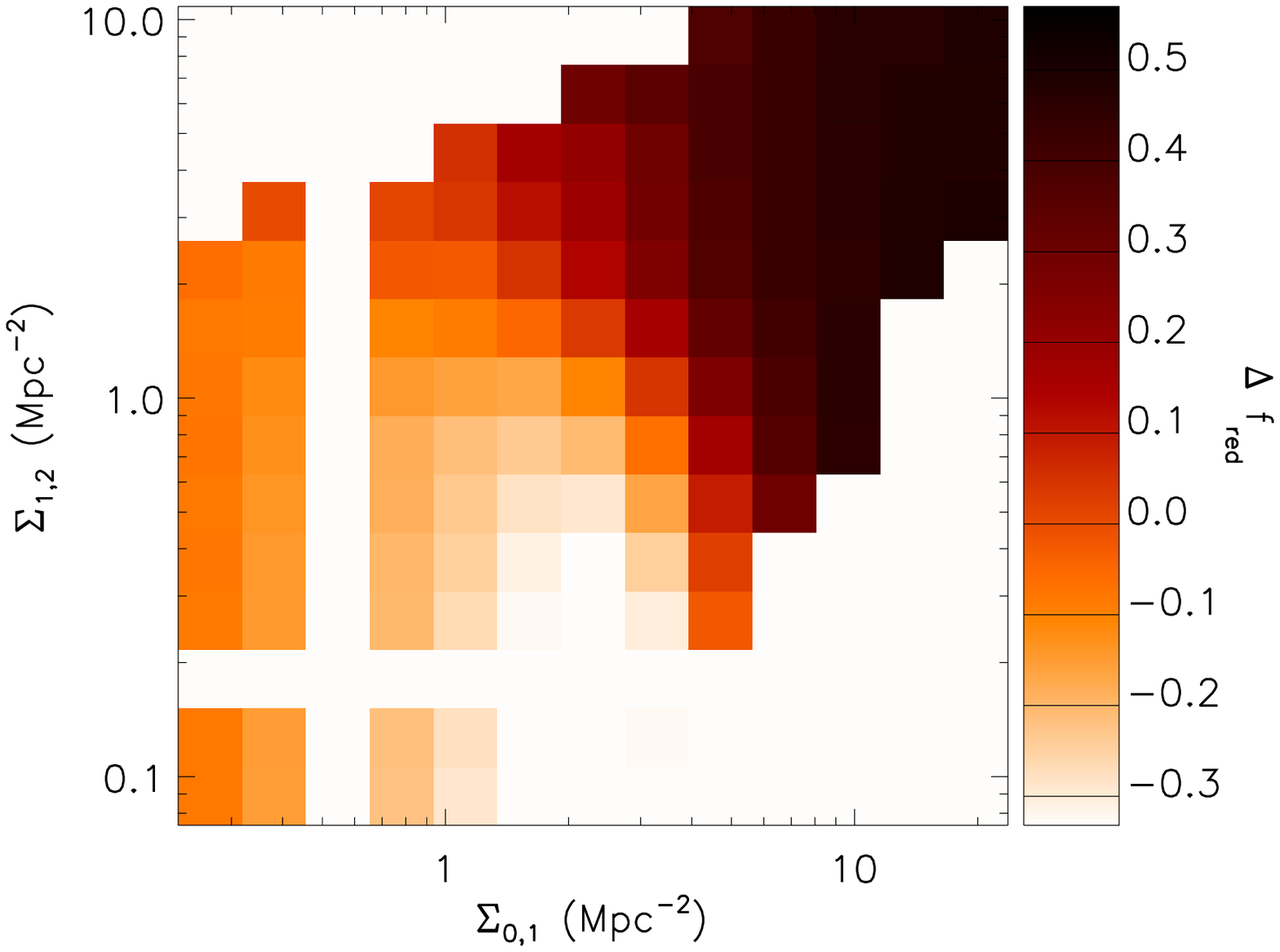,clip=t,width=8.cm}}
\centerline{\psfig{figure=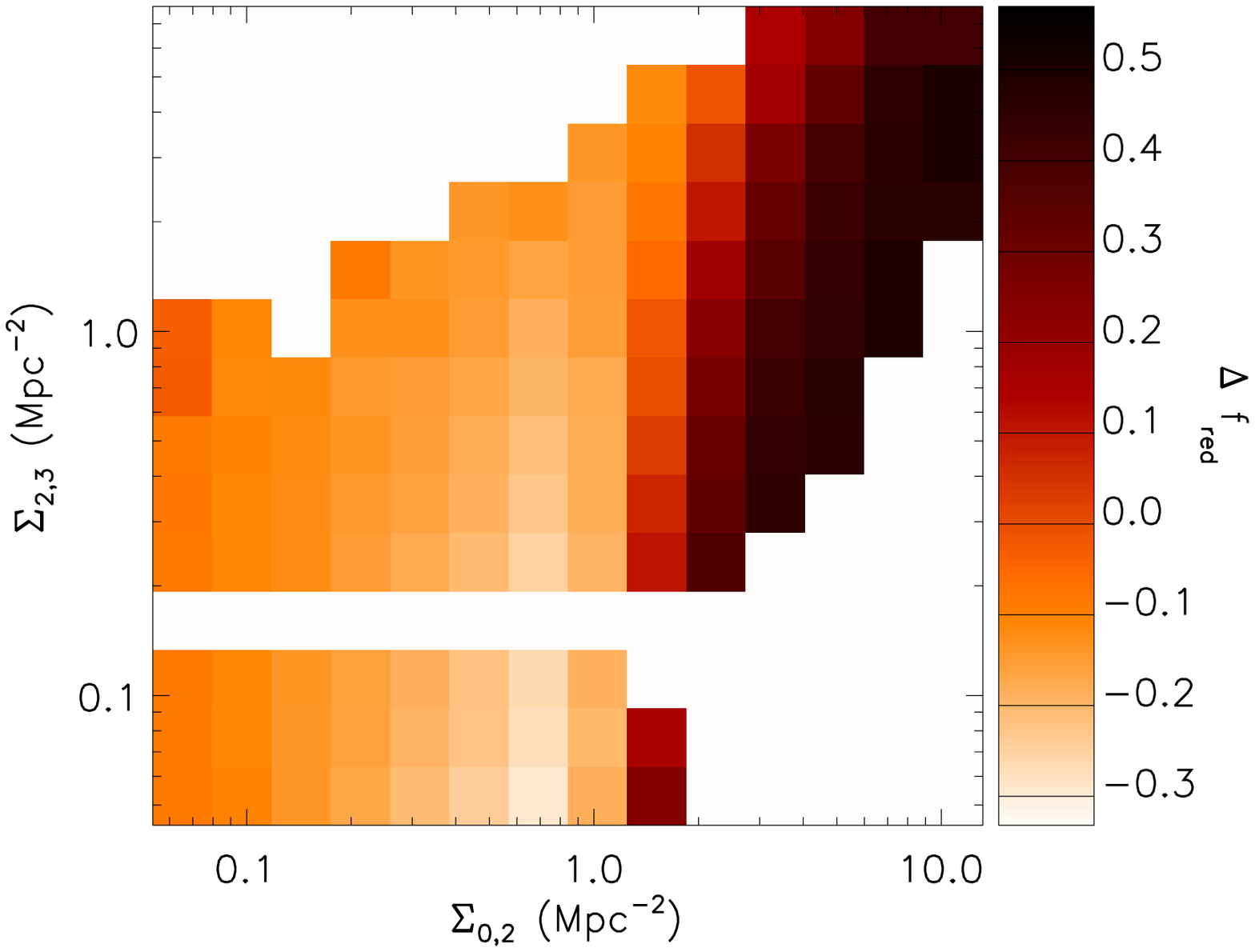,clip=t,width=8.cm}\psfig{figure=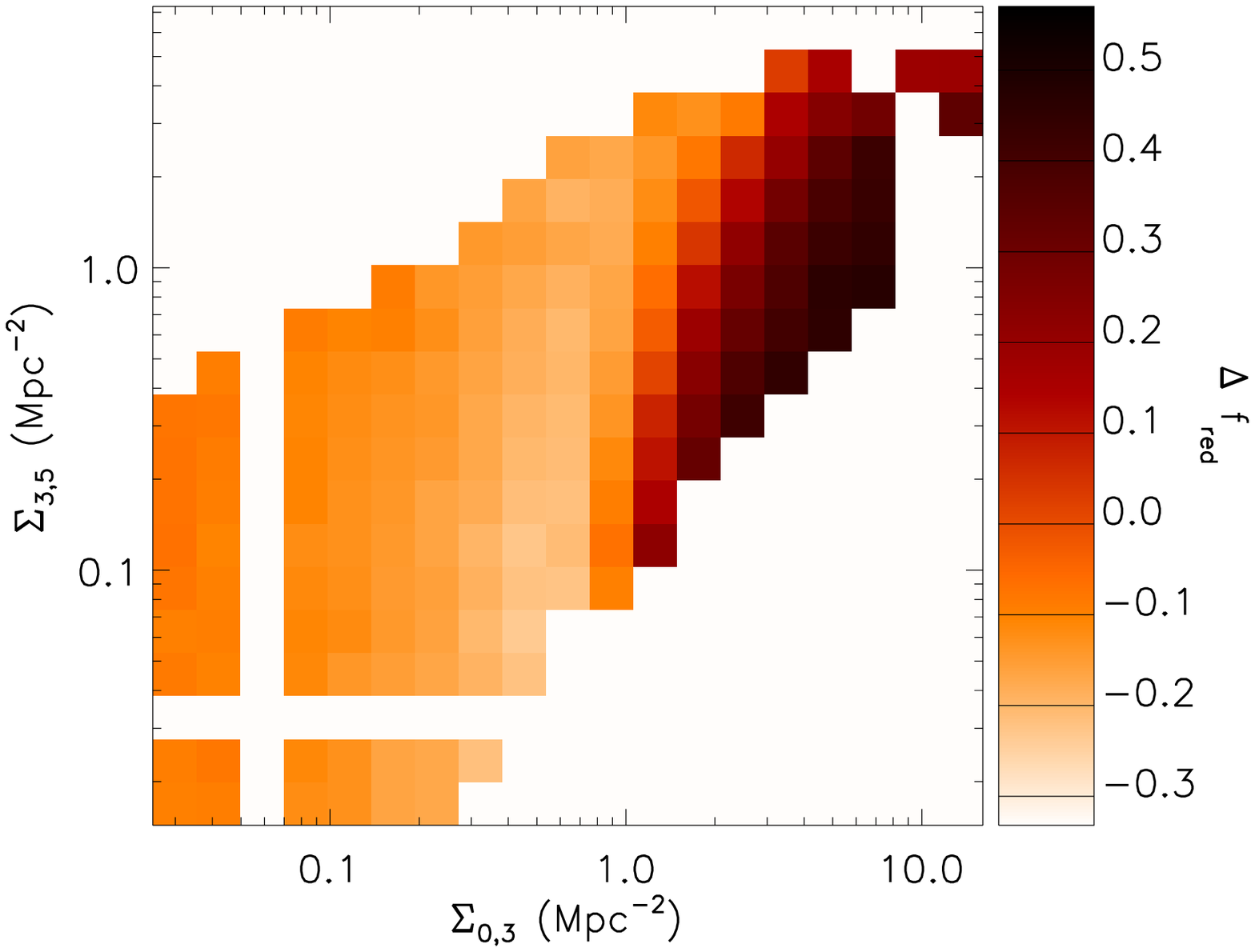,clip=t,width=8.cm}}
\caption[ ]{ Distribution of the red galaxy fraction for a model with a sharp transition ($\Gamma=0.001$) in the fraction of red satellites with halo mass, producing the correct total fraction of red galaxies. Areas which are populated by galaxies in the model but not in the observations are omitted.}\label{densdensfredsharpsats}
\end{figure*}

\begin{figure*}
\centerline{\psfig{figure=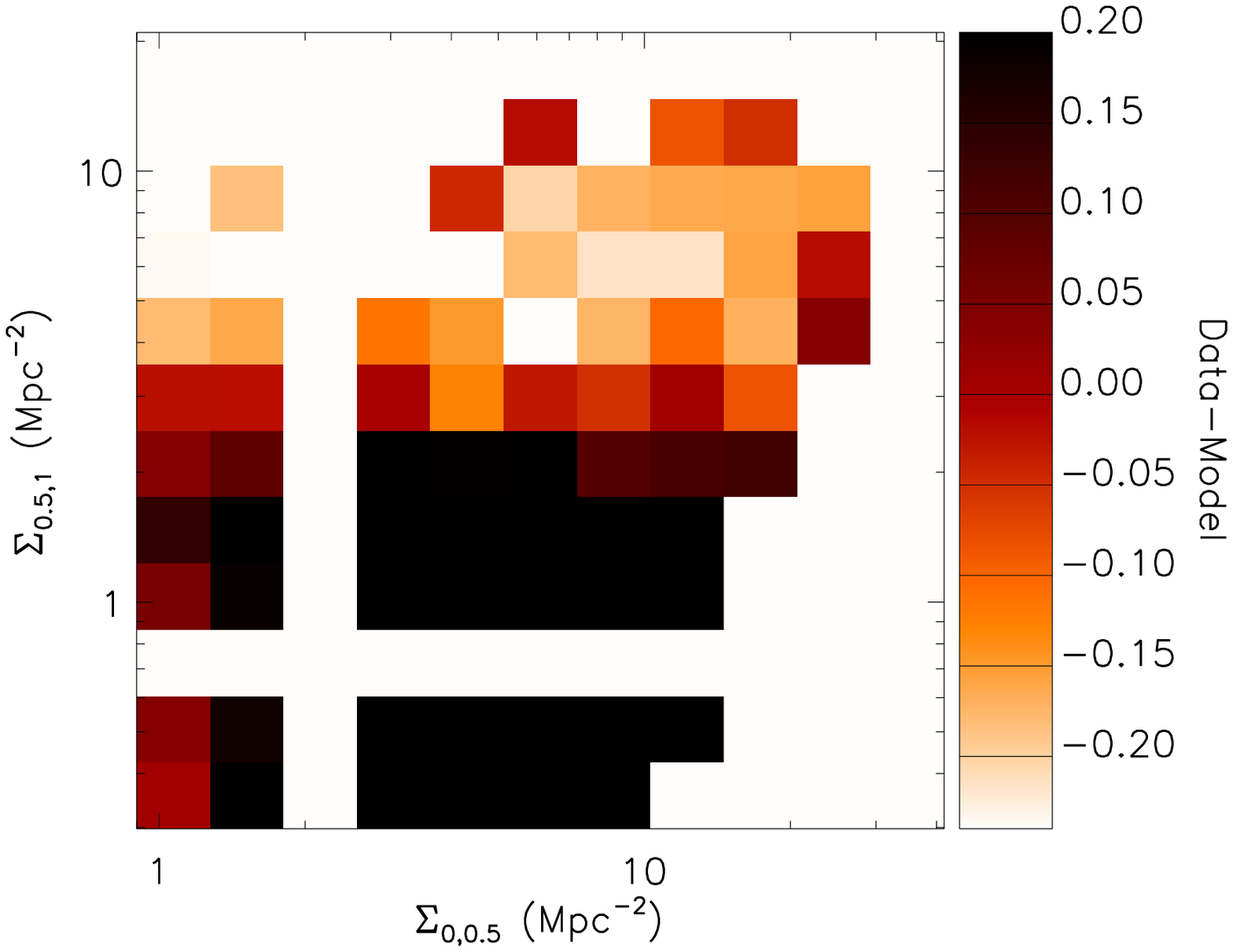,clip=t,width=8.cm}\psfig{figure=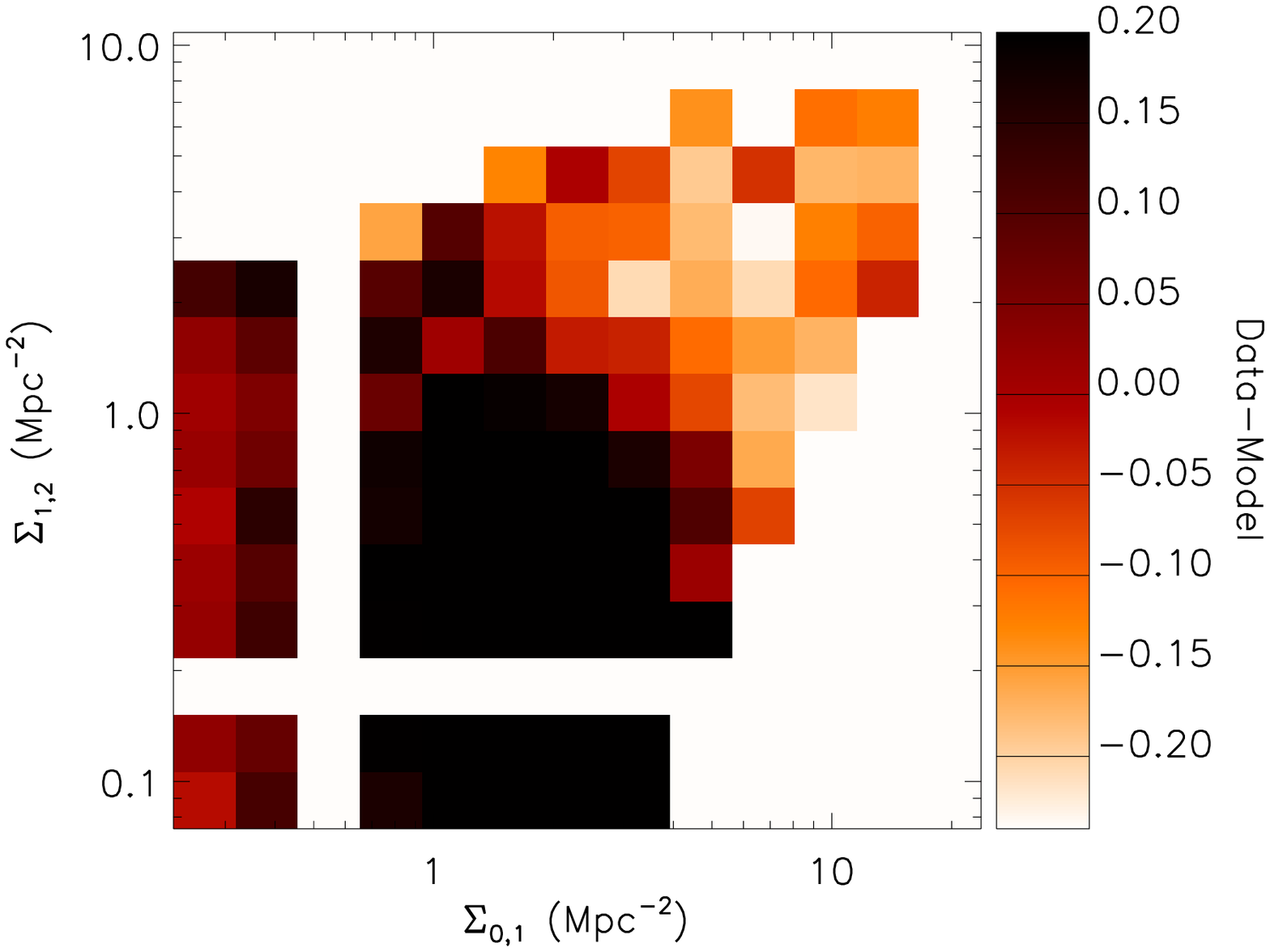,clip=t,width=8.cm}}
\centerline{\psfig{figure=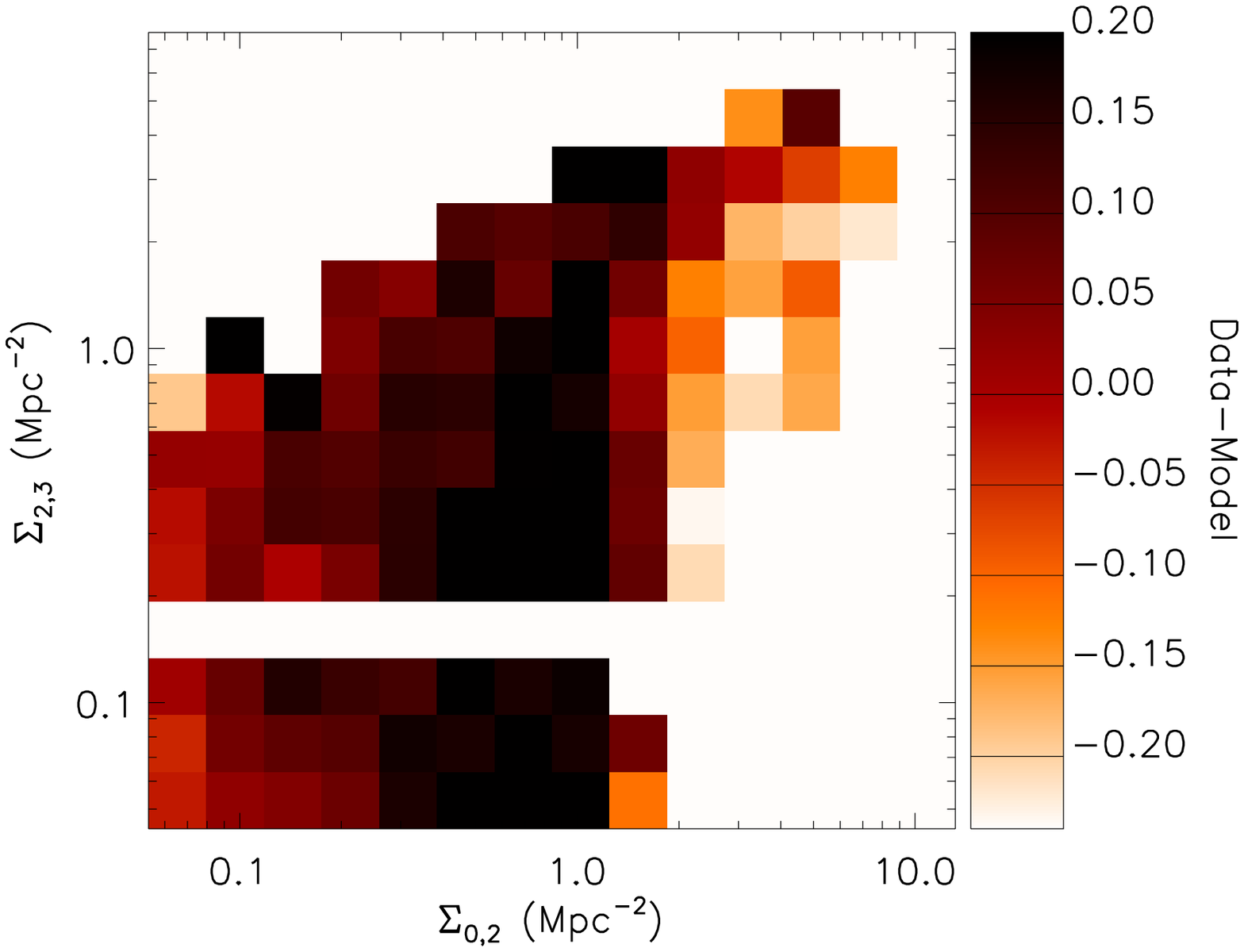,clip=t,width=8.cm}\psfig{figure=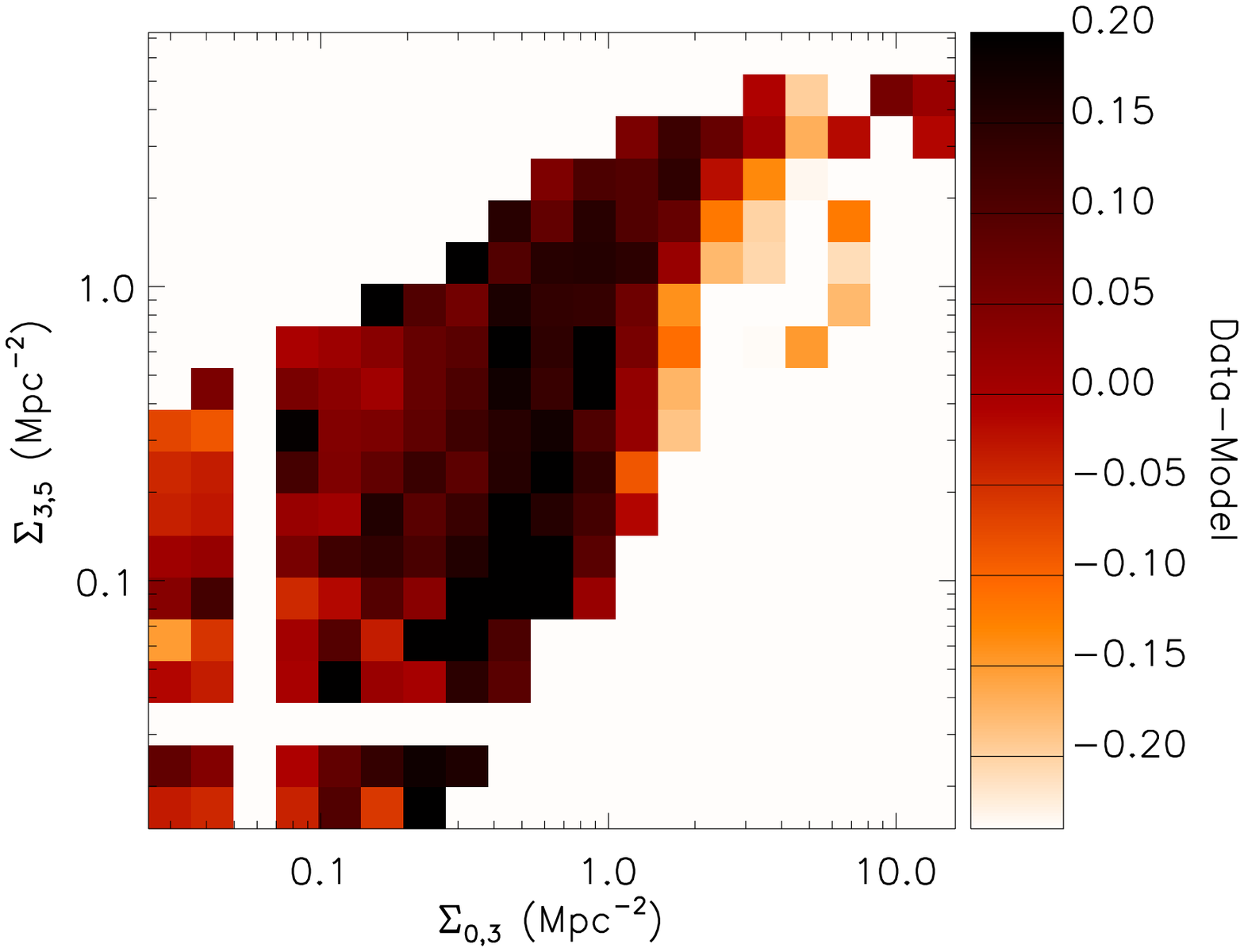,clip=t,width=8.cm}}
\caption[ ]{Distribution of the difference between the observed red fraction and that produced in the simulation for the sharp transition model, for which the red fraction is depicted in figure~\ref{densdensfredsharpsats}.}\label{fredsharpsatsresiduals}
\end{figure*}

We investigate this scenario with $\Gamma=0.001$ and $\log M_t=13.5$ -- selected such that the total red fraction is preserved. Fig.~\ref{densdensfredsharpsats} shows the dependence of the red galaxy fraction on multiscale density, while figure \ref{fredsharpsatsresiduals} shows the difference between observed and simulated red fractions in the same parameter space. These figures can be compared to those for the best-fit parameters: figures~\ref{fredmodel} and~\ref{fredresiduals}.

The residuals are much larger than for the shallow case everywhere, now not only in the highest densities on small scales, but also on intermediate densities on all scales, where the steep gradient in $f_{red}$ vs density produced in this case fails to match the more gradual slope seen in observations. The lowest densities are dominated by central galaxies in low mass halos, so these regions are not affected by the value of $\Gamma$. Intermediate regions are more mixed, containing satellites of halos of all mass with some dependence on location within the halo, as well as some central galaxies of more massive halos. Not only is the dependence of $f_{red}$ on halo mass much sharper than the best-fit case, but so is the dependence on density within this region. This cannot fit the data, as demonstrated by the very large residuals. Sudden transitions are not favoured by our data, and this is depicted by the gradual dependence of $f_{red}$ on density ({\it{Natura non facit saltus!}}).

\subsubsection{A constant fraction of red satellites?}\label{flatsatellites}

\begin{figure*}
\centerline{\psfig{figure=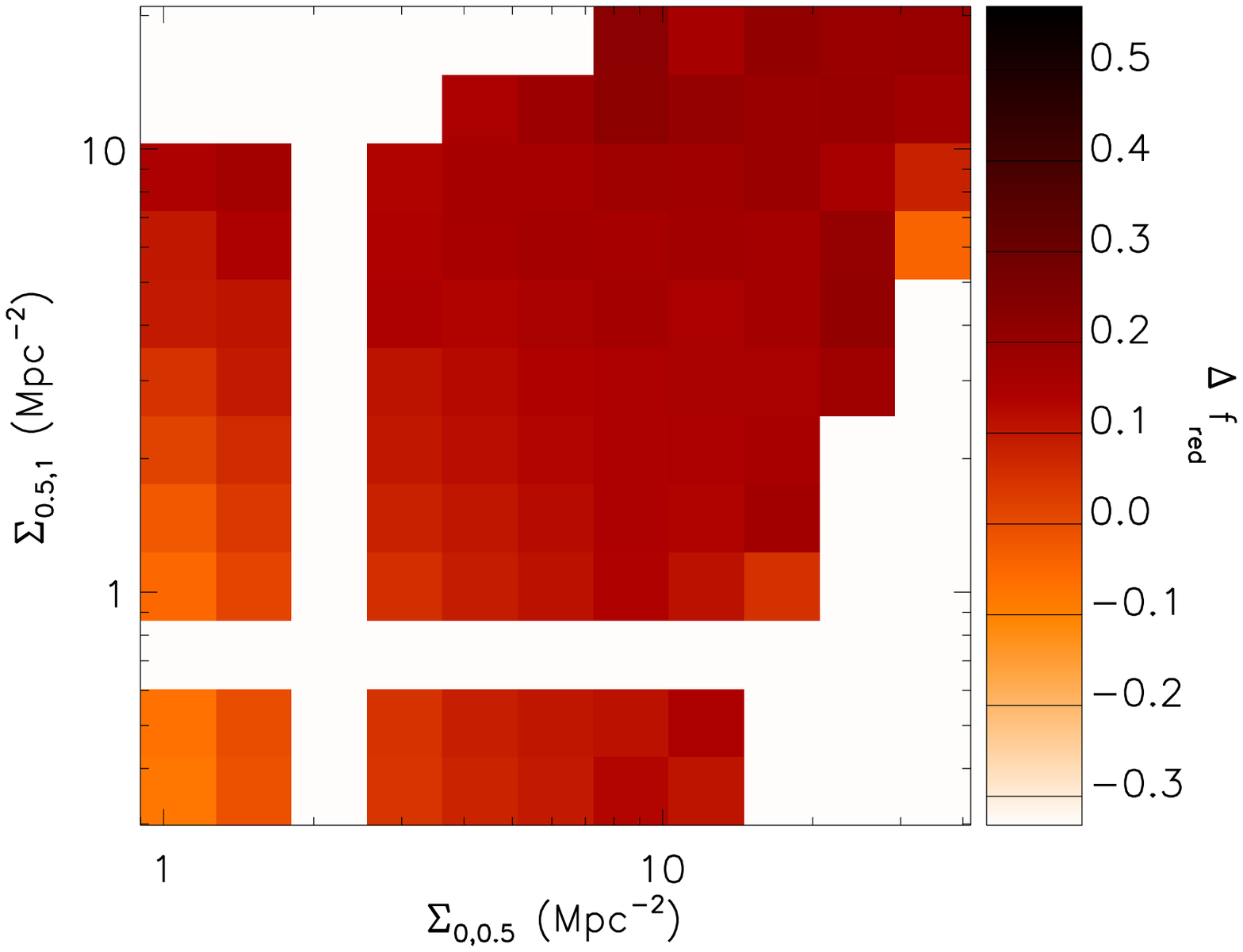,clip=t,width=8.cm}\psfig{figure=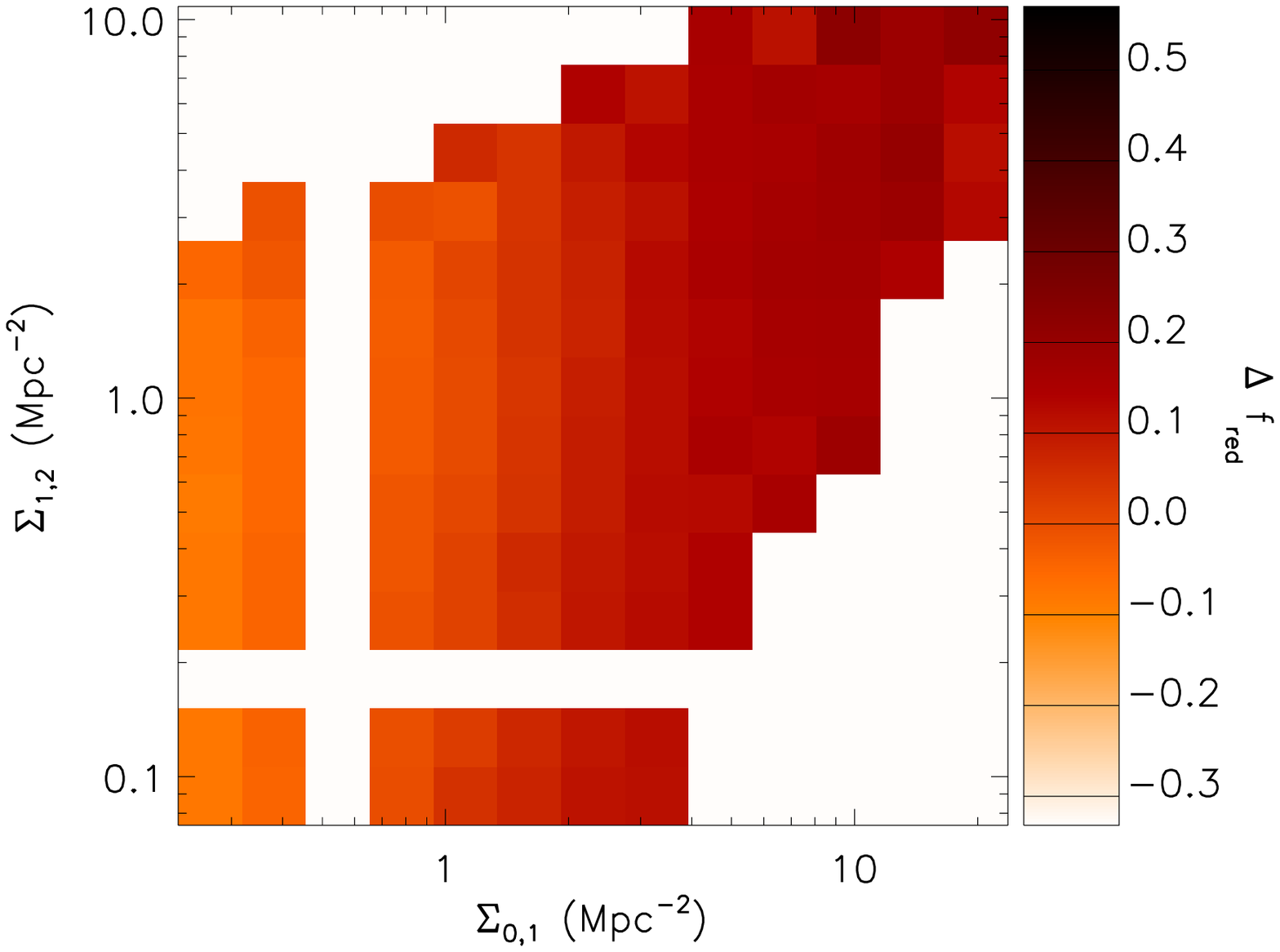,clip=t,width=8.cm}}
\centerline{\psfig{figure=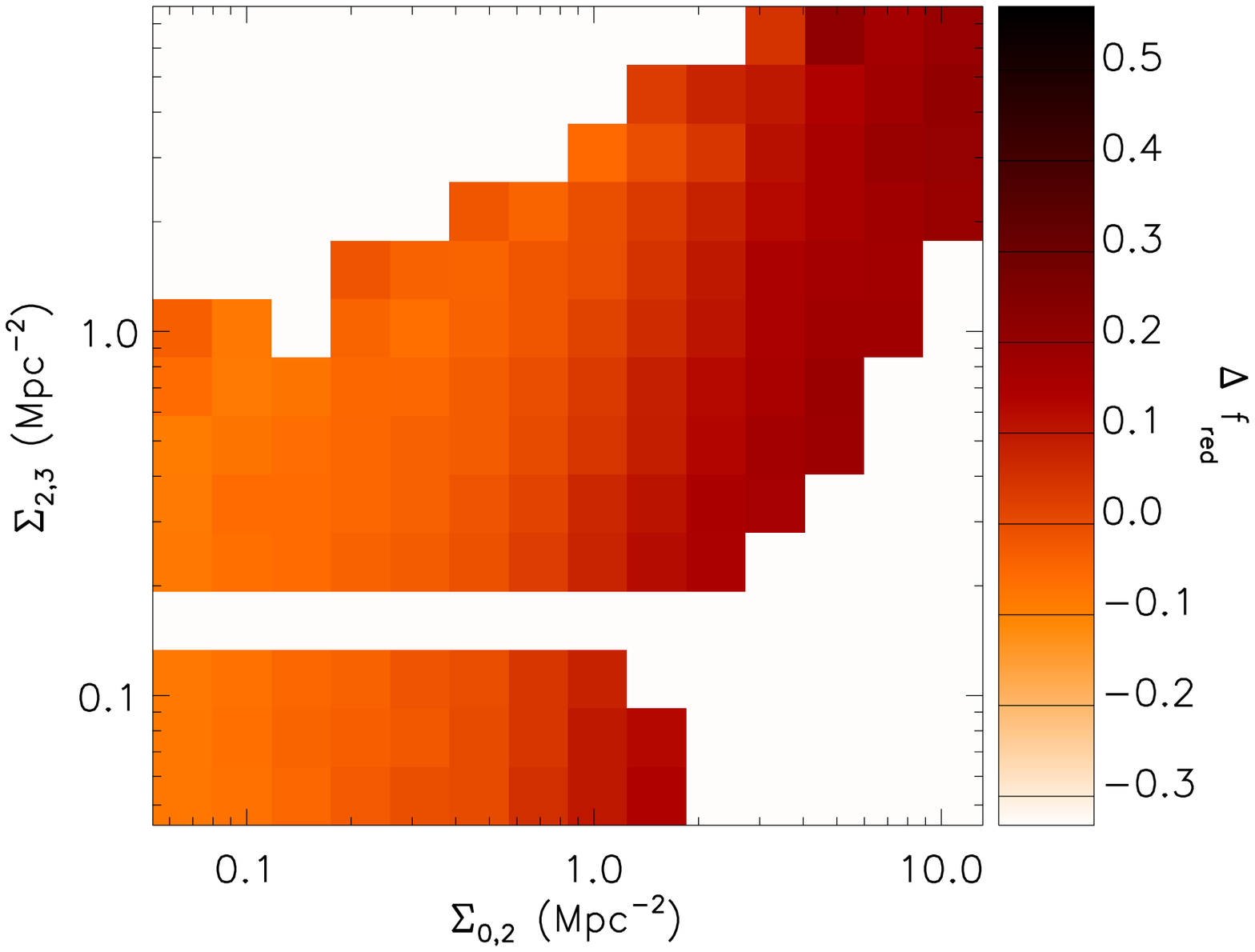,clip=t,width=8.cm}\psfig{figure=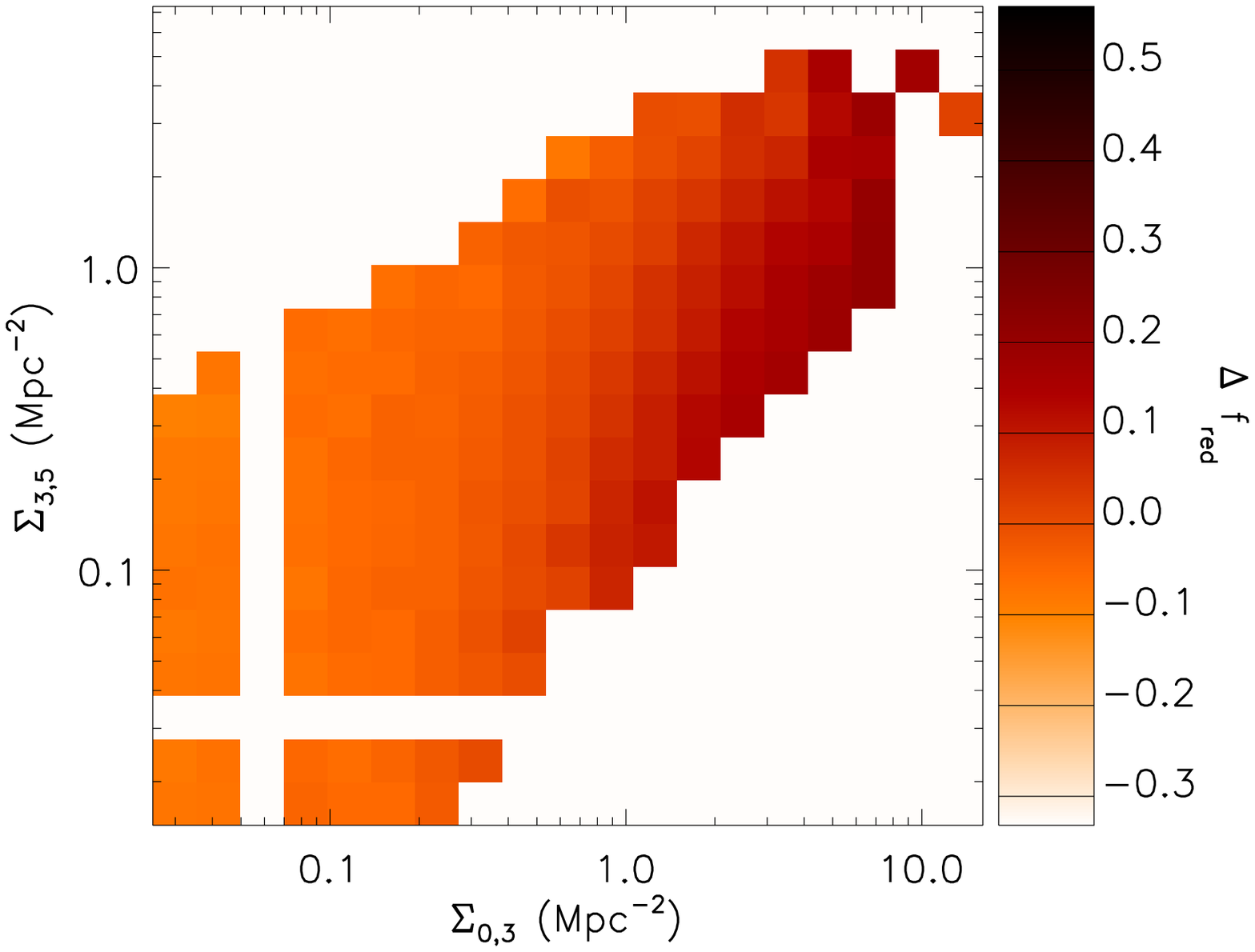,clip=t,width=8.cm}}
\caption[ ]{ Distribution of the red galaxy fraction for a model with a constant fraction of red satellites ($f^{red}_{sat}=0.575$), producing the correct total fraction of red galaxies. Areas which are populated by galaxies in the model but not in the observations are omitted.}\label{densdensfredconstsats}
\end{figure*}

\begin{figure*}
\centerline{\psfig{figure=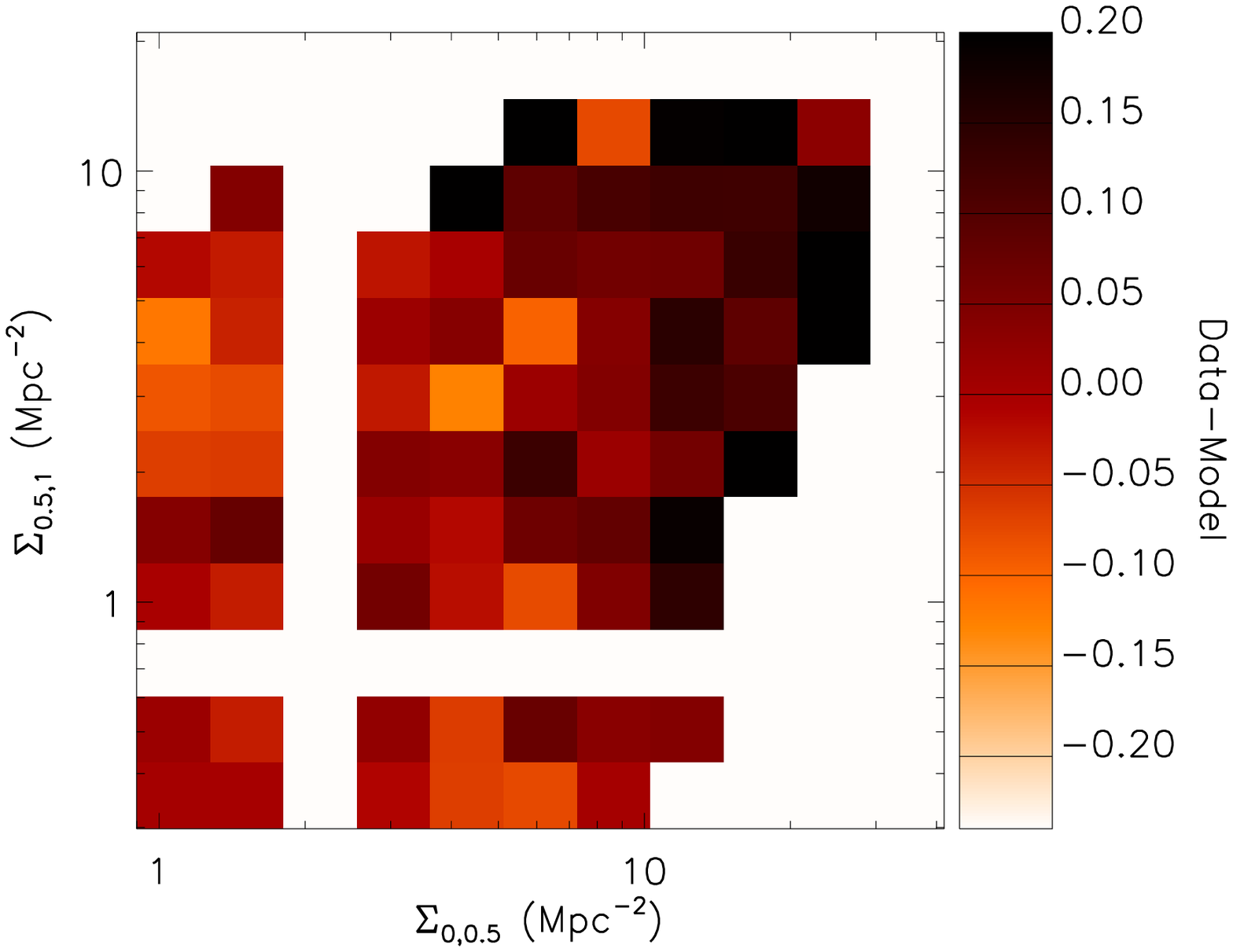,clip=t,width=8.cm}\psfig{figure=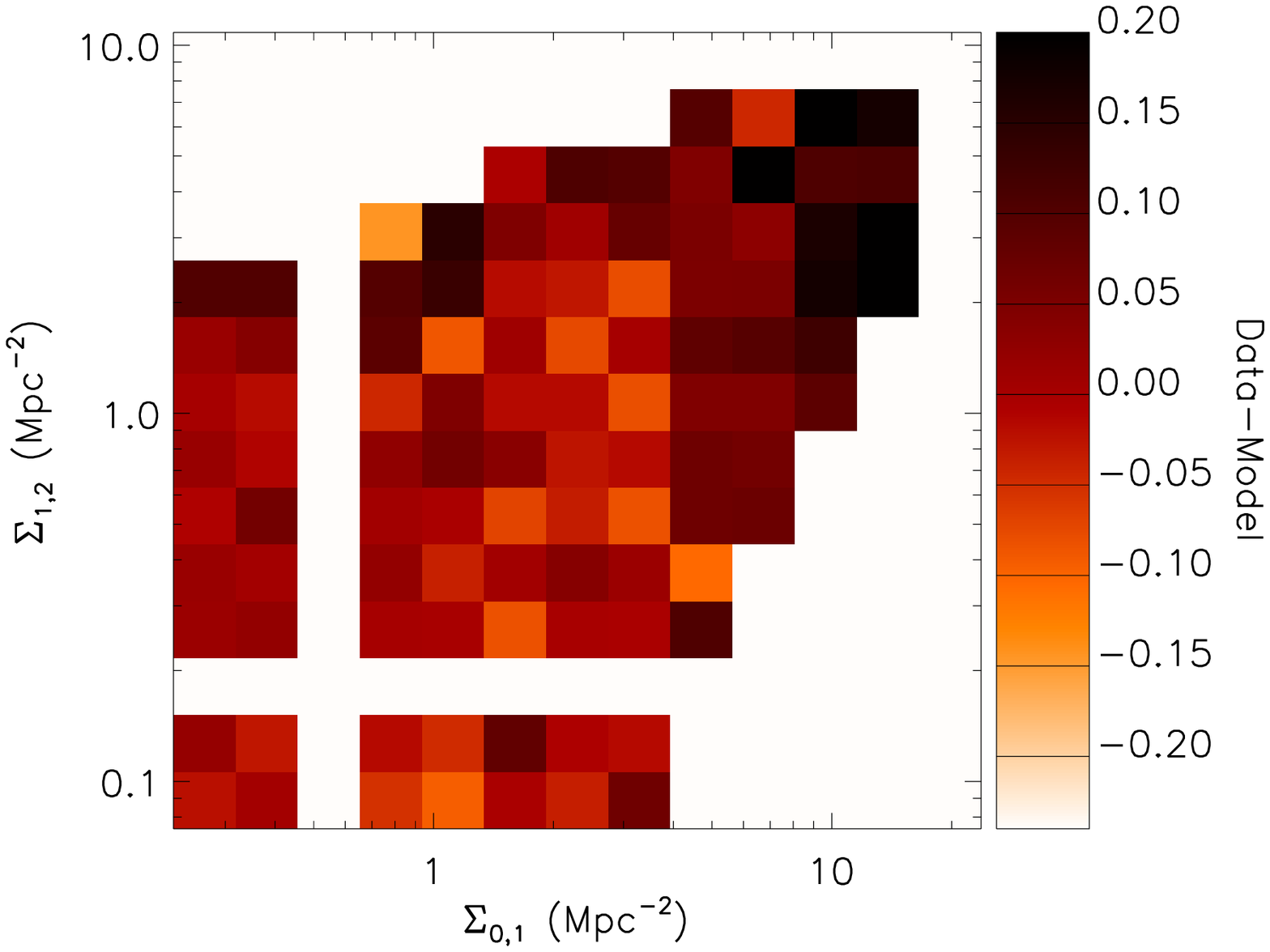,clip=t,width=8.cm}}
\centerline{\psfig{figure=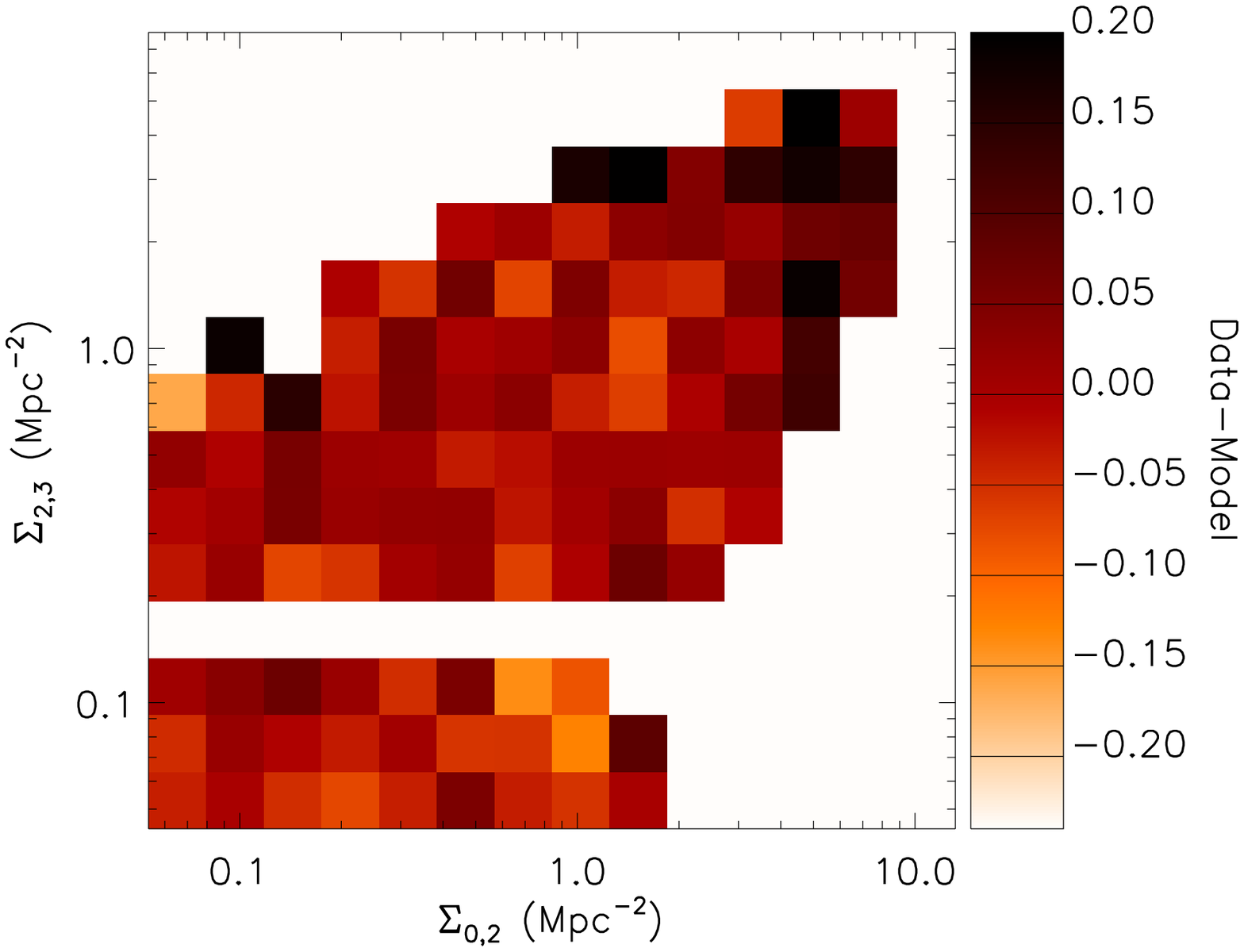,clip=t,width=8.cm}\psfig{figure=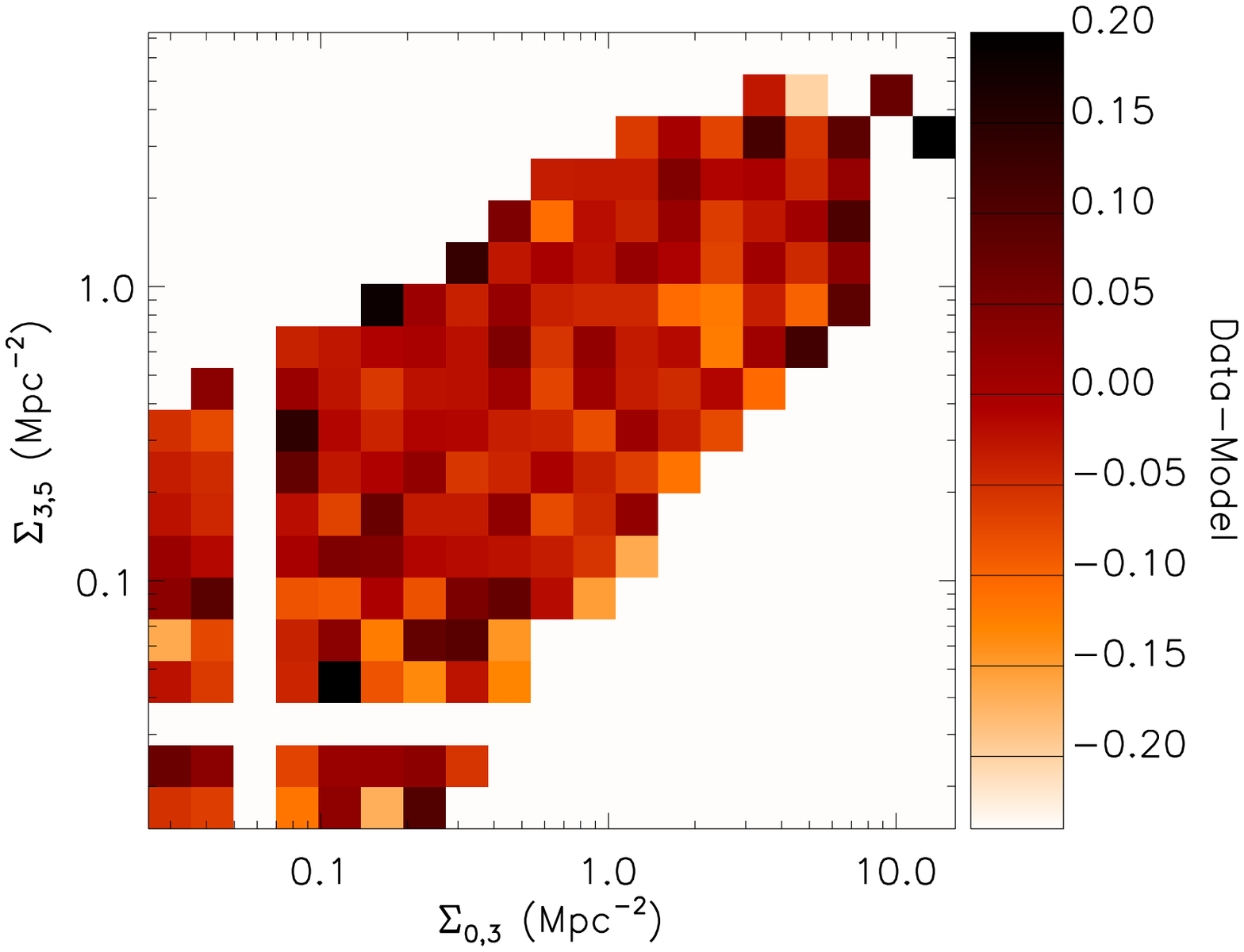,clip=t,width=8.cm}}
\caption[ ]{Distribution of the difference between the observed red fraction and that produced in the simulation for a constant $f^{red}_{sat}=0.575$, for which the red fraction is depicted in figure~\ref{densdensfredconstsats}.}\label{fredconstsatsresiduals}
\end{figure*}

Since we have seen that the slope describing the increase of the fraction of red satellite galaxies is very shallow, we also tested a scenario in which $f^{red}_{sat}$ is constant (but the total red fraction the same as before, so  $f^{red}_{sat}=0.575$). Fig.~\ref{densdensfredconstsats} shows the distribution of the red galaxy fraction in this model, and Fig.~\ref{fredconstsatsresiduals} the difference between the observed red fraction and that produced in the simulation. The values of $\chi^2$ we calculate for the fits of the four combinations of scales for this model (132.95, 41.83, 37.60,and 10.44, respectively) are slightly larger than the minimum $\chi^2$ we get from the fit with Equation \ref{redcenfraceq} (101.15, 30.59, 32.69, and 11.09, respectively). While a constant red satellite fraction can not be totally ruled out by the values of $\chi^2$, this model can not explain the observed high red galaxy fractions in high densities on small scales; again colour segregation would have to be included to improve the fit.

\subsubsection{A sharp transition mass of the red central fraction?}\label{sharpcentrals}
We have so far assumed a constant red central galaxy fraction across the limited range of halo mass for the select sample. However, a number of authors (see e.g. \citealp{DekelBirnboim06,KhochfarOstriker08}) have postulated that central galaxies cease to accrete gas when it is shock-heated above a well-defined halo mass, and/or that AGN-feedback suppresses further cooling above this mass \citep{Cattaneo06}. This could lead to a decreased supply of gas for star formation and to galaxy colours turning red. 

\begin{figure*}
\centerline{\psfig{figure=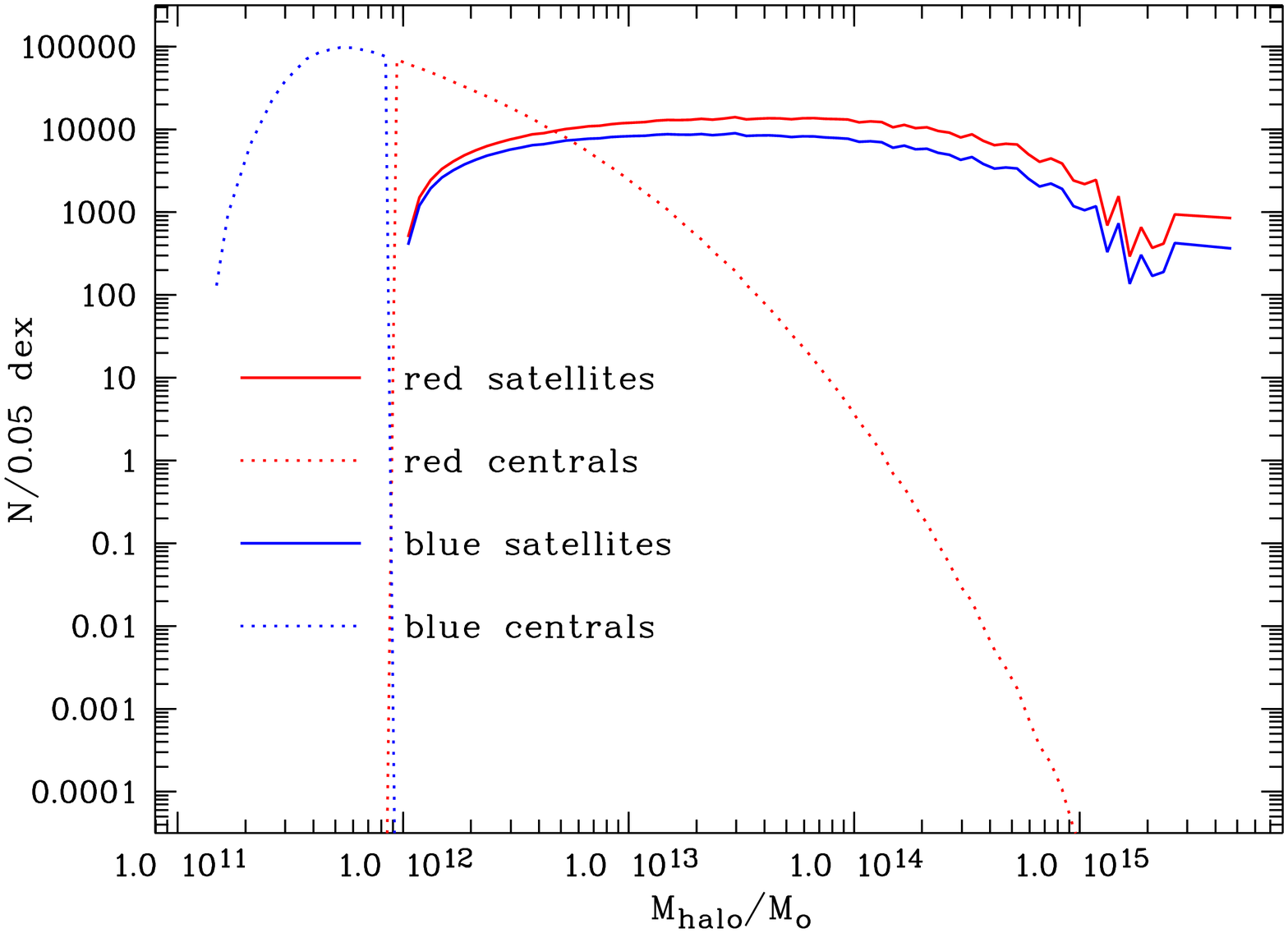,angle=270,clip=t,width=9.cm}}
\caption[ ]{The total number of red and blue galaxies as a function of halo mass (with the mass function measured from the Millennium simulation), given the HOD parameters determined from the measurement of \citet{Zehavi11} and the values of $f^{red}_{max}$, and $f^{red}_{min}$ as determined from the data. Here we assume a step function for the red central fraction. The transition mass $M_{t,cen}= 9.035\cdot 10^{11} M_\odot$, at which all centrals are red (while for $M_{halo}< M_{t,cen}$ all central galaxies are blue) has been chosen such that the total red central fraction is the same as in the constant case, $\bar{f^{red}_{cen}}=0.38$.}\label{Nredstepcen}
\end{figure*}

To examine this hypothesis, we test the most extreme model possible: a sharp transition in the red central fraction such that $f^{red}_{cen}=0.$ for $M<M_{t,cen}$, and $f^{red}_{cen}=1.$ for $M \ge M_{t,cen}$. $M_{t,cen}$ is selected such that the total  of red central galaxies is the same as in our model in which $f^{red}_{cen}=const=0.38$: we find $M_{t,cen}=9.035\cdot 10^{11} M_\odot$. Fig.~\ref{Nredstepcen} shows the number of red galaxies as a function of halo mass for this scenario. This can be compared to figure \ref{Nred} for our default model with a constant red central fraction.

\begin{figure*}
\centerline{\psfig{figure=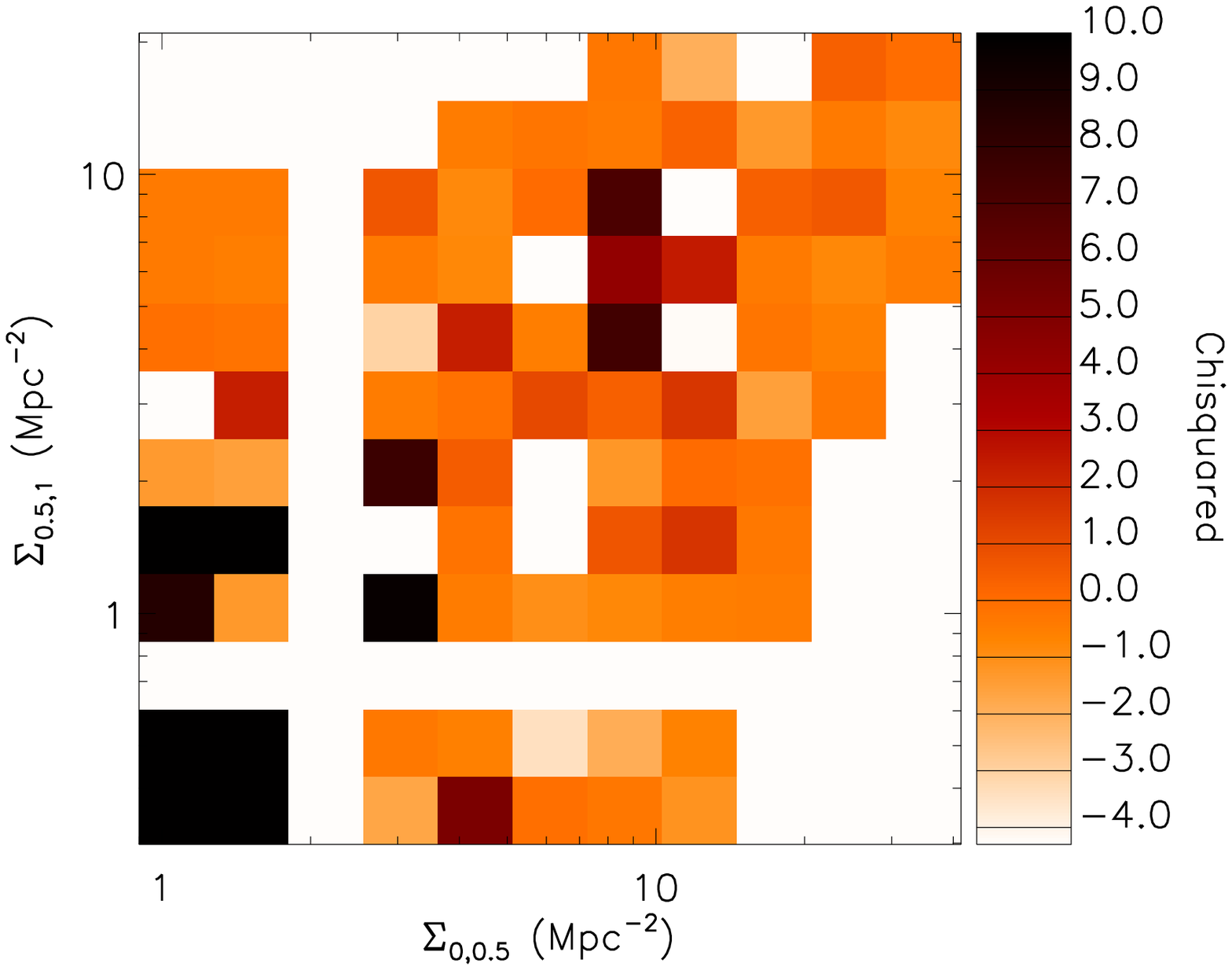,clip=t,width=8.cm}\psfig{figure=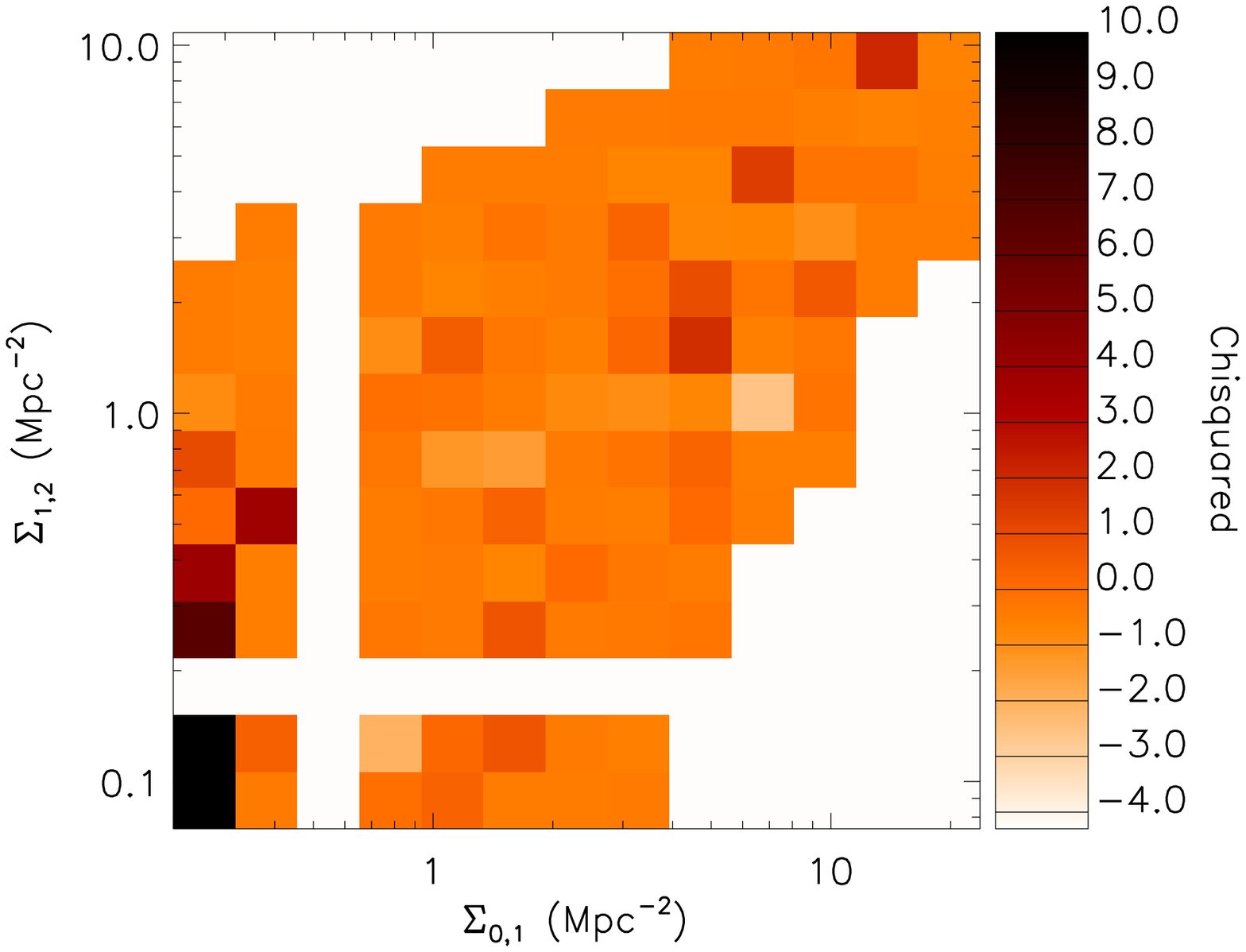,clip=t,width=8.cm}}
\centerline{\psfig{figure=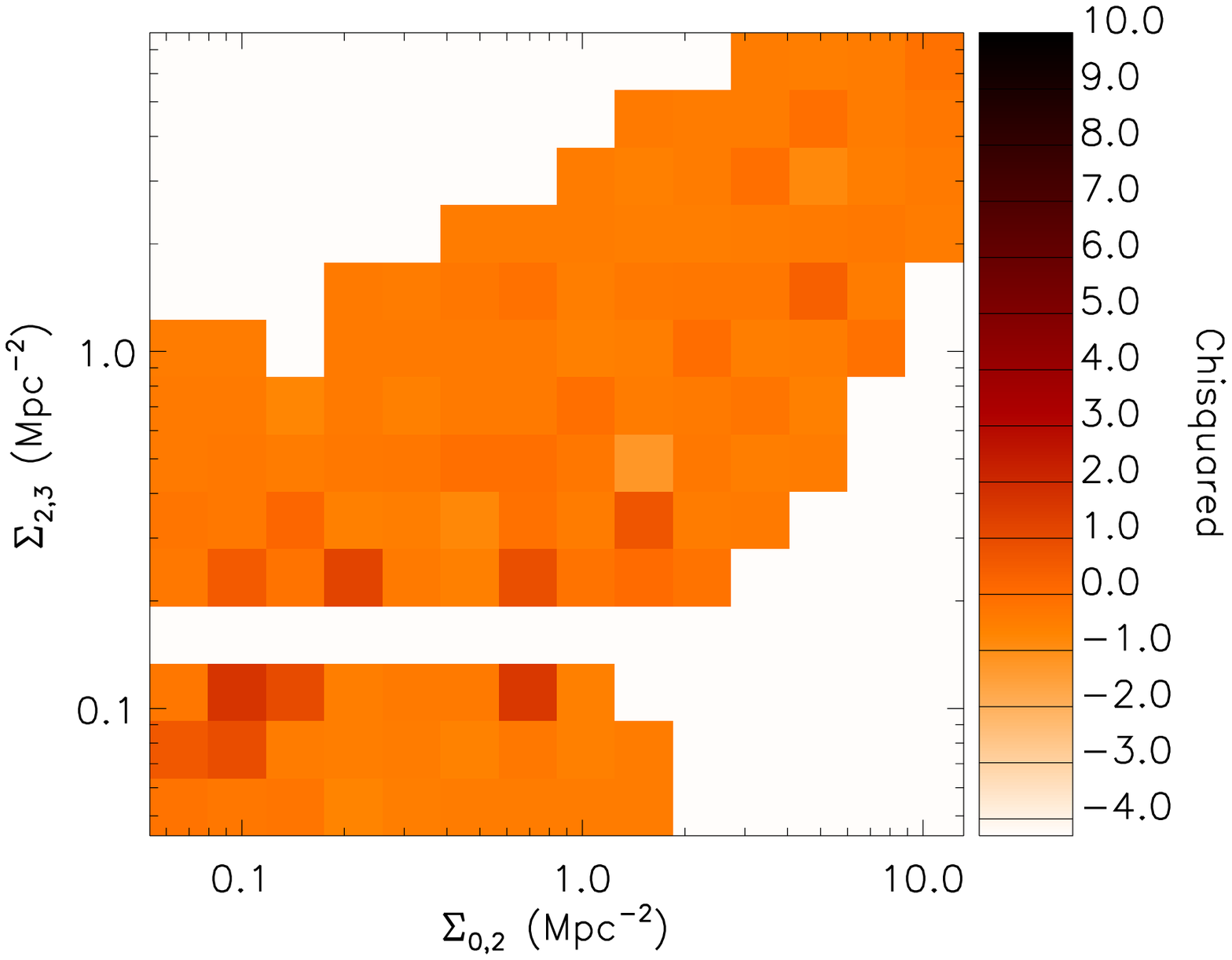,clip=t,width=8.cm}\psfig{figure=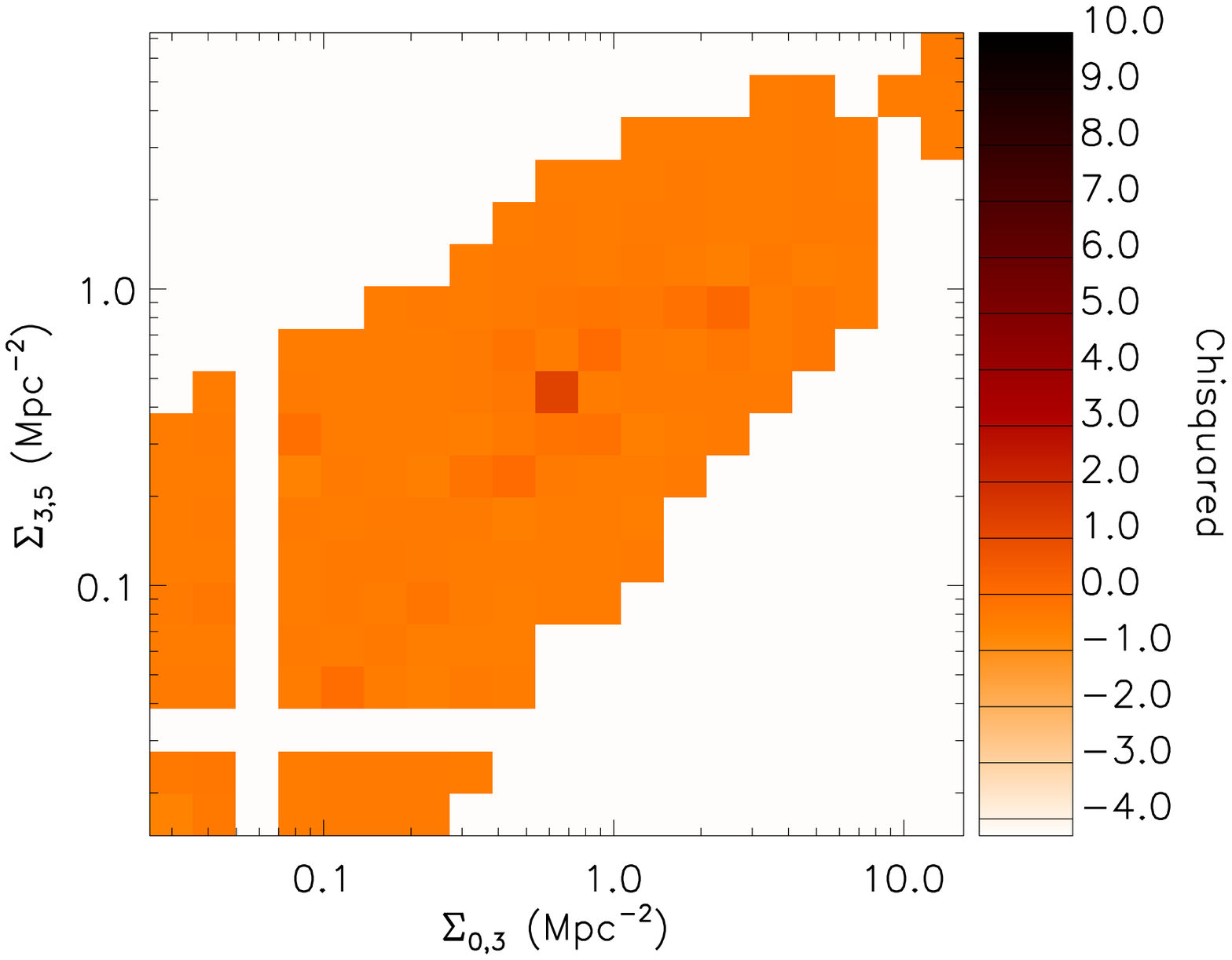,clip=t,width=8.cm}}
\caption[ ]{Distribution of $\chi^2$ for a step function for the red central fraction}\label{chisquaresharpfredcen}
\end{figure*}

Since with a sharp transition there are almost no red central galaxies for halo masses $M_{halo}< M_{t,cen}=9.035\cdot 10^{11} M_\odot$, and no satellite galaxies at all (neither red nor blue, see Fig.~\ref{Nredstepcen}) at these low masses, this halo mass range is now totally dominated by blue central galaxies. We refit $\log M_t$ and $\Gamma$ for the red satellite fraction, and find that the $\chi^2$ distributions of our four combinations of scales ) overlaps at $\log M_t= 14.4$ and $\Gamma\ga 2.5$.  Fig.~\ref{chisquaresharpfredcen} shows $\chi^2$ for single density bins (calculated taking the covariances between all the other bins into account). At the lowest densities, on small scales, $\chi^2\ga 150$ (well beyond the scale of the plot). This happens because the only galaxies with no neighbours on small scales are the centrals of low mass halos which, in the sharp central transition scenario, are all blue -- in direct contrast to observations. Therefore a scenario in which there are no red central galaxies in low-mass halos is not able to explain the observed distribution of red galaxies in multiscale density parameter space. We therefore conclude that our analysis rules out such a very steep transition of the fraction of red central galaxies with halo mass. Less extreme, shallower transitions, or transitions outside the range of halo mass for centrals covered by our select sample, are nonetheless still possible.

\section{Discussion}\label{Discussion}
\subsection{Assessment and the literature}
How galaxy properties depend on their environment has been investigated extensively using correlation functions as a means to quantify the clustering strength as a function of galaxy type. The HOD model of galaxy clustering can then be used to infer typical halo masses from the measured two-point correlation functions (\citealp{Zehavi04,Zehavi05,Phleps06,Zheng09,Zehavi11} and references therein). Correlation functions have the advantage that they are relatively easy and quick to compute analytically. However, the correlation function can only be calculated as a function of galaxy properties, while the multiscale density parameter space shows the properties as a function of density on all scales. The strength of the method lies in its ability to disentangle direct correlations of galaxy properties with density on each scale from the indirect correlation originating in the correlation between densities of different scales. This parameter space also make it possible to understand where individual or sub-popuations of satellite and central galaxies live, since each galaxy can be allocated a multiscale density. The correlation function in contrast merely measures the median clustering, and all galaxies contribute to many data points, with information about centrals and satellites distributed over the full range of scales. 

The mechanism(s) leading to the suppression of star formation in galaxies and the transition from blue to red colours have been the focus of much research in the last decades. While there is no definitive answer to date, many different methods have been employed to approach the question and build evidence. While most results are not directly comparible with ours due to a different selection, we can compare with a couple.

Using a very similar HOD description, \citet{Tinker10} assume a constant fraction of red satellites and a variable fraction of red centrals, parameterized with a function which can be steep or shallow, depending on a combination of three parameters. By fitting the correlation function of red galaxies at different redshifts ($0.4<z<2.0)$, they exclude a sharp transition scenario for central galaxies, consistent with our results. They also find that ``at all redshifts, including z=0, the data are consistent with $\sim60$\% of the satellite galaxies being red''. Although their definition of ``red'' is not exactly the same as ours, and they assume the red satellite fraction to be independent of host halo mass, we find a comparible total red satellite fraction of $f_{red}^{sat}=0.58$. 
 
\subsubsection{Comparison with a group catalogue}

Group catalogues provide a complementary method of assessing the environment of individual galaxies. This has the advantage that halo membership and proxies for halo mass are computed on a group by group basis. The disadvantage with respect to multiscale density parameter space is that every galaxy has to be assigned to a single halo / group, which is sensitive to the halo finder algorithm and ignores second order effects such as unvirialized halos and sub-halo clustering. Multiscale densities account for this because the measured distribution of densities on multiple scales should be the same in mock catalogues and in the real Universe. Thus they provide a statistical method to account for all environments.

\citet{Weinmann06} presented perhaps the most directly comparible results to ours, utilising the group catalogue from \citet{Yang05} based on SDSS DR2. In similar bins of luminosity they find a fairly gradual increase in red fraction with halo mass for both satellite and central galaxies. However their method of categorising galaxies by colour and specific star formation rates differs from ours.

To perform a better comparison, we take the DR7 version of the group catalogue (updated from the published DR4 version \citet{Yang07}) and cross-correlate with our sample. We take all galaxies defined by \citet{Yang07} as satellites and bin in luminosity and halo mass. In order to minimise systematics when comparing to our luminosity-limited sample (see below), we choose the version which uses luminosity rather than stellar mass to define the central galaxy (the most luminous one) and halo mass (rank ordered on total galaxy r-band luminosity, see \citet{Yang07} for more details). We then bin the satellite galaxies in luminosity and halo mass and apply exactly the same double gaussian fitting procedure as described in section~\ref{redgalfracdatasection}. The red fraction is defined as the total fraction integrated over the three luminosity bins (i.e. weighted by the number of galaxies in each bin).

The dependence on halo mass of the resultant red satellite fraction is overplotted in Fig.~\ref{frmaxdemo}. It is immediately clear that the trend of red satellite fraction with halo mass is neither close to a sudden transition nor as shallow as our solution based on multiscale densities. A best fit gives $\Gamma=1.6$ and log $M_t=14.4$.

The reasons for this apparent discrepancy are complex, yet important to understand. Firstly we note that at realisitically high halo masses the two solutions converge to a value $f^{red}_{sat} \sim 0.65$, consistent with the value seen at high density on large scales. It seems reasonable that this is truly a typical {\it global} value for massive halos, while segregation can dominate local values on smaller scales. The two methods diverge from each other at low halo mass, where a much lower $f^{red}_{sat}$ is found using the group catalogue than is inferred from our fits to the multiscale densities.

The steep fit to the group catalogue is inconsistent with the 90$\%$ confidence limits even on larger scales (Fig.~\ref{chisquared_logMtGamma}) and produces too few red satellite galaxies in total (see Fig.~\ref{constraintsfromfredtot}). This is only possible because the red fraction of {\it central} galaxies is higher than our choice. Indeed, in halos of log $M_{halo} \sim 11.5-12.5$ which host the central galaxies of our sample, the red fraction of group catalogue centrals is almost identical to that of group catalogue satellites, and significantly higher than that seen in the lowest bins of density (the most isolated galaxies). This illustrates the main problem with allocating every galaxy to a halo in a group catalogue: it works on average, but, especially in low mass halos, there is no way to accurately assess whether each galaxy belongs to the same halo as the next one without full six-dimensional phase space information (never observable). Therefore some halos will be artificially divided while others will be artificially joined. Satellites of low mass halos within our luminosity range will be: a) of similar luminosity to the ``central'' galaxy, and b) often not really part of the same halo. Together these effects work to force the red fraction of central and satellite populations to be roughly equal, but larger than the more conservatively estimated ``isolated'' population of true centrals in our lowest density bin.

Finally, we note that our prescription does not allow the red fraction of satellites to be all of: very shallow (as in our best fit); reach the maximum value seen on large scales; {\it and} reach $f^{red}_{min}$ within populated halos ($\gtrsim 10^{12}\Msol$). This gives our functional form the freedom to become very shallow by reaching its minimum value at some low mass of halo which is never encountered. Our fits appear to be primarily constrained by the following requirements: that a sensible maximum red fraction is reached; that the transition with density is not sudden; and that the total red fraction is roughly correct. This final requirement, together with the low red fraction of central galaxies (as observed for isolated galaxies), requires a shallow slope. Together with the first two constraints, this means that $f^{red}_{min}$ (and likewise the low red satellite fractions derived from the group catalogue at low halo mass) {\it cannot} be obtained. Indeed, there is no guarantee that the red satellite fraction {\it should} be as low as the red fraction of centrals: our condition $f^{red}_{min} = f^{red}_{cen}$ is merely designed to apply a sensible lower limit.

It is nonetheless clear that, regardless of the very different biases of the two methods, that an at least {\it fairly} shallow slope is preferred over a sharp transition.

The gradual increase of red satellite fraction with halo mass is not consistent with a scenario in which galaxies only become red in halos above a given mass. Rather, it favours a mechanism which acts hierarchically, such that galaxies in more massive halos simply have a larger probability of having experienced an event which suppresses star formation. The gradual dependence of red fraction on density also encourages the idea that this does not require a dense sub-clump within a halo of whatever mass, but can happen (though with lower probability) in a low density region of a lowish mass halo. Central, isolated galaxies also become red, and so it seems not inconceivable that the satellite galaxies in low density regions and low mass halos experience, with enhanced probability, the same or similar mechanism as the central galaxies.

\subsection{Limitations and potential future improvements}
This paper presents for the first time a simple HOD-based fit to multiscale density parameter space. It is designed to test whether a simple prescription of red fraction which only depends upon halo mass and central vs satellite classification can describe the full dependence of red fraction on multiscale density -- and it does indeed describe this full parameter space very well, considering its simplicity.  Clearly there is nonetheless much room for improvement or further adaptations of both model and method -- a few examples are:\\

\noindent{\it Improvement of the modelling:}
\begin{itemize}
\item Our allocation of galaxies -- one per subhalo and the remainder to follow a NFW profile -- could be improved using a subhalo HOD ($\geq 1$ galaxy per subhalo), in particular in the halo accretion regions. While this may be more realistic, it has the disadvantage that it introduces additional free parameters.
\item On scales smaller than halos, galaxies are segregated such that the red fraction increases with decreasing halo-centric radius, especially in clusters, e.g. (\citealp{Melnick77,Balogh00,Girardi03,deLucia12}). This could be matched by adding a dependence of the colour of a galaxy on e.g. subhalo mass or/and halo-centric distance to the HOD recipe. 
\item Our current model assumes that the minimum red fraction of the satellites at small halo masses is equal to  the total fraction of red centrals {\it at redshift $z=0$}, $f^{red}_{min}=f^{red}_{cen}$. The assumption is based on the fact that all satellites have been centrals before they fell into a more massive halo and became satellites. However, at the time of accretion the fraction of red central galaxies may have been different from that seen in the local Universe. Nonetheless, we can regard $f^{red}_{min}$ as the fraction of red satellites which we would find if the satellites evolved in exactly the same way as the centrals. An improvement of the model would be to take the time since accretion into account, e.g. by tracking of individual galaxies via their host haloes in the simulation, while self-consistently track their colours.
\item Using a more realistic numerical N-body simulation (with lower value of $\sigma_8$) as input, with larger volume and higher resolution will be a better description of the clustering and mass function of the dark matter halos.
\end{itemize}

\noindent{\it Improvement of the analysis:}
\begin{itemize}
\item At the moment we fit four two-dimensional pairs of scale. The measurements are correlated, because the same galaxies and their neighbours can appear in all four pairs of scale. A larger data set would make it possible to fit a higher (than two-) dimensional parameter space (preferentially a single multi-dimensional space where all scales are independent). However, since the number of galaxies falling into a bin decreases rapidly with the dimension of the histogram (and the noise increases), this only  makes sense for very large numbers of galaxies, or otherwise broader binning is required.
\item Constraints on galaxy evolutionary processes can be better constrained by examining how multiple, complimentary galaxy properties (derived star formation rates or spectral indices tracing stellar populations of different ages, colours, morphology etc) depend simultaneously on multiscale density and galaxy mass. This expansion of the parameter space can only be contemplated with sufficient statistics, and once the methodology is well understood in a simpler case such as the one presented here.

\end{itemize}

These improvements are at the current stage either impossible (e.g. neither is there a larger high-resolution simulation, nor a larger volume-limited spectroscopic sample of galaxies with a representative population mix and a large sampling rate available), or far beyond the scope of this paper, which has the main purpose of introducing the method, proving that it is able to give stable results and illustrating complimentary information from different scales. Possible improvements as well as applications to higher redshifts (e.g. using the VIPERS survey) can be addressed in the future.

\section{Summary and Conclusion}\label{Conclusion}
In the framework of the HOD description of the spatial galaxy distribution we have developed a model to describe the red fraction of central and satellite galaxies as a function of halo  mass. This is constrained via a novel method -- the number and red fraction of observed SDSS galaxies as a function of the density measured on multiple scales -- the {\it ``multiscale density framework''} as first introduced by WZB10. We limit our analysis to galaxies in the SDSS DR8 (DR7 spectroscopy) in the range $-21.5\leq M_r \leq -20.0$ (for which there are enough galaxies to bin the sample finely in luminosity and density on two independent scales simultaneously.)

We use the Millenium Simulation (scaled to $\sigma_8\approx0.8$) to provide a realistic set of halos and subhalos with appropriate clustering statistics and peculiar velocities. Halos are populated with central and satellite galaxies: the number of galaxies per halo is picked randomly from a poisson distribution with a given expectation value which is taken from a functional parametrisation of halo mass. Satellites are first assigned to subhalos, then (if there are too few subhalos) the remainder are distributed according to a NFW profile, with peculiar velocities randomly drawn from a halo mass dependent velocity dispersion. 

HOD parameters are inferred from a previous measurement by \citet{Zehavi11} and are used to model  histograms of neighbour galaxies, binned in four separate pairs of scale. Each pair includes an inner aperture and outer annulus (with an inner radius corresponding to the radius of the inner aperture).  The  neighbours are selected within each aperture / annulus and within $\pm 1000 kms^{-1}$ and $M_r\leq -20.0$. Complementary scales are selected with inner and outer projected radii for the four pairs of scales (in Mpc) equal to $[r\leq0.5, 0.5<r\leq1], [r\leq1., 1<r\leq2] , [r\leq2, 2<r\leq3], [r\leq3, 3<r\leq5]$. 

The central galaxies of the select sample live in halos with mass $M\ga 1.1\times 10^{11}$, the expectation value $\ncenexp$ reaches its maximum (unity) at $M\sim 5.7\times10^{11}$, and then decreases again, as the central galaxies of more massive halos have absolute magnitude $M_r<-21.5$. Satellites are found predominantly in more massive halos and, while less common overall, dominate the whole multiscale density parameter space other than the lowest density bins.

We examine the fraction of red galaxies. This is parameterized in the data as by WZB10 and first by \citet{Baldry04} using a double gaussian fit to the (u-r) colour distribution. Given our limited halo mass range and dynamic range in density of central galaxies, we assume a constant red central fraction. With the much greater dynamic range in halo mass and density of satellites, we fit the satellite red fraction using a simple symmetric (hyperbolic tangent) function with only two free parameters -- a transition mass $M_t$ and a slope $\Gamma$. For a grid in these parameters, we fit the red fraction as a function of density for our four pairs of scales using a $\chi^2$ minimization technique. We calculated the full covariance matrix for each model we tested from the Millennium simulation, which we split into 64 independent boxes. Instead of fitting the fractions, we assumed that the HOD reproduces the total numbers of galaxies exactly, so we can fit $N_{red}$ and $N_{blue}$, and take not only correlations between different bins, but also between red and blue galaxies into account.

The shape of the  $\chi^2$ distribution is slightly different for each of our four combinations of scale. The 68\% confidence areas of all four overlap more or less only at one point, with $M_t \sim 10^{14.2}\Msol$, and $\Gamma\approx6.35$. However, in all cases we find that the red satellite galaxy fraction rises very slowly with halo mass, while the transition mass is $M_t \ga 10^{13.8}\Msol$. This means that while high red satellite fractions only occur in galaxy clusters ($M \gtrsim 10^{14}\Msol$) there is an enhanced red fraction over the field (central or low halo mass) value down to at least $M \sim 10^{12}\Msol$, consistent with group versus field comparisons. While the transition mass is quite well constrained just by the need to get the total red fraction correct, $\Gamma$ is mostly constrained by the overall gradient in red fraction versus density on different scales. In the case of a sharp transition at the transition mass in the red satellite fraction, the gradient in red fraction with density would be much steeper than observed, and can be ruled out.

The fact that the $\chi^2$ distributions are different for the four combinations of scales, in particular that the one on the smallest scales deviates from the others, can well be explained by the lack of colour segregation within the haloes in our model. If in fact red galaxies are concentrated towards the centres of the haloes, the red fraction on very small scales (smaller or comparable to the halo sizes) created by the model would be larger. On larger scales this effect would not be measured, and indeed the larger scales yield better fits even without colour segregation. If we assume that the real maximum red satellite fraction on small scales is even larger than the value we can measure ($f^{red}_{max}=0.9$), e.g. because blue field galaxies can be scattered into the cylinders, we derive nevertheless a very similar course of the function within the range of observed halo masses.

This example demonstrates the power of the method: the analysis of the galaxy properties on different scales (and combinations thereof) give insight into different aspects of how galaxies populate dark matter haloes, and the influence of the environment on those scales on their properties. At the same time by investigating the limitations of the model it is possible to understand which observations cannot be described by a simple HOD model in which galaxy properties depend only on halo mass (e.g. colour segregation, or satellite quenching in infall regions). 

A constant fraction of red satellites cannot be ruled out, but cannot explain the high red fractions in high density regions on the smallest scales. Again this is a hint that there is a decrease of the red satellite fraction with galacto-centric distance.

We also examined the difference between a constant red {\it central} fraction and a sharp transition in the red central fraction (such that the total red central fraction is conserved). Within our limited range of halo mass, this would be possible with a transition at $M_{t,cen}=9.035\cdot 10^{11} M_\odot$. This would be consistent with a scenario suggested by some authors (e.g. \citealp{DekelBirnboim06,Cattaneo06,KhochfarOstriker08}) in which newly accreted gas onto the halo is shock-heated above a critical halo mass, and the cooling of that gas onto a central galaxy is inefficient. The extreme sharp transition solution can be ruled out because regions of low densities on small scales are populated by isolated galaxies which are the central galaxies of halos with mass below $M_{t,cen}$. Observations tell us that there are indeed $\sim 38\%$ red galaxies in these regions, while with a sharp transition at $M_{t,cen}$ there would be none.

\appendix
\section{The HOD parameters}\label{Appendix}
\begin{figure*}
\centerline{\psfig{figure=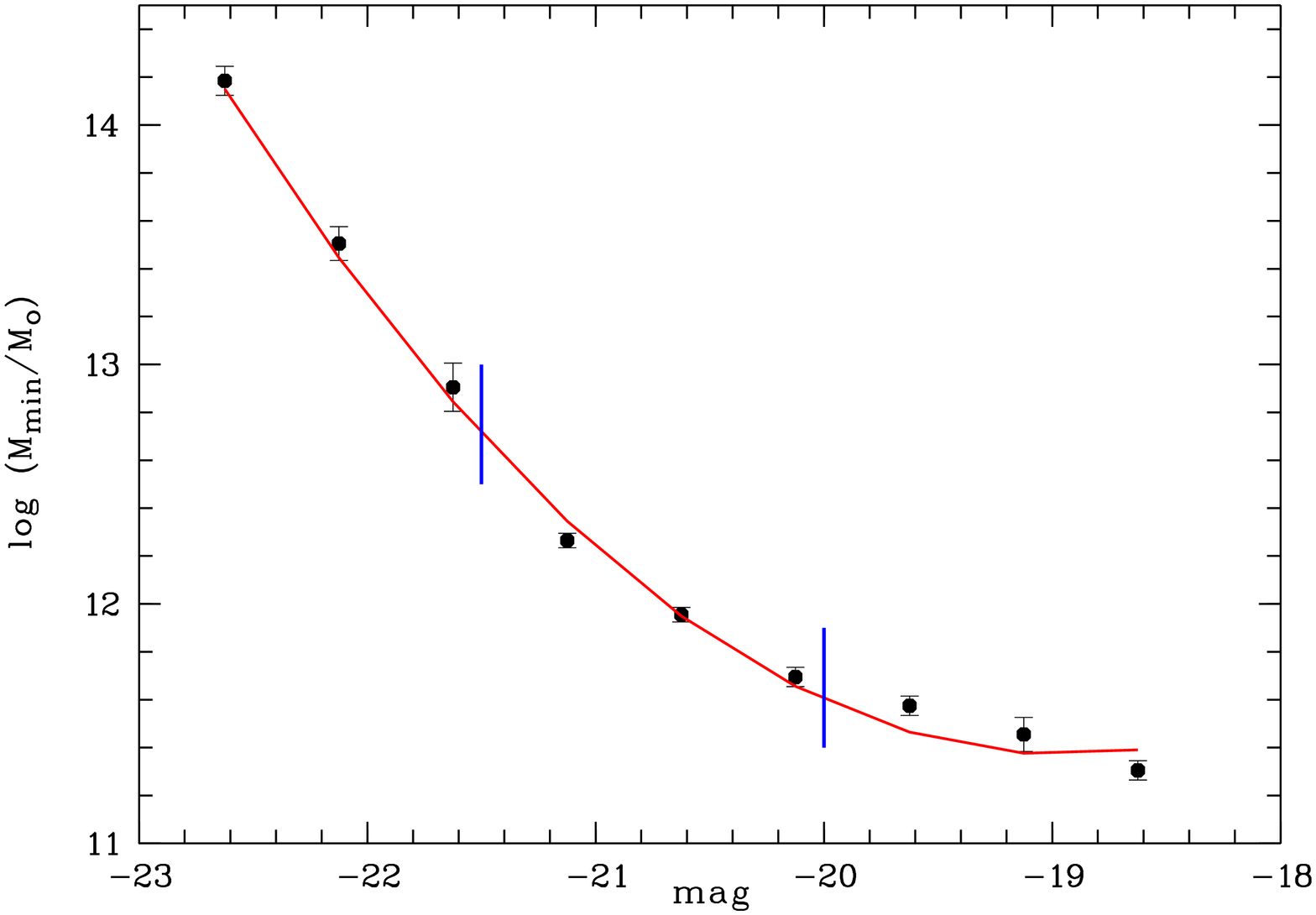,width=6.3cm,angle=270}}
\centerline{\psfig{figure=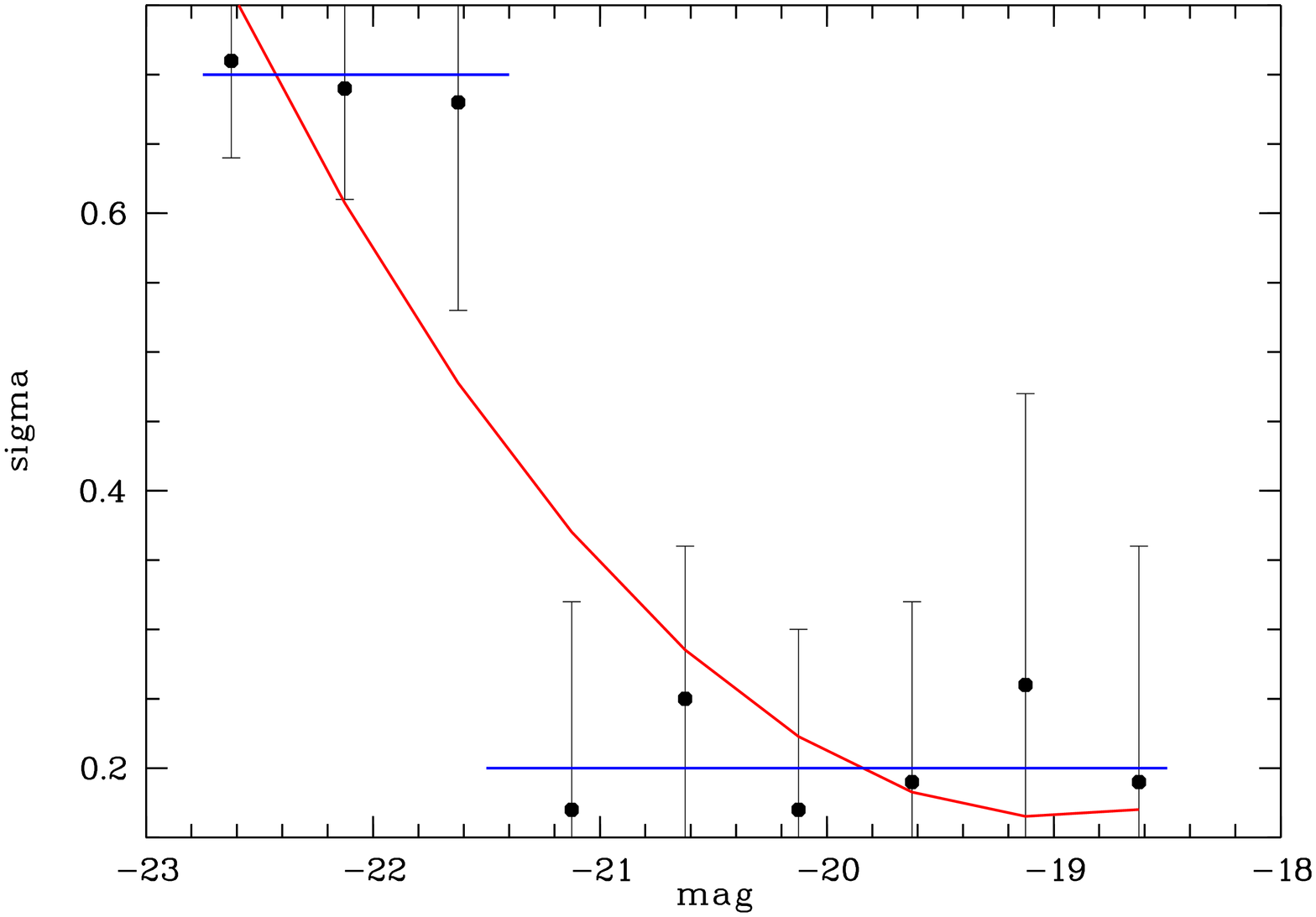,width=6.3cm,angle=270}}
\centerline{\psfig{figure=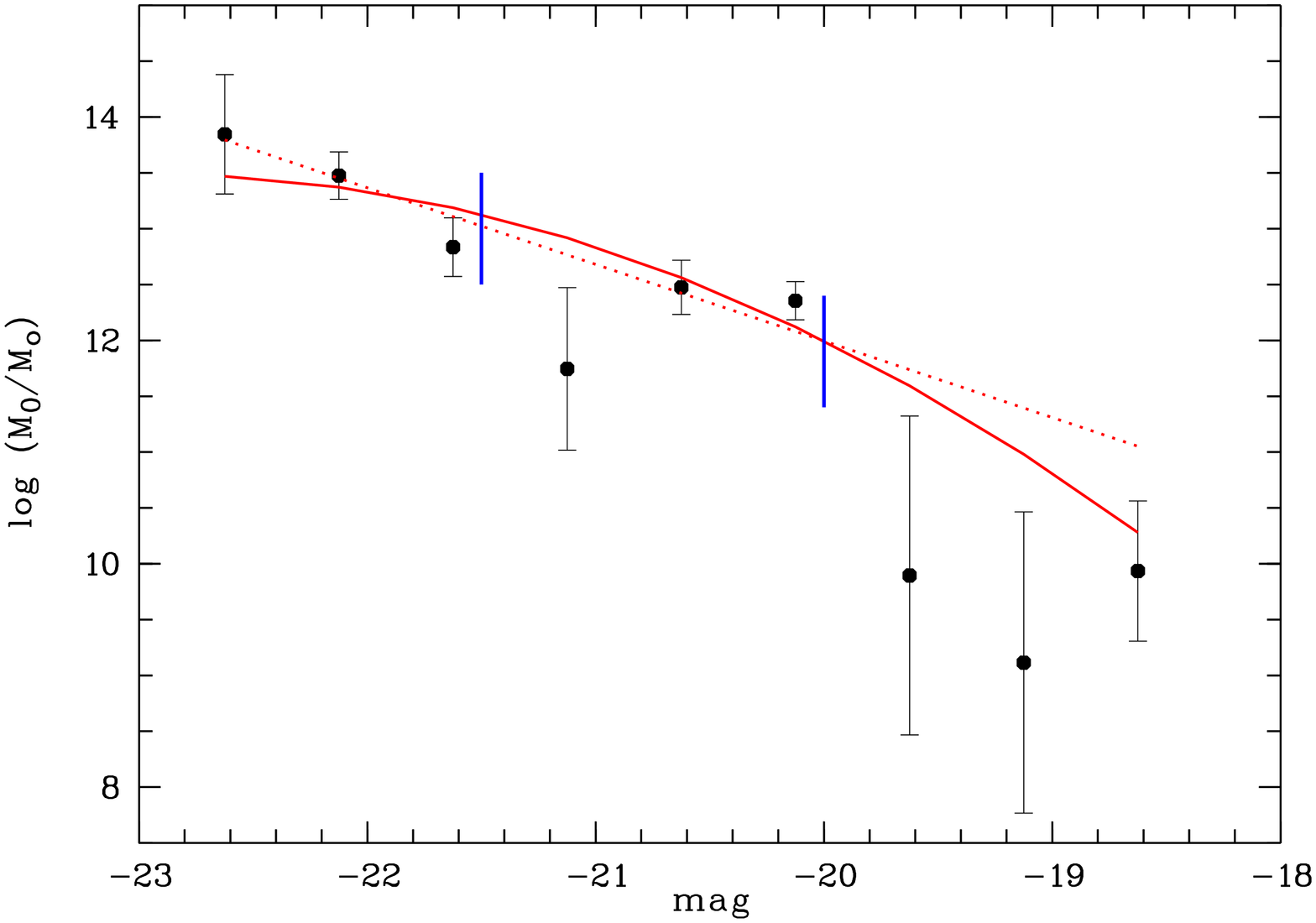,width=6.3cm,angle=270}}
\centerline{\psfig{figure=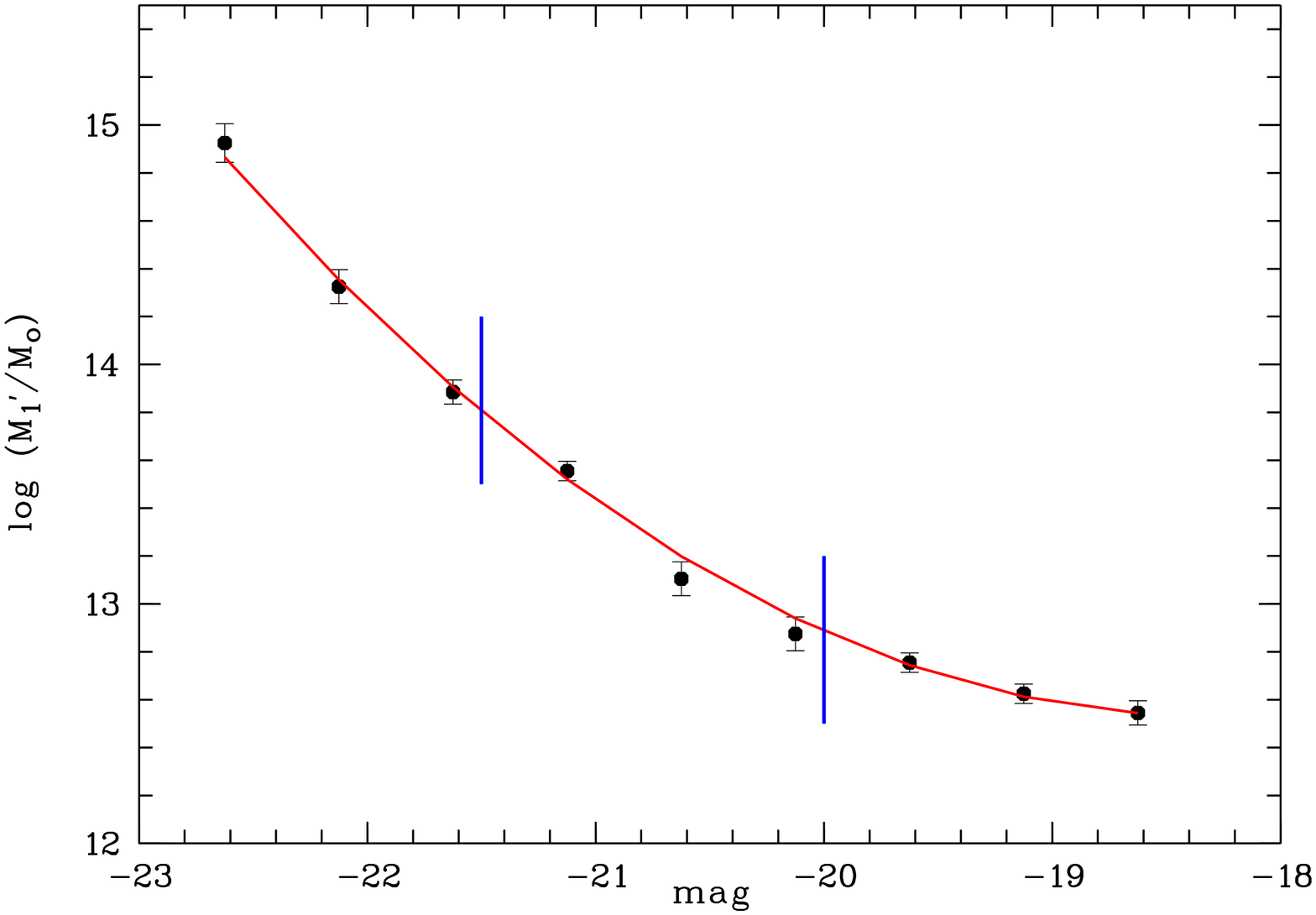,width=6.3cm,angle=270}}
\centerline{\psfig{figure=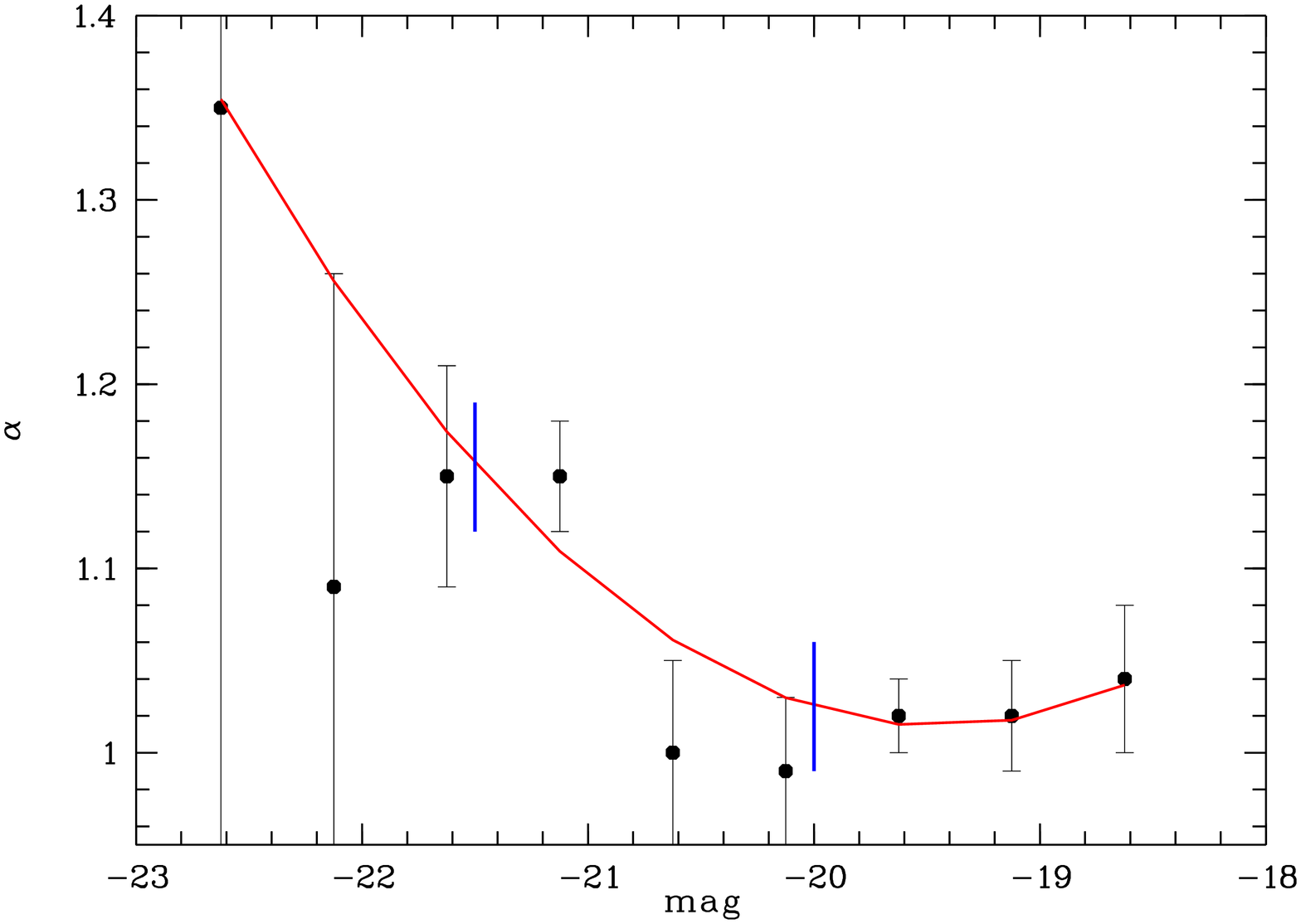,width=6.3cm,angle=270}}
\caption[ ]{The HOD parameters determined by \citet{Zehavi11} and our fits (red lines). Top to bottom: $\log M_{min}$, $\sigma$, $\log M_0$, $\log M^\prime_1$, and $\alpha$. Since we use $h=0.75$, masses and luminosities given in their table 3 have been shifted accordingly. The vertical blue lines indicate the lower luminosity limits of the select and the bright sample. The horizontal blue lines in the plot showing the parameter $\sigma$ indicate the mean values in these ranges (which we assumed instead of the values we would derive from the polynomial fit. The dotted red line in the plot showing $\log M_0$ is a fit with a straight line, the values which would be derived from this fit would not be very different from those derived from the polynomial fit, however, its $\chi^2$ is slightly larger.}\label{Zehaviparams}
\end{figure*}

In order to populate the dark matter haloes of the Millennium simulation with galaxies, we use the HOD description by \citet{Zehavi11}, see Equations \ref{HODcen} and \ref{HODsat}. In their paper, they use the projected correlation function of volume-limited samples of galaxies constructed from the main sample of the SDSS DR7 to derive HOD parameters as a function of luminosity (and color). Since the upper redshift limits are different for different luminosity thresholds in their analysis, our samples do not comprise exactly the same galaxies, but as the authors show, cosmic variance is not a problem for their investigation of the HOD parameters as a function of luminosity, even though each measurement has been carried out in a different volume (nor can redshift evolution be, considering the maximum redshift limit of $z_{max}=0.3$). Hence we can use their measurement of the HOD parameters to describe the multiscale densities in our sample. Their halo masses are in units of $h^{-1} M_\odot$ and their absolute luminosities calculated assuming $h=1$, while we use $h=0.75$ throughout the analysis. Therefore we shift the magnitudes and parameters accordingly, then fit them, in order to derive the appropriate values for our full and bright samples. The values and fits are shown in Fig.~\ref{Zehaviparams}.  While $\log M_{min}$ and $\log M^\prime_1$ can be well described by a third order polynomial, the best choice of a parametric form of $\sigma$, $\log M_0$ and $\alpha$ is not so obvious. In the case of $\sigma$ it seems that the value is constant and high ($\sigma\approx 0.7$) for bright galaxies ($M<-21.4$) and is constant and low ($\sigma\approx 0.2$) for fainter galaxies. We adopt these values, which are indicated by blue lines in Fig.~\ref{Zehaviparams}, for our bright and full samples.  $\log M_0$  looks as if a fit with a straight line would be appropriate. However, taking the error bars into account, the resulting fit (red line in Fig.~\ref{Zehaviparams}) is not satisfying. The third order polynomial we use instead (blue line) has a slightly smaller $\chi^2$. Although the value of $\alpha$ is not well determined for very bright galaxies, again a parametrisation with a third order polynomial seems not to be an unreasonable choice.

\begin{figure*}
\centerline{\psfig{figure=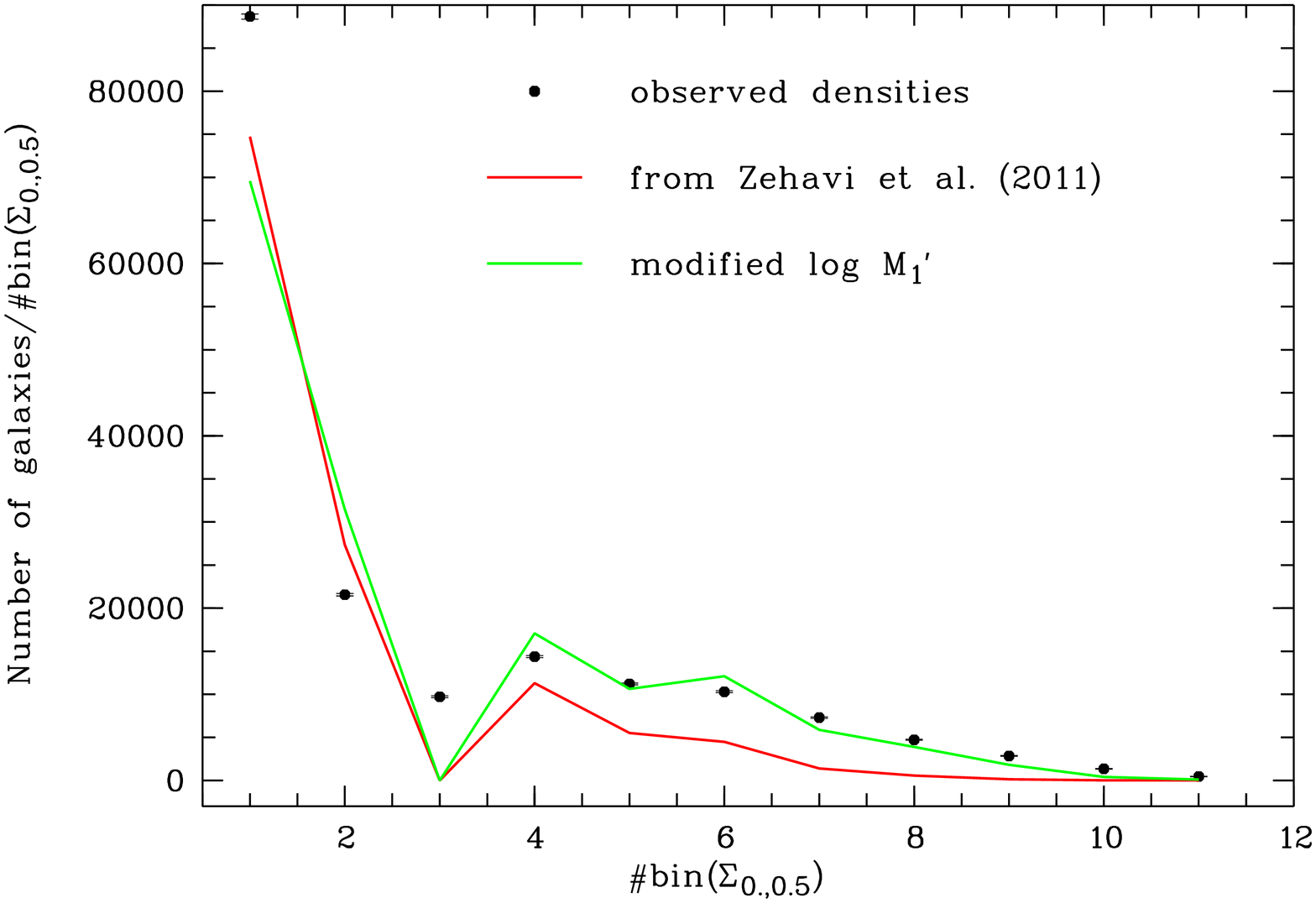,angle=270,clip=t,width=9.cm}}
\caption[ ]{The number of galaxies per density bin $\Sigma_{0,0.5}$: black dots are the observed densities in the full sample, the red line is the result of the mock when we use the HOD parameters we determined from the measurement of \citet{Zehavi11}, the red line is for $\log M^\prime_1=12.6$ instead of $\log M^\prime_1=12.88$. The dip at bin \#3 is due to the ``quantisation'' of the numbers of neighbours.}\label{dens_Zehavi}
\end{figure*}

From these fits we find for the full sample ($M_r\leq-20$): $\log M_{min}=11.60$, $\sigma=0.2$, $\log M_0=12.00$, $\log M^\prime_1=12.88$, and $\alpha=1.02$; and for the bright sample ($M_r\leq -21.5$): $\log M_{min}=12.71$, $\sigma=0.7$, $\log M_0=13.13$, $\log M^\prime_1=13.80$, and $\alpha=1.16$. However, we found that if we use exactly these parameters, we do not reproduce our observed densities in the full sample, as can e.g. be seen from Fig.~\ref{dens_Zehavi}: the plot shows the number of observed galaxies per density bin $\Sigma_{0,0.5}$ in comparison with the densities which are measured from the simulated galaxies when we use the above HOD parameters to populate the Millennium.\footnote{The dip at the third bin, where the observed data point is low and the model zero, occurs because in the simulation the number of neighbours is always integer, and these do not fall into all logarithmically spaced bins, while the observed numbers have been completeness corrected (see Section \ref{completenesscorrection}), hence some (fewer) galaxies cross into these columns/rows (see also Section \ref{Multiscales}).} In particular on high densities the observed numbers of galaxies falling into the bins are significantly higher by a factor which increases with density as compared to the ones predicted by the HOD parameters derived from the fits. We achieve a much better description of the measured numbers of galaxies particularly at high densities if we use $\log M^\prime_1=12.6$ instead of $\log M^\prime_1=12.88$, the effect of which is to create more satellite galaxies and hence more galaxies at higher densities.

The reason our measurement demands a slightly lower value of $\log M^\prime_1$ may be cosmic variance (we are not investigating exactly the same volume as \citet{Zehavi11}), the way we populate the haloes (one per subhalo and the remainder to follow a NFW profile in a spherical halo), or the methodology (while the correlation function is a convolution of the overdensity field with itself (an expectation value), information about the distribution of local densities is lost, which could make it more difficult to distinguish the two cases presented in Fig.~\ref{dens_Zehavi}). In any case, in order for the simulated galaxies to describe the observations better, we used this altered parameter for the analysis of the dependence of the red fraction on the multiscale densities throughout this paper.

\begin{figure*}
\centerline{\psfig{figure=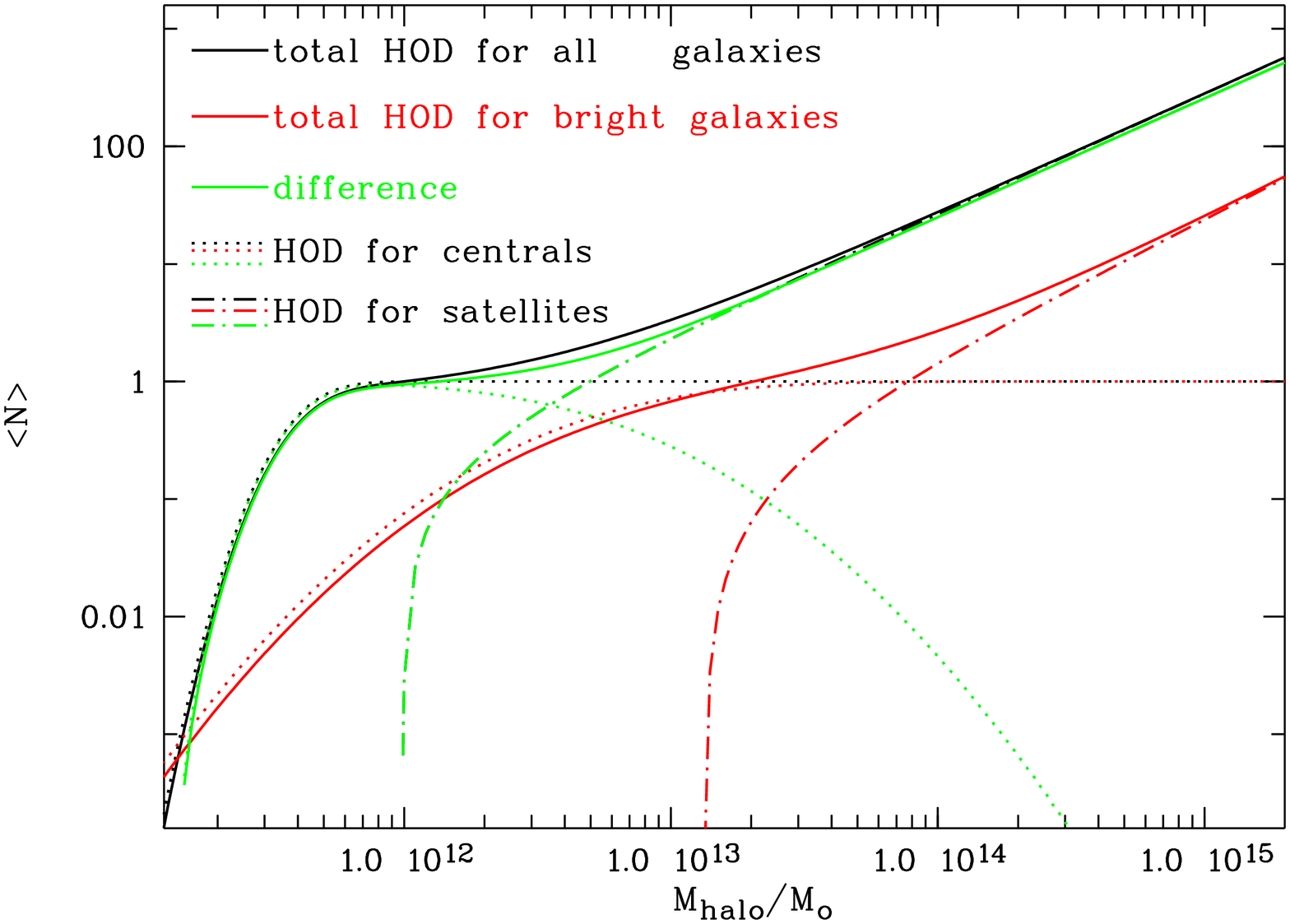,angle=270,clip=t,width=9.cm}}
\caption[ ]{The HOD for the full sample of galaxies (black solid line), for the bright sample (red solid line), and the resulting HOD for the select sample (the difference between the two, green solid line). The HODs for the central and satellite galaxies are also shown (dotted and dot-dashed lines, respectively).}\label{HODplot}
\end{figure*}

Fig.~\ref{HODplot} shows the corresponding HODs for the full sample (black solid line) and the bright sample (red solid line). Contributions of central and satellite galaxies are shown by the dotted and dot-dashed lines, respectively. We define the difference between the full and bright samples (representing galaxies in the luminosity range $-21.5\leq M_r\leq -20.0$) as the {\it select sample}. This is shown as the green solid line (and dashed and dot-dashed lines for the central and satellite contributions).

\section*{Acknowledgments}

We are greatly indepted to our anonymous referee for several points which had not received sufficient attention in the original manuscript. This led to a substantial improvement of the paper.

We also thank Xiaohu Yang for providing his unpublished DR7 version of the SDSS group catalogues.

DJW is supported by the DFG grant WI 3871/1-1. SP and DJW also thank the Max-Planck Gesellschaft. The Dark Cosmology Centre is funded by the Danish National Research Foundation.

Funding for the SDSS and SDSS-II has been provided by the Alfred P. Sloan Foundation, the Participating Institutions, the National Science Foundation, the U.S. Department of Energy, the National Aeronautics and Space Administration, the Japanese Monbukagakusho, the Max Planck Society, and the Higher Education Funding Council for England. The SDSS Web Site is {\tt http://www.sdss.org/}. The SDSS is managed by the Astrophysical Research Consortium for the Participating Institutions. The Participating Institutions are the American Museum of Natural History, Astrophysical Institute Potsdam, University of Basel, University of Cambridge, Case Western Reserve University, University of Chicago, Drexel University, Fermilab, the Institute for Advanced Study, the Japan Participation Group, Johns Hopkins University, the Joint Institute for Nuclear Astrophysics, the Kavli Institute for Particle Astrophysics and Cosmology, the Korean Scientist Group, the Chinese Academy of Sciences (LAMOST), Los Alamos National Laboratory, the Max-Planck-Institute for Astronomy (MPIA), the Max-Planck-Institute for Astrophysics (MPA), New Mexico State University, Ohio State University, University of Pittsburgh, University of Portsmouth, Princeton University, the United States Naval Observatory, and the University of Washington.

Funding for SDSS-III has been provided by the Alfred P. Sloan Foundation, the Participating Institutions, the National Science Foundation, and the U.S. Department of Energy Office of Science. The SDSS-III web site is {\tt http://www.sdss3.org/}. SDSS-III is managed by the Astrophysical Research Consortium for the Participating Institutions of the SDSS-III Collaboration including the University of Arizona, the Brazilian Participation Group, Brookhaven National Laboratory, University of Cambridge, Carnegie Mellon University, University of Florida, the French Participation Group, the German Participation Group, Harvard University, the Instituto de Astrofisica de Canarias, the Michigan State/Notre Dame/JINA Participation Group, Johns Hopkins University, Lawrence Berkeley National Laboratory, Max Planck Institute for Astrophysics, Max Planck Institute for Extraterrestrial Physics, New Mexico State University, New York University, Ohio State University, Pennsylvania State University, University of Portsmouth, Princeton University, the Spanish Participation Group, University of Tokyo, University of Utah, Vanderbilt University, University of Virginia, University of Washington, and Yale University.

\label{lastpage}

\end{document}